\documentclass[3p]{elsarticle}

\usepackage{amsmath}
\usepackage{amssymb}
\usepackage{xspace}

\usepackage{graphicx}
\usepackage{xcolor}

\graphicspath{{./}}

\usepackage{aas_macros}
\biboptions{sort&compress}
\usepackage{hyperref}
\usepackage{lineno}

\makeatletter
\def\l@subsubsection#1#2{}
\makeatother

\newcommand{\Fermic}{\emph{Fermi}}
\newcommand{\Fermi}{\Fermic\xspace}
\newcommand{\FermiLATc}{\Fermic~LAT}
\newcommand{\FermiLAT}{\FermiLATc\xspace}

\newcommand{\gr}{\mbox{$\gamma$ ray}\xspace}
\newcommand{\grs}{\mbox{$\gamma$ rays}\xspace}
\newcommand{\gray}{\mbox{$\gamma$-ray}\xspace}
\newcommand{\grays}{\mbox{$\gamma$-rays}\xspace}

\usepackage{color}
\newcommand{\newText}[1]{#1}

\newcommand{\sigmav}{\ensuremath{\langle \sigma v \rangle}\xspace}
\newcommand{\mchi}{\ensuremath{m_\chi}\xspace}
\newcommand{\bb}{\ensuremath{b\bar{b}}\xspace}
\newcommand{\tautau}{\ensuremath{\tau^+\tau^-}\xspace}
\newcommand{\ee}{\ensuremath{e^+e^-}\xspace}
\newcommand{\mumu}{\ensuremath{\mu^+\mu^-}\xspace}

\newcommand{\beff}{\ensuremath{b_{\mathrm{eff}}}\xspace}
\newcommand{\nsig}{\ensuremath{n_{\mathrm{sig}}}\xspace}
\newcommand{\fsig}{\ensuremath{f_{\mathrm{sig}}}\xspace}
\newcommand{\fstat}{\ensuremath{f_{\mathrm{stat}}}\xspace}
\newcommand{\fsyst}{\ensuremath{f_{\mathrm{syst}}}\xspace}

\newcommand{\unit}[1]{\ensuremath{\mathrm{\,#1}}\xspace}
\newcommand{\meV}{\unit{meV}}
\newcommand{\eV}{\unit{eV}}
\newcommand{\MeV}{\unit{MeV}}
\newcommand{\GeV}{\unit{GeV}}
\newcommand{\TeV}{\unit{TeV}}
\newcommand{\degree}{\unit{^{\circ}}}

\newcommand{\cm}{\unit{cm}}

\newcommand{\kpc}{\unit{kpc}}
\newcommand{\Gpc}{\unit{Gpc}}
\newcommand{\fb}{\unit{fb}}
\newcommand{\second}{\unit{s}}

\newcommand{\photons}{\unit{ph}}
\newcommand{\photon}{\unit{ph}}

\newcommand{\Msolar}{\unit{M_\odot}}
\newcommand{\Msun}{\Msolar}
\newcommand{\Lsolar}{\unit{L_\odot}}
\newcommand{\Lsun}{\Lsolar}

\newcommand{\roughly}{\ensuremath{ {\sim}\,} }

\newcommand{\evtclass}[1]{\texttt{#1}}
\newcommand{\irf}[1]{\texttt{#1}}

\newcommand{\acronym}[3][]{\noindent\makebox[70pt]%
  {\ifx\\#1\\#2\else#1\fi\hfill}#3\\}

\begin{document}

% Line number every fifth line
\modulolinenumbers[5]
%\linenumbers

% Title page 
\title{Sensitivity Projections for Dark Matter Searches with the \Fermi Large Area Telescope}

\date{\today}

\author[add2]{E.~Charles}\ead{echarles@slac.stanford.edu} 
\author[add3,add4]{M.~S\'anchez-Conde} 
\author[add3,add4]{B.~Anderson}
\author[add7]{R.~Caputo} 
\author[add10,add11]{A.~Cuoco}
\author[add2]{M.~Di~Mauro} 
\author[add12]{A.~Drlica-Wagner}
\author[add14,add15]{G.~A.~Gomez-Vargas}
\author[add3,add4]{M.~Meyer}
\author[add21]{L.~Tibaldo} 
\author[add2]{M.~Wood} 
\author[add22,add23]{G.~Zaharijas} 
\author[add24,add3,add4]{S.~Zimmer}
\author[add1]{M.~Ajello}
\author[add2]{A.~Albert}
\author[add5,add2]{L.~Baldini}
\author[add6]{K.~Bechtol} 
\author[add2]{E.~D.~Bloom} 
\author[add2]{F.~Ceraudo} 
\author[add9]{J.~Cohen-Tanugi}
\author[add2]{S.~W.~Digel}
\author[add13]{J.~Gaskins}
\author[add16]{M.~Gustafsson}
\author[add17,add18]{N.~Mirabal}
\author[add19,add20]{M.~Razzano} 

\address[add2]{W. W. Hansen Experimental Physics Laboratory, Kavli Institute for Particle Astrophysics and Cosmology, Department of Physics and SLAC National Accelerator Laboratory, Stanford University, Stanford, CA 94305, USA}
\address[add3]{Department of Physics, Stockholm University, AlbaNova, SE-106 91 Stockholm, Sweden}
\address[add4]{The Oskar Klein Centre for Cosmoparticle Physics, AlbaNova, SE-106 91 Stockholm, Sweden}
\address[add7]{Santa Cruz Institute for Particle Physics, Department of Physics and Department of Astronomy and Astrophysics, University of California at Santa Cruz, Santa Cruz, CA 95064, USA}
\address[add10]{RWTH Aachen University, Institute for Theoretical Particle Physics and Cosmology, (TTK),, D-52056 Aachen, Germany}
\address[add11]{Istituto Nazionale di Fisica Nucleare, Sezione di Torino, I-10125 Torino, Italy}
\address[add12]{Center for Particle Astrophysics, Fermi National Accelerator Laboratory, Batavia, IL 60510, USA}
\address[add14]{Instituto de Astrof\'isica, Facultad de F\'isica, Pontificia Universidad Cat\'olica de Chile, Casilla 306, Santiago 22, Chile}
\address[add15]{Istituto Nazionale di Fisica Nucleare, Sezione di Roma ``Tor Vergata", I-00133 Roma, Italy}
\address[add21]{Max-Planck-Institut f\"ur Kernphysik, D-69029 Heidelberg, Germany}
\address[add22]{Istituto Nazionale di Fisica Nucleare, Sezione di Trieste, and Universit\`a di Trieste, I-34127 Trieste, Italy}
\address[add23]{Laboratory for Astroparticle Physics, University of Nova Gorica, Vipavska 13, SI-5000 Nova Gorica, Slovenia}
\address[add24]{University of Geneva, D\'epartement de physique nucl\'eaire et corpusculaire (DPNC), 24 quai Ernest-Ansermet,
CH-1211 Gen\`eve 4, Switzerland}
\address[add1]{Department of Physics and Astronomy, Clemson University, Kinard Lab of Physics, Clemson, SC 29634-0978, USA}
\address[add5]{Universit\`a di Pisa and Istituto Nazionale di Fisica Nucleare, Sezione di Pisa I-56127 Pisa, Italy}
\address[add6]{Dept.  of  Physics  and  Wisconsin  IceCube  Particle  Astrophysics  Center, University  of  Wisconsin, Madison,  WI  53706, USA}
\address[add9]{Laboratoire Univers et Particules de Montpellier, Universit\'e Montpellier, CNRS/IN2P3, F-34095 Montpellier, France}
\address[add13]{GRAPPA, University of Amsterdam, Science Park 904, 1098XH Amsterdam, Netherlands}
\address[add16]{Georg-August University G\"ottingen, Institute for theoretical Physics - Faculty of Physics, Friedrich-Hund-Platz 1, D-37077 G\"ottingen, Germany}
\address[add17]{NASA Goddard Space Flight Center, Greenbelt, MD 20771, USA}
\address[add18]{NASA Postdoctoral Program Fellow, USA}
\address[add19]{Istituto Nazionale di Fisica Nucleare, Sezione di Pisa, I-56127 Pisa, Italy}
\address[add20]{Funded by contract FIRB-2012-RBFR12PM1F from the Italian Ministry of Education, University and Research (MIUR)}

\begin{abstract}
The nature of dark matter is a longstanding enigma of
physics;  it may consist of particles beyond the Standard Model that
are still elusive to experiments.  Among indirect search
techniques, which look for stable products from the annihilation or decay of dark matter particles, 
or from axions coupling to high-energy photons, observations of the \gray sky have
come to prominence over the last few years, because of the excellent 
sensitivity of the Large Area Telescope (LAT) on the {\it Fermi Gamma-ray 
Space Telescope} mission.   The LAT energy range from 20\MeV to above 300\GeV
is particularly well suited for searching for products
of the interactions of dark matter particles.  In this report we
describe methods used to search for evidence of dark matter
with the LAT, and review the status of searches performed with 
up to six years of LAT data.   We also discuss the factors that determine the
sensitivities of these searches, including the magnitudes of the signals 
and the relevant backgrounds, considering both statistical and 
systematic uncertainties.  We project the expected sensitivities
of each search method for 10 and 15~years of LAT data taking.
In particular, we find that the sensitivity of searches targeting
dwarf galaxies, which provide the best limits currently,
will improve faster than the square root of observing time.   Current
LAT limits for dwarf galaxies using six years of data reach the thermal relic level
for masses up to 120\GeV for the \bb annihilation channel 
for reasonable dark matter density profiles.  With projected discoveries of additional dwarfs, these
limits could extend to about 250\GeV.  With as
much as 15~years of LAT data these searches would be
sensitive to dark matter annihilations at the thermal relic cross section 
for masses to greater than 400\GeV (200\GeV) in the
\bb (\tautau) annihilation channels.
\end{abstract}

\maketitle

\tableofcontents

\section{Introduction}
\label{sec:intro}

Since its launch in June 2008, the \Fermi~{\it Gamma-ray Space Telescope} 
(\Fermi) has surveyed the high-energy sky from 
20\MeV to above 300\GeV with its main instrument, 
the Large Area Telescope (LAT)~\cite{Atwood:2009ez}. 
The LAT has opened new experimental windows into physics beyond the Standard Model.
In this report we consider one
such window, the indirect search for dark matter via observations of high-energy \grs. 
Annihilation or decay of massive dark matter (DM) particles could produce \grs, and 
the coupling of light axions or axion-like particles (ALPs) to photons in a magnetic field would alter the \gray spectra of distant sources.  For these reasons, \gray observations could provide compelling indirect evidence for particle DM.
Analyses of LAT data have provided 
hints of a potential signal of DM annihilation from the Galactic 
center~\cite[e.g.,][]{2011PhLB..697..412H,2014PhRvD..90b3526A,2014arXiv1402.6703D}
(and see \S\ref{sec:dm_milky_way_current}) and in observations of other regions have set the strongest
constraints on DM annihilation cross sections~\cite{2015PhRvL.115w1301A}.

Several comprehensive and detailed reviews of the
particle physics and particle astrophysics related to DM 
searches with the LAT and other instruments have been published 
recently---see, e.g.,
\cite{2010ARA&A..48..495F,2011ARA&A..49..155P,2015JETP..121.1104C},
expanding on earlier summaries, e.g.,
\cite{2005PhR...405..279B}.  In this paper we will 
focus in particular on the details that are relevant for
understanding the sensitivity of indirect DM searches with the LAT data
and quantifying the expected sensitivity gains with additional data taking.

This paper is organized as follows.  In \S\ref{sec:dm_theory} we
summarize the phenomenology of DM searches in \gray data.
In \S\ref{sec:backgrounds} we describe the astrophysical backgrounds 
for these searches, and discuss both the statistical and systematic 
uncertainties arising from those backgrounds. 
In \S\ref{sec:dm_targets} we present the primary astrophysical 
targets for DM searches with the LAT, providing the status of current
searches and estimates of the sensitivity for DM searches projecting
to 10 and 15~years of data taking.   In \S\ref{sec:discussion} we
examine the projected search sensitivities and compare them 
with indirect DM searches based on observations at other energies and
of cosmic rays and neutrinos.  We also compare them with current and
projected constraints on DM derived from underground detector searches (``direct-detection'' DM 
searches) and from searches for new physics at the Large Hadron
Collider  (LHC, ``production'' searches).  We then summarize our 
findings and present the expected sensitivity reach for DM searches with the continuing 
\Fermi mission \S\ref{sec:summary}.  Finally, we provide technical details in 
several appendices.

% LocalWords:  DM axions axion ALPs Hadron Collider LHC

\section{Dark Matter and Dark Matter Particle Candidates}
\label{sec:dm_theory}

Overwhelming evidence indicates that the observed structure of the Universe 
cannot be explained by the Standard Model of particle 
physics alone.  In particular, measurements of galactic rotation curves~\cite[e.g.,][]{Rubin:1980zd} 
and galaxy cluster dynamics~\cite[e.g.,][]{Zwicky:1933gu}, precision measurements of the 
cosmic microwave background~\cite[e.g.,][]{Ade:2013zuv,2015arXiv150201589P}, observations of the primordial
abundances of heavy isotopes produced by Big Bang nucleosynthesis~\cite[e.g.,][]{Olive:2003iq},
and other lines of evidence all point to a substantial fraction of
the Universe's energy density being in a form of matter
without significant interaction with the Standard Model particles.
Numerical simulations of large-scale 
structure also support this conclusion; such simulations require
non-relativistic DM in order to be consistent with
observations~\cite[e.g.,][and references therein]{2012AnP...524..507F}.   

From the observational evidence, we infer that DM interacts gravitationally, 
is non-relativistic (i.e., ``cold'') during the formation of large-scale structure~\cite[e.g.,][]{Ade:2013zuv,2015arXiv150201589P}, 
and does not have large scattering cross sections 
with either itself~\cite[e.g.,][]{Markevitch:2003at} or Standard Model particles~\cite[e.g.,][]{Akerib:2013tjd}.  
No particle in the Standard Model meets the requirements. 
We have no other solid experimental or theoretical understanding of the 
fundamental nature of DM.

In the standard cosmology, because DM is cold, its distribution is structured over
a wide range of mass scales.   Small, dense DM clumps
form first and later merge to form larger structures.  
Each of these structures, such as the DM {\it halo} in which the Milky Way resides,
retains a population of less massive {\it subhalos}.  In this report we 
consider searches that target DM halos and subhalos at distance scales
from \kpc (e.g., \S\ref{sec:dm_milky_way}) to \Gpc (\S\ref{sec:dm_cosmo}).

Many particles have been posited as DM candidates.  Here we
focus in particular on weakly interactive massive particles (WIMPs) and
axions or ALPs, as these are well-motivated 
candidates that may leave detectable signatures in the LAT data.

\subsection{Weakly Interacting Massive Particles}
\label{sec:dm_wimps}

A heavy ($\mchi \gtrsim 1$\GeV) particle with a finite
annihilation cross section into Standard Model particles
would be theoretically well motivated as DM.
The canonical example of such DM is a non-relativistic thermal 
relic that froze out of equilibrium with the particle bath 
in the early Universe.  While significant annihilation would cease during 
freeze-out, if the DM pair annihilation is due to an $s$-wave process and 
therefore velocity independent, low rates of annihilation would 
continue to the present day. 
Such a DM particle could yield the measured DM energy 
density today, $\Omega h^2 = 0.1199\pm 0.0027$~\cite{Ade:2013zuv,2015arXiv150201589P}, provided the
annihilation cross section averaged over the velocity distribution is near 
$\sigmav \sim 3\times 10^{-26}$~cm$^3$~s$^{-1}$~\cite{2012PhRvD..86b3506S}\footnote{Unless 
  the $s$-wave amplitude is $\lesssim 10$ times smaller than the $p$-wave amplitude,
  the current annihilation rate should be within an order of magnitude of the
  required rate for thermal relic DM~\cite{2010PhRvD..82i5007C}.}.
This can be realized in models with supersymmetry~\cite[e.g.,][]{2015PhR...555....1B,2012AnP...524..479B}, 
though other models can also work~\cite[e.g.,][]{Feng:2008ya}. 

However, relatively simple modifications of the canonical thermal relic theory 
can result in present-day annihilation cross sections that differ by orders 
of magnitude from the standard assumption of 
$\sigmav \sim 3\times 10^{-26}$~cm$^3$~s$^{-1}$~\cite{ArkaniHamed:2008qn}.
Furthermore, in some theoretical frameworks, the DM particles
may decay with lifetimes that are large compared to the age 
of the Universe, at rates that may be observable 
(see~\cite{2013IJMPA..2830040I} 
for a recent review).  
Therefore in our searches we do not presuppose anything about the 
intensity of the expected DM signals.

The end products of such DM annihilation can be searched for as signals 
correlated with astrophysical objects known to have large DM content.  
Because \grs are not deflected by magnetic fields, regions where 
DM is densest are directions with the greatest intensity
of \grs from DM annihilations. 
However, these signals must be measured relative to backgrounds from products of Standard Model 
astrophysical processes (\S\ref{sec:backgrounds}). 

Final states including \grs are a generic expectation of
DM annihilation and can result from many different annihilation
channels.  Two-body final states (with one or both a photon)
produce \grs that have a characteristic line spectrum (see \S\ref{sec:dm_lines}).
DM may also convert into pairs (or a larger multiplicity) of quarks,
leptons, gluons, or $SU(2)_L$ gauge bosons, 
all of which will decay or hadronize, resulting in a continuum
spectrum of \grs and other stable particles (see \S\ref{sec:dm_spectrum}).
The LAT is also able to measure spectra of electrons~\cite{2009PhRvL.102r1101A,2010PhRvD..82i2004A,2010PhRvD..82i2003A}, 
positrons~\cite{2012PhRvL.108a1103A}, and other cosmic rays, but we
will not discuss sensitivity projections for those measurements in this report.

\subsubsection{Characterizing a WIMP Signal in the Gamma-Ray Sky}
\label{sec:dm_signal}

The \gray flux from DM annihilation (or decay) depends 
on the product of factors related to the particle physics 
and the spatial distribution of the DM.   

\newText{We do not know the mass and annihilation (or decay) channels of DM particles. 
However, if DM annihilates into a pair of Standard Model particles other than
neutrinos, be they gauge bosons, quarks, or charged leptons, then (with the exception of the stable particles
such as $e^{\pm}$), those particles must decay or hadronize. This leads to a cascade of Standard Model 
particles, decaying down to electrons, protons, their anti-partners, and a 
large multiplicity of photons with \gray energies.  Photons are also 
emitted as final-state radiation from charged particles, including \ee
pairs.   Finally, secondary \grs can be emitted as the final state
particles propagate from the interaction point by 
such processes as inverse-Compton scattering (from $e^{\pm}$ interactions
with radiation fields) or pion production (from hadronic interactions
with interstellar gas and dust).} 

The differential flux of
\newText{prompt} \grs from DM annihilation within a solid 
angle $\Delta \Omega$ is typically expressed as:

\begin{equation}\label{eq:general_flux}
\frac{d\phi(\Delta \Omega)}{dE_\gamma} = \left( \frac{x \sigmav}{8\pi}\frac{dN_\gamma}{dE_\gamma}\frac{1}{\mchi^2}\right) \left( \int_{\Delta\Omega} d\Omega \frac{dJ_{\rm ann}}{d\Omega} \right).
\end{equation}

\noindent where $x=1$ if DM is its own antiparticle and $x=1/2$ 
if it is not (in this paper we will assume that $x=1$), \sigmav is 
the velocity-averaged annihilation cross section, $\mchi$ is the WIMP mass, 
and $dN_\gamma/dE_\gamma$ is the differential spectrum of \grs from 
annihilation of a pair of DM particles (see
\S\ref{sec:dm_spectrum}).  \newText{Eq.~\ref{eq:general_flux} 
does not include the secondary \grs, which can modify both the 
spectrum and spatial distribution of the emission.}  

The ``$J$~factor'' depends on the details of the spatial distribution of the DM:

\begin{equation}\label{eq:J_factor}
  \frac{dJ_{\rm ann}}{d\Omega} = \int_\text{l.o.s.}\!\!\!\!\!
 d\ell \rho(\vec{\ell})^2
\end{equation}

\noindent The integration is performed along the line of sight, and $\rho(\vec{\ell})$ 
is the density of the DM particles~\cite{Ullio:2002pj}.  For DM decay (as opposed to annihilation) 
the factor of $\left( \frac{x \sigmav}{8\pi}\frac{1}{\mchi^2}\right)$
becomes $\left( \frac{1}{4\pi}\frac{1}{\tau \mchi}\right)$, where 
$\tau$ is the lifetime of the DM particle, and $\frac{dJ_{\rm dec}}{d\Omega}  = \int_\text{l.o.s.}\!\!\!\!\!
  d\ell \rho(\vec{\ell})$.  As we are concerned primarily with exploring the
case for additional data-taking with the LAT in this paper
we will focus on DM annihilation for concreteness.  However, 
\newText{as we will discuss in \S\ref{sec:dm_targets}} 
our estimates of the scaling of the detection sensitivity would 
be very similar for the case of DM decay.

The $J$~factor is an astrophysical quantity that is target dependent. Finding an 
astronomical object that maximizes this quantity is a key step in 
designing a sensitive search for indirect signals of DM. (It is also important
to consider the astrophysical backgrounds associated with potential
targets, see \S\ref{sec:backgrounds}.)
Typically the total $J$~factor integrated over the
solid angle of a region of interest (ROI, $\Delta\Omega$) is quoted for a given target:
\begin{equation}
J(\Delta\Omega) \equiv \int_{\Delta\Omega} d\Omega \int_\text{l.o.s.} d\ell ~\rho^2_\chi(\vec{\ell}\,).
\end{equation}
Note that $J$ 
depends implicitly 
on the distance to the DM target.  The density profiles 
of DM halos must be determined 
from a combination of observation and simulation, and depend
on the search target.  

\begin{figure}[!htbp]
  \begin{center}
    \includegraphics[width=0.49\columnwidth]{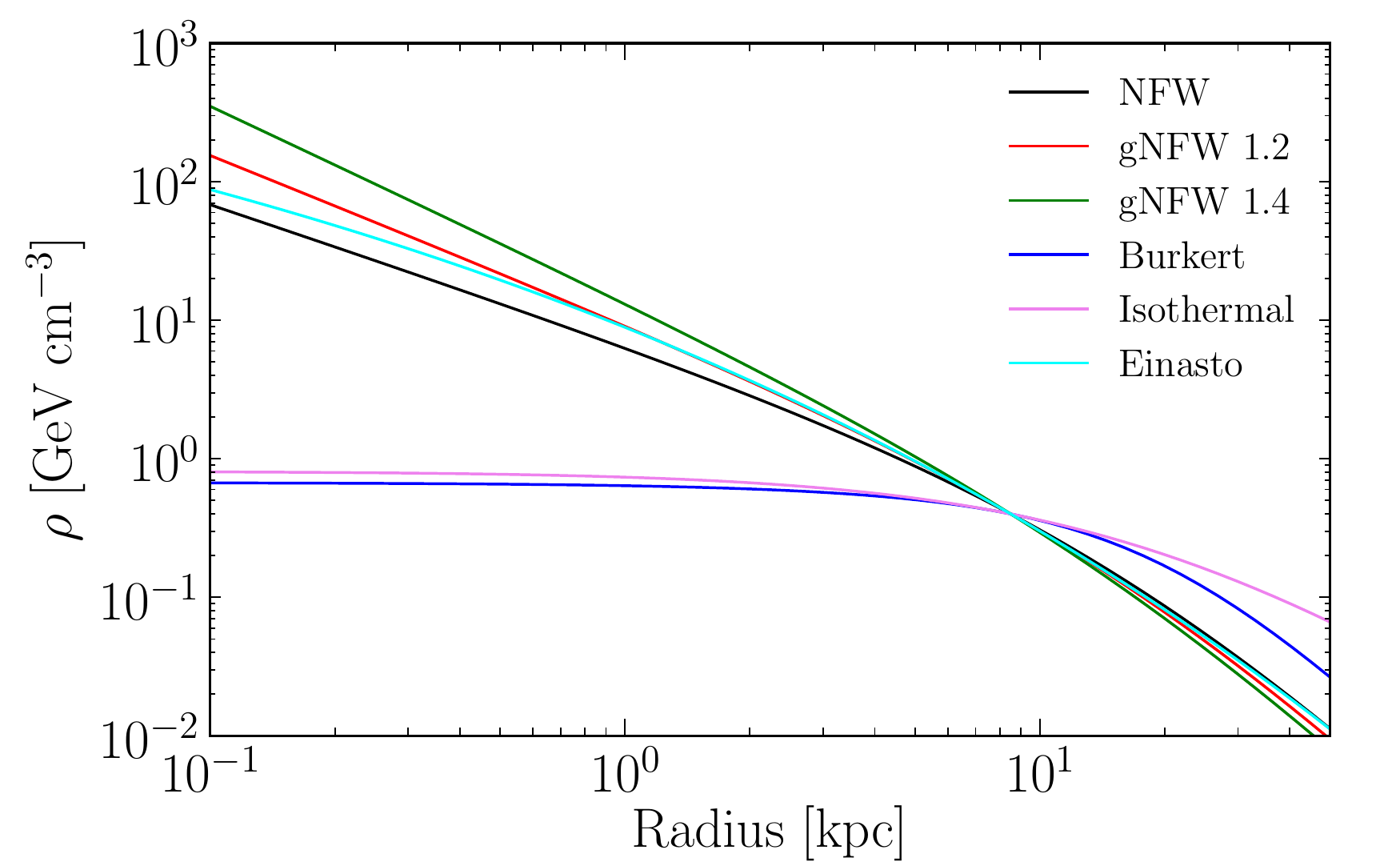}
    \includegraphics[width=0.49\columnwidth]{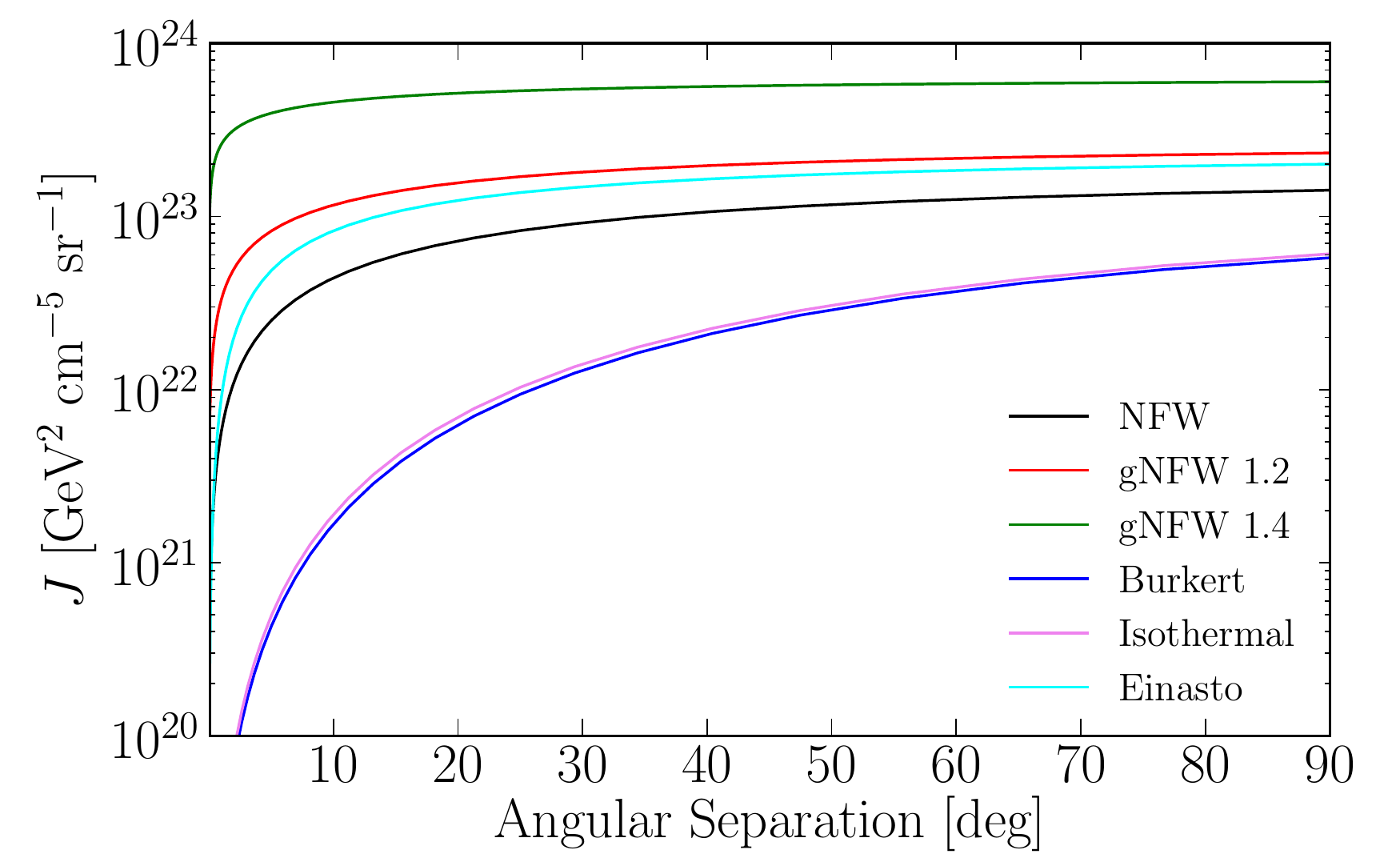}
  \end{center}
\caption{The density as a function of Galactocentric radius (left) and 
  the integrated $J$~factor as a function of angular separation from the 
  Galactic center for several widely considered radial profiles of the DM halo of the Milky Way.
  \label{fig:gc_radial}}
\end{figure}

As an example, the densities and resulting integrated $J$~factors for various potential DM density profiles
for the Milky Way are shown in Fig.~\ref{fig:gc_radial}.
For this comparison we adopted a
scale radius of $r_s = 20\kpc$, and selected $\rho_0$ such that
the local DM density (at $8.5\kpc$ from the Galactic center)
is $0.4\GeV\cm^{-3}$ \newText{(see,
e.g.,~\cite{2010JCAP...08..004C,2012ApJ...756...89B,2015JCAP...12..001P})}.  
The parameters were chosen to be consistent with dynamical
constraints~\cite[e.g.,][]{2013JCAP...07..016N,2012PASJ...64...75S,2010A&A...509A..25W}.
Details for the various radial profile models are given in App.~\ref{app:dark_matter_profiles}.

\newText{With the exception of annihilation to leptons in relatively dense environments, such 
as near the Galactic center~\cite{Buch:2015iya,Kaplinghat:2015gha, Lacroix:2015wfx, 2013JCAP...10..029G},
secondary processes are sub-dominant for \gray emission and we will not consider them
in this paper. (See, however, \S\ref{sec:disc_other_efforts} for discussion
of constraints on lepton-dominated channels from cosmic-ray and 
radio observations.)}

\subsubsection{Spectrum of Gamma Rays from WIMP Interactions}
\label{sec:dm_spectrum}

\begin{figure}[!htbp]
  \begin{center}
    \includegraphics[width=0.49\columnwidth]{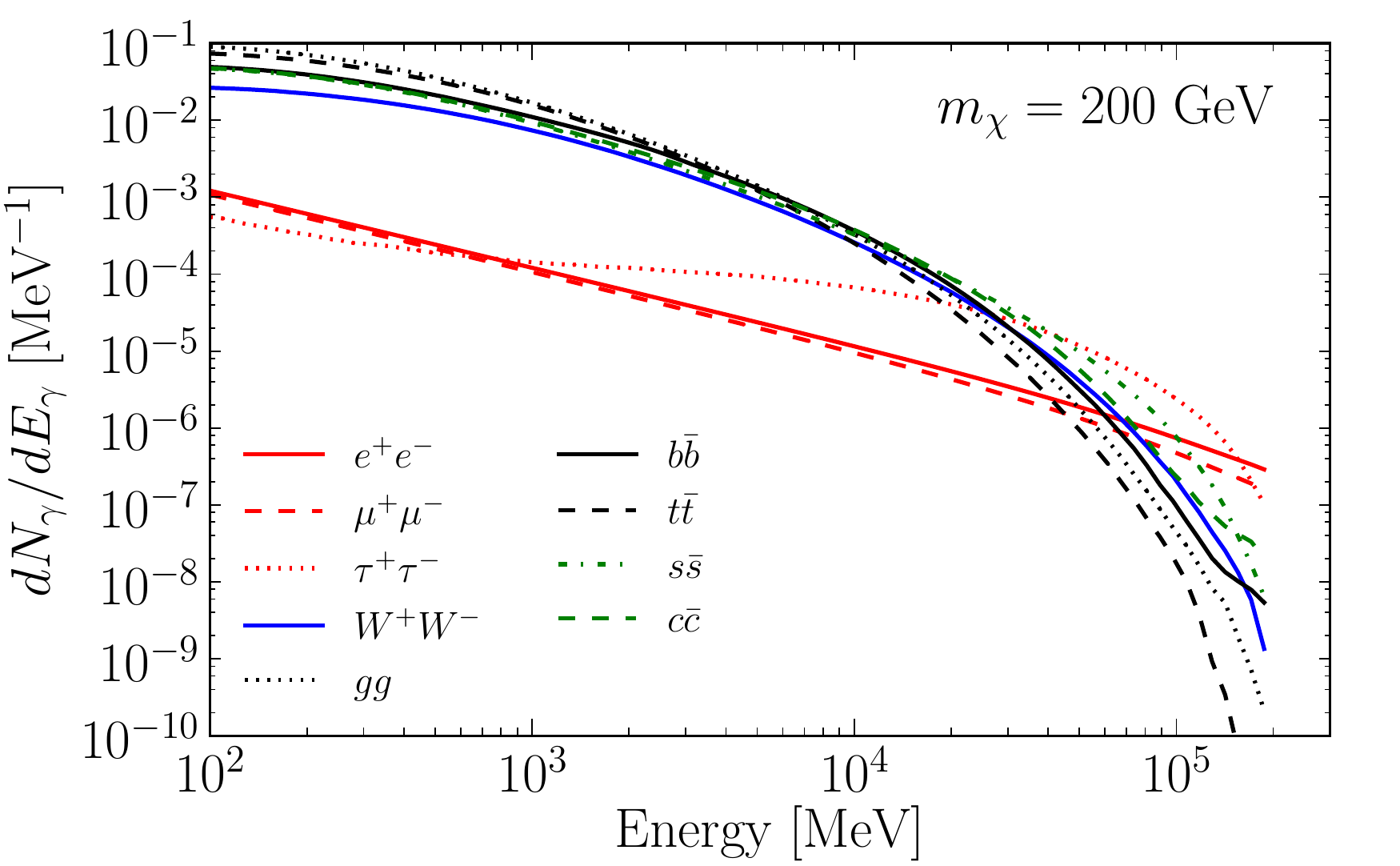}\includegraphics[width=0.49\columnwidth]{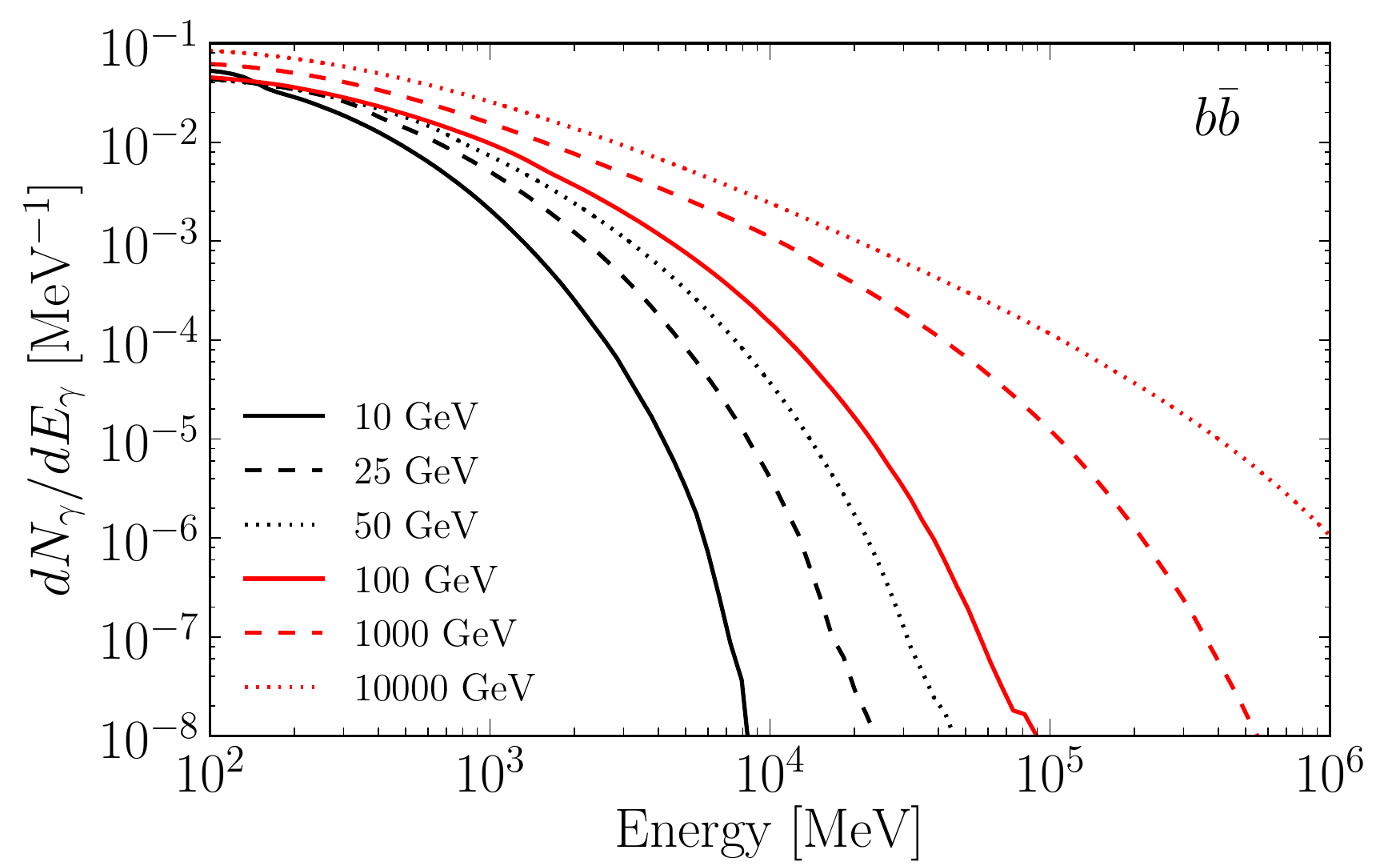}\\
    \includegraphics[width=0.49\columnwidth]{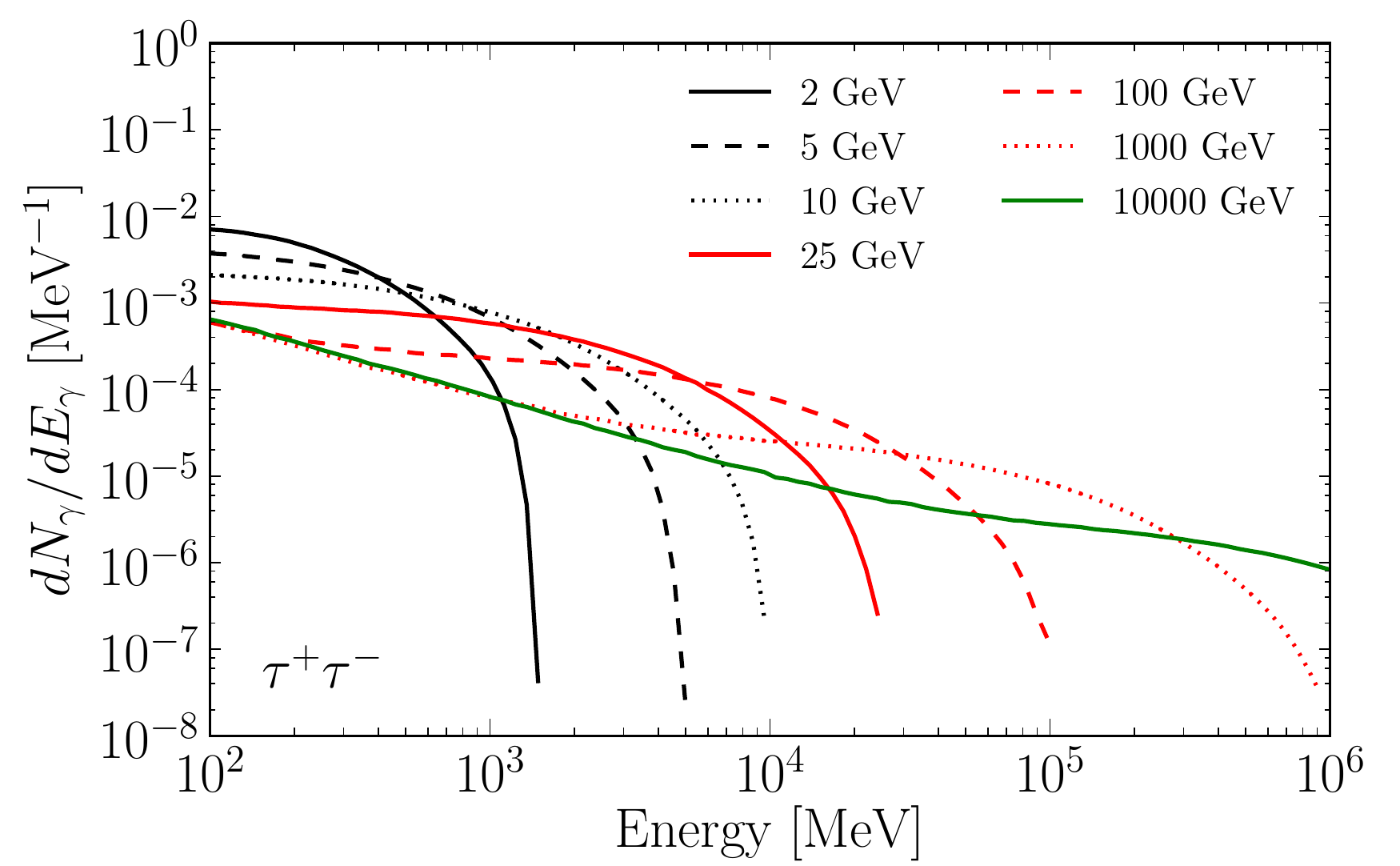}\includegraphics[width=0.49\columnwidth]{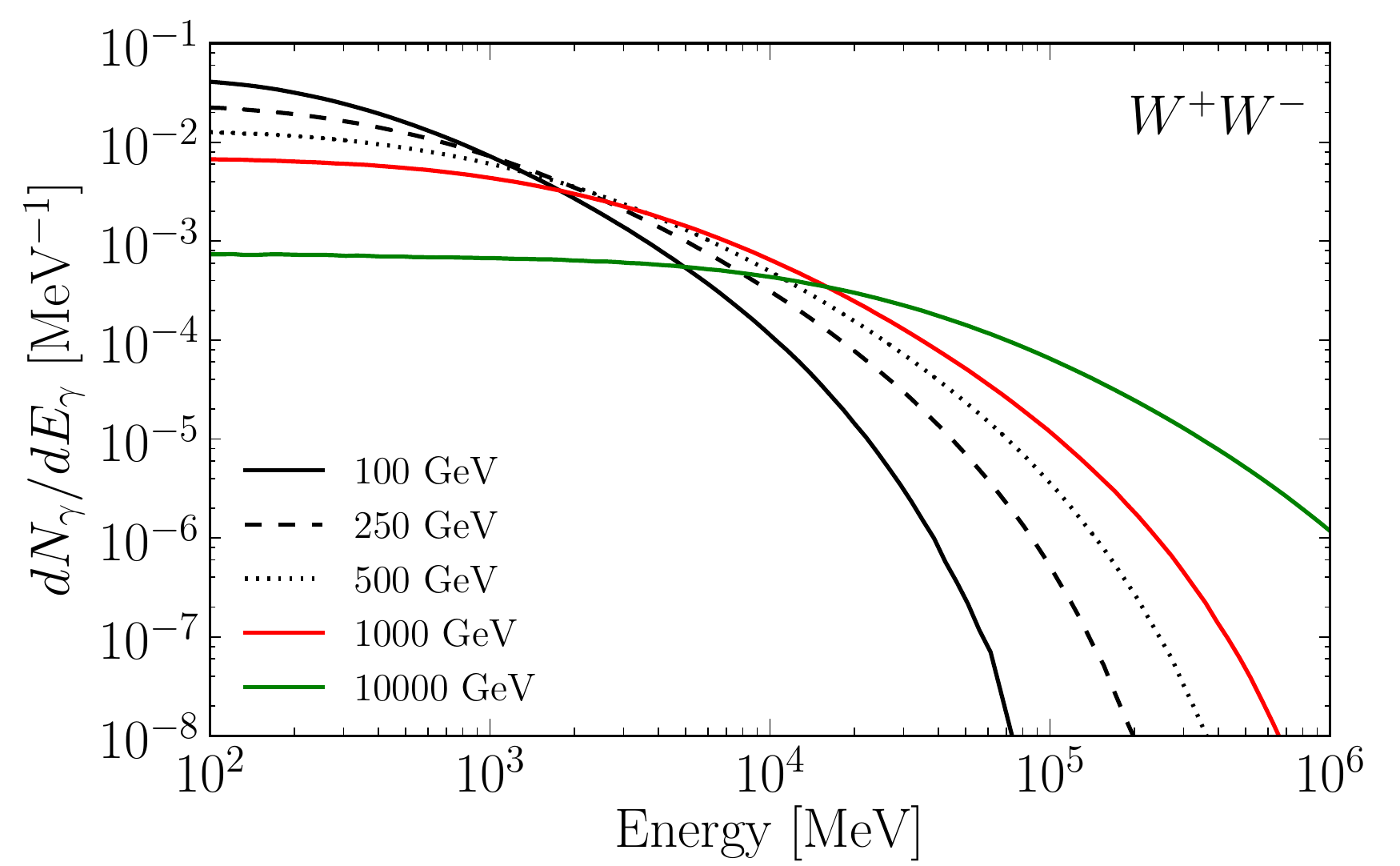}
  \end{center}
  \caption{Spectra, $dN_\gamma/dE_\gamma$, of \newText{prompt} \grs per DM pair annihilation for different
    annihilation channels and DM masses. (Upper left) Annihilation spectra of
    200\GeV DM  into various annihilation channels.  Annihilation
    spectra into \bb (upper right), \tautau (lower left), and $W^+W^-$
    (lower right) for a range of DM  masses.
    See Ref.~\cite{Jeltema:2008hf} and
    App.~\ref{app:dark_matter_spectra} 
    for details of the calculation of these spectra.   \newText{These
      spectra do not include secondary emission of \grs, which
      will enhance the emission at lower energies in the leptonic
      channels and can be important in dense environments.}
 \label{fig:annihilation_spectra}}
\end{figure}

Fig.~\ref{fig:annihilation_spectra} shows the prompt-emission 
spectra, $dN_\gamma/dE_\gamma$, per pair annihilation, for a variety of 
channels and DM masses~\cite{Jeltema:2008hf}.  We typically 
present results for the \bb channel in this report.   Similar results
apply to other channels, but in some cases we 
highlight how the results depend on the channel considered,
in particular for the \tautau channel, for which  $dN_\gamma/dE_\gamma$ 
peaks at higher energies.   

\subsection{Axions and Axion-Like Particles}
\label{sec:dm_axions}

Alternative classes of DM candidates include axions and ALPs.  
Axions were originally considered to be pseudo-Nambu-Goldstone bosons (pNGBs)
that arise when an additional symmetry of the standard-model Lagrangian 
is spontaneously broken~\cite{weinberg1978,wilczek1978}.
This additional shift symmetry was introduced to solve the strong CP problem 
in quantum chromodynamics~\cite{1977PhRvL..38.1440P}.
The axion mass $m_a$ is inversely proportional 
to the breaking scale $f_a$ of the additional symmetry, $m_a \sim
0.6\meV (10^{10}\GeV / f_a)$~\cite{jaeckel2010}. 
For ALPs, which are commonly predicted in string theories~\cite[e.g.][]{witten1984,ringwald2014review}, these two parameters are independent.
If produced non-thermally in the early Universe via the so-called misalignment mechanism, the coherent oscillations of the pNGB field 
can act as cold DM~\cite{Abbott:1982af,preskill1983,1983PhLB..120..137D,marsh2011,arias2012}.

Axions and ALPs can be detected through their coupling to photons described by the Lagrangian~\cite[e.g.][]{raffelt1988},
\begin{equation}
\mathcal{L}_{a\gamma} = -\frac{1}{4}g_{a\gamma}F_{\mu\nu}\tilde{F}^{\mu\nu} a,
\end{equation}
where $F_{\mu\nu}$ ($\tilde{F}_{\mu\nu}$) is the (dual) electromagnetic field tensor, $a$ the pNGB field strength,
 and $g_{a\gamma}$ the coupling constant to photons. 
The coupling is related to the symmetry breaking scale via  
\begin{equation}
g_{a\gamma} = \frac{\alpha}{2\pi}\frac{\mathcal N}{f_a},
\end{equation}
with the fine-structure constant $\alpha$ and a model-dependent factor $\mathcal N$, usually assumed to be of the order one.
As a consequence of two-photon coupling,
ALPs and axions can be detected either through their decay 
to photons (as a \gray flux associated with objects
with extreme magnetic fields, such as neutron stars)
or an oscillation to photons in external magnetic fields
(as irregularities in the spectra of \gray sources). 

The pNGB lifetime for photon decay in vacuum is given by~\cite{preskill1983}
\begin{equation}
\tau_{a\gamma} = \frac{64\pi}{m_a^3g_{a\gamma}^2} = 1.3\times10^{27}\,\mathrm{s}\left(\frac{g_{a\gamma}}{10^{-11}\GeV^{-1}}\right)^{-2}
\left(\frac{m_a}{\eV} \right)^{-3}.
\end{equation}
For axions and ALPs to be cold DM, we require that they are sufficiently stable, i.e., $\tau_{a\gamma}$ must be larger 
than the age of the Universe $\sim 13.7\,\mathrm{Gyr}$. 
A further requirement is that the pNGB field should have 
started to oscillate no later than when the expanding early Universe reached matter-radiation equality.
Also, we require that mass not receive any further radiative corrections, so that the energy density $\Omega_a$ is 
diluted with the expanding Universe. 
For an axion or ALP making up a fraction $\Omega_a / \Omega$ of the total DM, 
these requirements lead to an upper limit on the photon coupling in terms of the mass~\cite{arias2012},
\begin{equation}
\frac{g_{a\gamma}}{10^{-11}\,\GeV^{-1}} \lesssim 2.20\times10^3 \theta_1 \mathcal{N} \sqrt{\frac{m_a}{\eV}} \sqrt{\frac{\Omega}{\Omega_a}},
\end{equation}
where $\theta_1$ is the initial misalignment angle of the pNGB field. 

% LocalWords:  DM gravitationally WIMPs axions axion ALPs Muon pion hadronic GC
% LocalWords:  baryonic AMS gluons bosons hadronize Eq ROI Frenk NFW gNFW gluon
% LocalWords:  Einasto isothermal eq ScienceTools DMFitFuction electroweak CP
% LocalWords:  Peccei QCD PVLAS ADMX AGNs TeV AGN GRBs QSOs nucleosynthesis
% LocalWords:  subhalos supersymmetry Galactocentric Nambu Goldstone pNGBs pNGB
% LocalWords:  chromodynamics radiative leptonic

\section{Astrophysical Backgrounds for Dark Matter Searches}
\label{sec:backgrounds}

\begin{figure}[!htbp]
\begin{center}
\includegraphics[width=0.49\columnwidth]{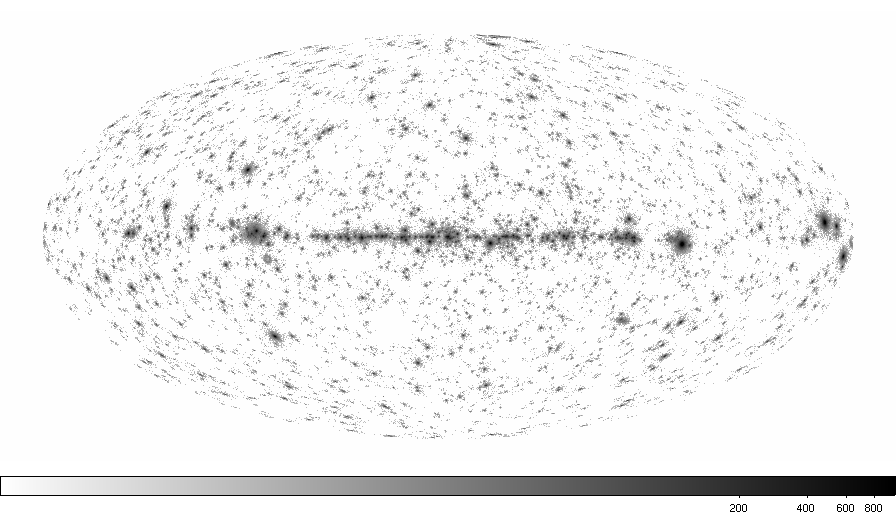} \\
\includegraphics[width=0.49\columnwidth]{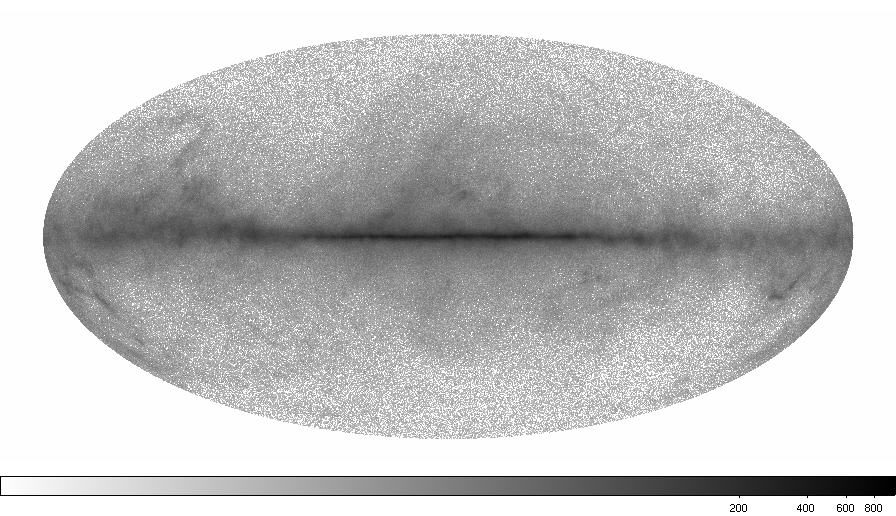}
\includegraphics[width=0.49\columnwidth]{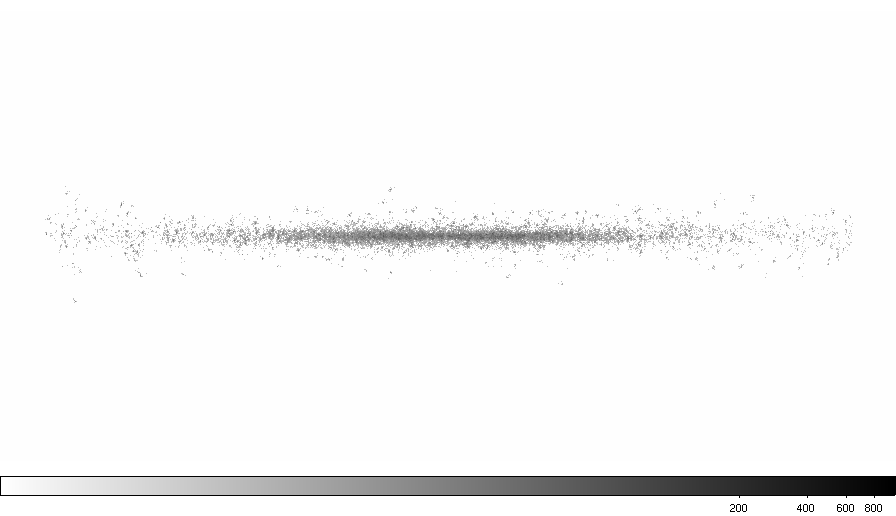} \\
\includegraphics[width=0.49\columnwidth]{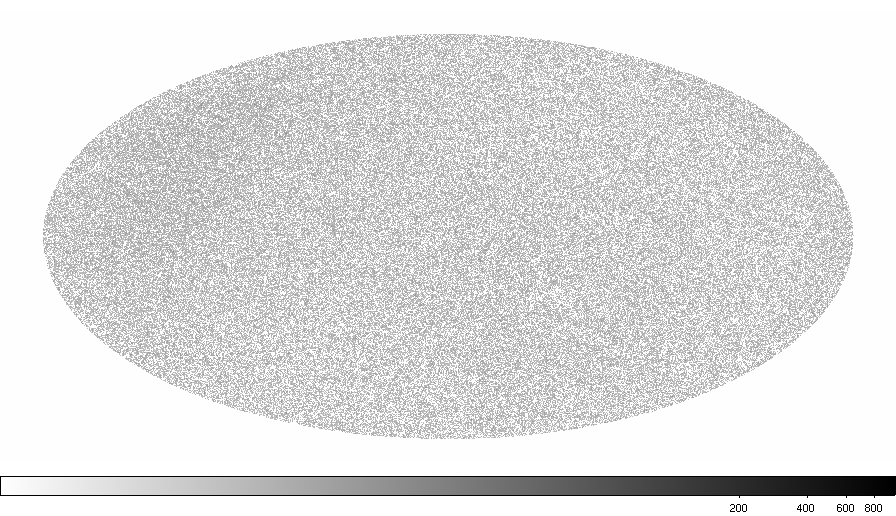}
\includegraphics[width=0.49\columnwidth]{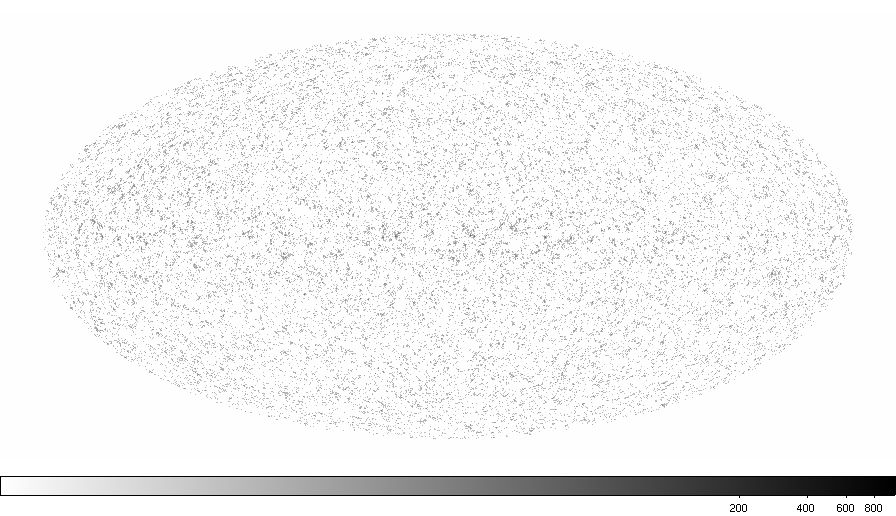}
\end{center}
\caption{Counts maps of all \grs with reconstructed energy above
  1\GeV for 15~years of simulated data for
  the \evtclass{P8R2\_SOURCE} event class.  The maps are shown in a Hammer-Aitoff projection in Galactic coordinates with 
  $1\degree \times 1\degree$ pixels.  The individual maps show cataloged sources 
  (\S\ref{sec:point_sources}, top), the diffuse Galactic emissions (\S\ref{sec:bkg_galactic_diffuse}, middle left),   
  unresolved Galactic sources (\S\ref{sec:bkg_galactic_sources}, middle right) isotropic background emission 
  (\S\ref{sec:bkg_isotropic} and \S\ref{sec:bkg_instrumental}, bottom left), and unresolved 
  extragalactic sources (\S\ref{sec:bkg_isotropic}, bottom right).
  The maps have the same gray scale to facilitate comparisons.}
\label{fig:backgrounds}
\end{figure}

All DM searches using the LAT must contend with backgrounds from 
a variety of astrophysical sources as well as instrumental backgrounds.
(Details about the LAT instrument and data are provided in App.~\ref{sec:lat}.)
The effect of the different backgrounds depends on the location and 
size of the DM search target as well as on the methodology of the
particular search technique.  In this section we summarize 
five major astrophysical backgrounds for DM searches.   Simulated all-sky maps
for 15~years of data for each background type are shown in Fig.~\ref{fig:backgrounds}.
(See App.~\ref{app:method_monte_carlo} for simulation details.)

\subsection{Cataloged Sources}
\label{sec:point_sources}

The third \FermiLAT source catalog (3FGL, \cite{2015ApJS..218...23A}) 
was constructed based on the analysis of the first four years of LAT data, and contains more than 
3000 individual
sources.  Of those, more than 2000 are associated with multi-wavelength 
counterparts of known \gray emitting source classes.  Although
follow-up studies on the earlier 1FGL~\cite{2010ApJS..188..405A} and
2FGL~\cite{2012ApJS..199...31N} catalogs found plausible
associations for \roughly 80\% of the sources in those catalogs~\cite{2015ApJS..217....2M}, the 
1010 unassociated sources in the 3FGL catalog could conceivably
include signals from DM subhalos (see \S\ref{sec:dm_unid}).

For analyses of DM targets that have spatial extents of less than a few degrees, 
we typically include all of the cataloged sources in the ROI 
in the baseline model (i.e., DM-free null hypothesis) and allow the spectral
parameters (including overall flux normalizations) of the nearest sources to vary when fitting for the DM targets.  
This avoids incorrectly setting limits that exclude DM signals that were 
mis-attributed as already-known sources.  If a DM target were placed at the position
of a cataloged source, the two would be at least partially degenerate in the likelihood fitting 
and any limits on the DM cross section would consequently worsen since some or all of the flux from the target location was potentially attributable to DM.   Allowing the spectral parameters of sources to 
vary also helps to account for any variability in the source,
particularly if an analysis uses a different integration time from the
catalog analysis.

For searches for DM signals from large regions 
of the sky, e.g., in a spectral line search (\S\ref{sec:dm_lines}, 
\newText{see also, e.g., Refs.~\cite{Weniger:2012tx,2013PhRvD..88h2002A,2015PhRvD..91l2002A}}), or
for \grs from WIMPs at cosmological distances (\S\ref{sec:dm_cosmo}, 
\newText{see also, e.g., Refs.~\cite{2012PhRvD..85h3007A,Xia:2015wka}})
we typically choose to mask regions around the brightest cataloged
sources.   This is done to reduce the background
and increase the signal-to-noise of the analysis, and to simplify the 
analysis by avoiding the need to re-optimize the spectral parameters of
thousands of sources.    

Therefore, when estimating the sensitivity for a DM search, depending
on the analysis considered, we either include the cataloged sources 
in the background model or account for the reduced search area 
(and reduced signal) from masking parts of the sky. 

The 4FGL catalog now under development will be based on seven years of
LAT data and is projected to include more than 4000 \gr sources.  For
sources at a given flux level, the deepening exposure of the LAT
observations also results in more precise localization of sources,
consequently increasing the power to classify sources based on
associations with counterparts at other wavelengths.  This helps to increase
the numbers of classified sources for population studies. 
The remaining unassociated sources also become `cleaner' targets in 
searches for undiscovered dark satellites of the Milky Way (\S\ref{sec:dm_unid}).

The collected emission from sources that are too faint to be
detected individually will contribute to both the Galactic (\S\ref{sec:bkg_galactic_sources}) and 
isotropic background (\S\ref{sec:bkg_isotropic}).  After
detailed accounting for the detection efficiencies, the intrinsic flux 
distributions $dN/dS$ of sources of a given class are essential for 
population studies that dissect these backgrounds.

\subsection{Diffuse Emission from the Milky Way}
\label{sec:bkg_galactic_diffuse}

Diffuse emission, i.e., emission not associated with discrete
sources, accounts for the majority of the \grs detected by
the LAT.  Diffuse \gray emission can be categorized as 
Galactic, associated with the Milky Way, or extragalactic, which has a nearly isotropic distribution over the
sky.   An unresolved source component may constitute less than 10\% of the
Galactic diffuse emission~\cite{2015ApJS..218...23A}, Therefore, the majority is due to interactions of
high-energy cosmic rays with interstellar gas and radiation fields in
the Milky Way. The production processes are inelastic collisions of 
cosmic-ray nuclei with nuclei in the interstellar medium, which produce \grays 
mainly through production and decay of neutral mesons like $\pi^{0}$, 
bremsstrahlung of cosmic-ray electrons and positrons off gas nuclei,
and inverse-Compton scattering of cosmic-ray electrons and positrons 
off interstellar photons from radio to optical/UV.  Predicting the
intensities of interstellar \gray emission requires knowing
the densities of cosmic rays, interstellar gas, and radiation fields
everywhere in the Milky Way. In practice, we use realistic, but
simplified, models of these quantities.   Details of the modeling 
procedure are provided in App.~\ref{app:galactic_diffuse_model}.

Detailed studies comparing models of the diffuse Galactic
emission~\cite{2012ApJ...750....3A} to LAT data have found that the models are fairly accurate,
with typical residuals of 10\% or better over much of the sky.   
This is compatible with, and in some cases somewhat better than, estimates of the uncertainties of the
modeling procedures (that are closer to 20 to 30\% for the local
emissions alone, see Ref.~\cite{2011A&A...531A..37D}).   However, as mentioned
above, large-scale residuals associated with specific structures and source populations
do exist, and in some parts of the sky and for some energy ranges the observed Galactic diffuse emission
can exceed the modeling predictions by 100\% or more.  

Similarly, the intensity 
of the Galactic diffuse emission, and hence its contribution to the background 
in particular analyses, varies significantly across the sky (see Fig.~\ref{fig:backgrounds}, middle left).
Therefore, treatment of the uncertainties arising from the model of the Galactic diffuse emission
depends on the search target.  In searches targeting discrete sources at 
high Galactic latitudes (such as known Milky Way satellites, see \S\ref{sec:dm_satellites}), 
the Galactic diffuse emission was found to be faint enough that the systematic uncertainties of the model can be neglected~\cite[e.g.,][]{2015PhRvL.115w1301A}.  For searches targeting larger regions
of the sky away from the Galactic plane it is often sufficient to use a small set
of representative Galactic diffuse emission models to quantify the uncertainties arising from the 
the Galactic fore- and backgrounds~\cite{2015ApJ...799...86A}.   Finally, searches targeting the Galactic center and 
inner Galactic halo must contend with very large backgrounds with large uncertainties, and
typically either mask the Galactic plane (e.g., Ref.~\cite{2012ApJ...761...91A}), perform detailed studies
of the uncertainties (e.g., Ref.~\cite{2015JCAP...03..038C}), undertake detailed fitting of the 
Galactic diffuse emission and point sources (e.g., Ref.~\cite{2015arXiv151102938T}) or do some of all of these.

In the context of developing a new model of the Galactic diffuse
emission to support Pass 8 analyses of seven-year and longer data
sets, we are investigating ways to increase the accuracy and reduce 
the systematic uncertainty of the model.  These include incorporating 
higher-resolution surveys of the interstellar gas in the Galactic
plane that became available recently~(e.g., \cite{GASS_HI,2016A&A...585A..41W}),
improving the derivation
of gas distributions along lines of sight toward the inner Galaxy
using recently published analyses of the distribution of interstellar 
dust in the Galactic plane based on reddening of stars, refining 
the evaluation of the dark neutral medium using {\it Planck} microwave 
data, and of course iterating the fitting with deep searches for point 
sources in the Pass 8 data.  Each of these is expected to increase the 
accuracy of the modeling in the challenging directions near the Galactic plane.

\subsection{Unresolved Galactic Source Populations}
\label{sec:bkg_galactic_sources}

\begin{figure}[!htbp]
  \begin{center}
    \includegraphics[width=0.75\columnwidth]{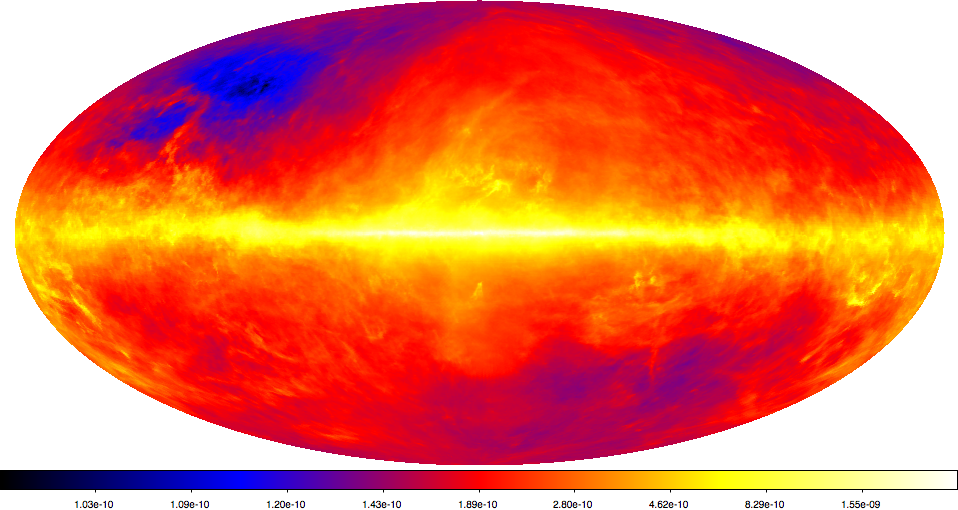}
  \end{center}
\caption{Map of the source detection threshold flux above 1\GeV (in $\cm^{-2}\second^{-1}$) for a point 
source with a power-law spectrum and a spectral index $\Gamma = 2.2$. The map is in 
Hammer-Aitoff projection in Galactic coordinates and is derived for the first four years of LAT observations.
\label{fig:detection_map}}
\end{figure}

Although unresolved Galactic sources make a relatively small
contribution to the total \gray luminosity of the Milky Way, their distribution is unlikely to follow the dust 
and gas that dominate the diffuse Galactic \gray distributions.  The 
template fitting procedure used to generate the model of diffuse Galactic
emission does have some freedom to absorb contributions from unresolved sources, but
exactly how much is difficult to predict as it depends on the correlation between
the morphology of the distribution of unresolved sources and the templates used to model the diffuse Galactic
\gray emission.  Furthermore, as can be seen in Fig.~\ref{fig:detection_map}, 
the flux threshold for the detection of a point source depends strongly on the diffuse Galactic
background.   The LAT sensitivity is also slightly better near the
celestial poles, as those regions have the largest exposure. 

\begin{figure}[!htbp]
  \begin{center}
    \includegraphics[width=0.5\columnwidth]{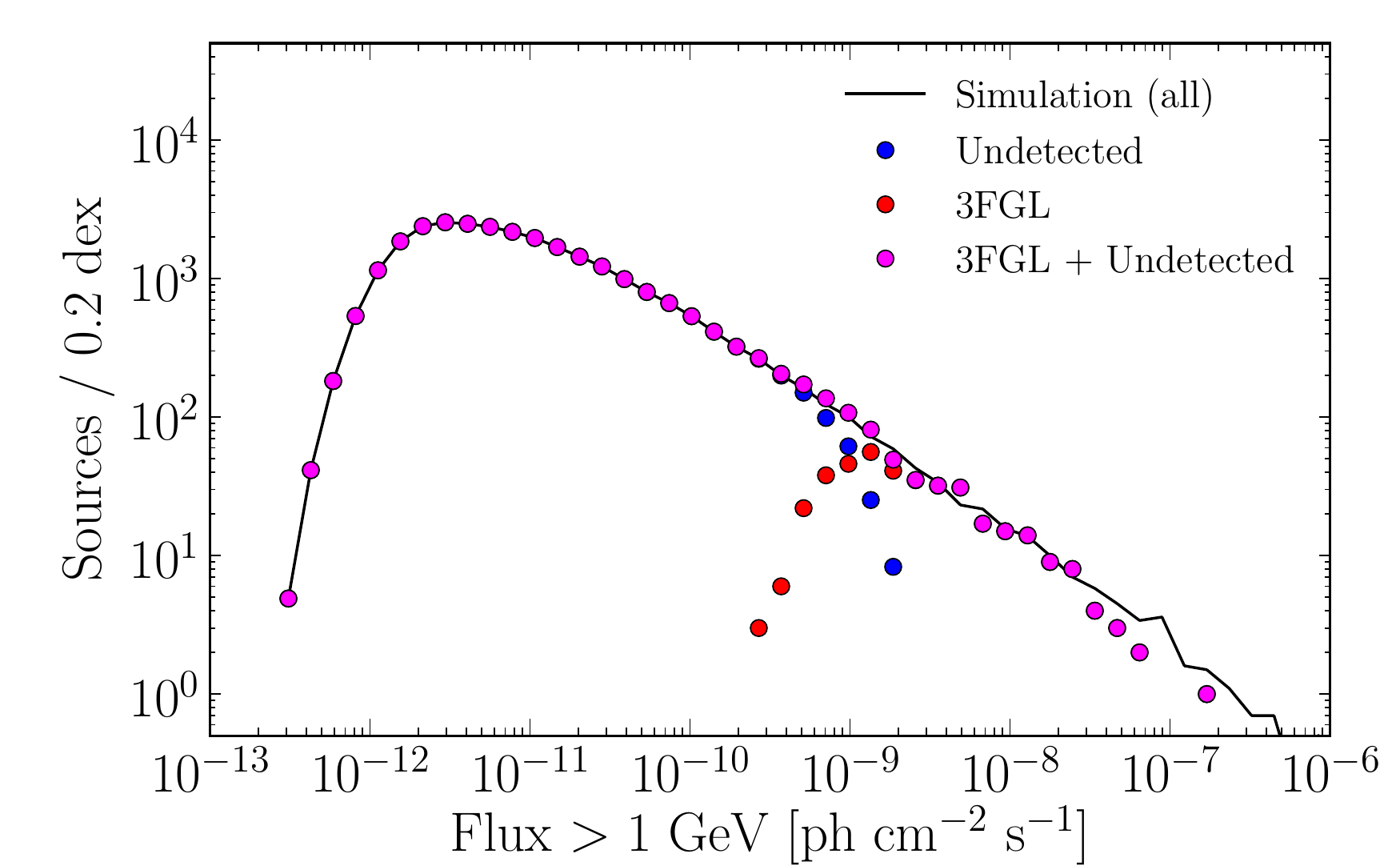}
  \end{center}
\caption{Integral photon flux above 1\GeV for Galactic sources.  The
  black line shows the total simulated distribution, the blue points
  show the simulated sources below the detection threshold for the
  3FGL catalog and  the red points show the distribution of 3FGL
  sources.  Finally, the magenta points show the sum of the simulated
  distribution below the detection threshold and the 3FGL catalog
  sources.
\label{fig:galactic_source_flux}}
\end{figure}

The population of unresolved Galactic sources was studied in detail in
\S~6 of the 3FGL catalog paper~\cite{2015ApJS..218...23A}. 
A mock catalog (with ten times the number of expected Galactic \gray sources)
was produced for that work.  We have simulated populations of both detected and undetected 
Galactic sources by applying thresholding using the map shown in Fig.~\ref{fig:detection_map}.
Fig.~\ref{fig:galactic_source_flux} shows the simulated
source counts as a function of integral photon flux above 1\GeV for both detected and undetected 
Galactic sources, illustrating that the total simulated distribution of
sources is well modeled by the sum of the simulated undetected sources
and the 3FGL catalog of detected sources.  (We applied the thresholding in energy flux, 
rather than photon flux, to reduce the dependence on the spectral index of the source.) 

The unresolved Galactic sources are not an important background for
most LAT searches.  Search targets typically are at 
high Galactic latitudes (such as the dwarf galaxies,
\S\ref{sec:dm_satellites}), and often low Galactic latitude regions are masked.
The exception is searches targeting the Galactic center, where
unresolved Galactic sources are potentially
a very challenging background (see \S\ref{sec:dm_milky_way}).
Pulsars pose a particular challenge, as their \gray spectra are 
similar to the expected DM signal for some parts of the parameter 
space we are searching, e.g., $\sim 50\GeV$ ($20\GeV$) DM annihilating to 
$b$-quarks ($\tau$-leptons)~\cite{2010ApJ...722.1939M,2013MNRAS.436.2461M,2013PhRvD..88h3009H,2015arXiv150402477O,2015JCAP...02..023P,2015JCAP...06..043C}.

\subsection{Isotropic Background and Extragalactic Source Populations}
\label{sec:bkg_isotropic}

The isotropic gamma-ray background (IGRB) comprises all extragalactic
emissions too faint or too diffuse to be resolved in a given survey,
as well as any residual Galactic foregrounds that are approximately 
isotropic (see, e.g.,~\cite{Fornasa:2015qua} for a recent
review).   In contrast, the extragalactic \gray background (EGB) 
includes the IGRB as well as the emission from resolved extragalactic point sources.

\begin{figure}[!htbp]
  \begin{center}
    \includegraphics[width=0.5\columnwidth]{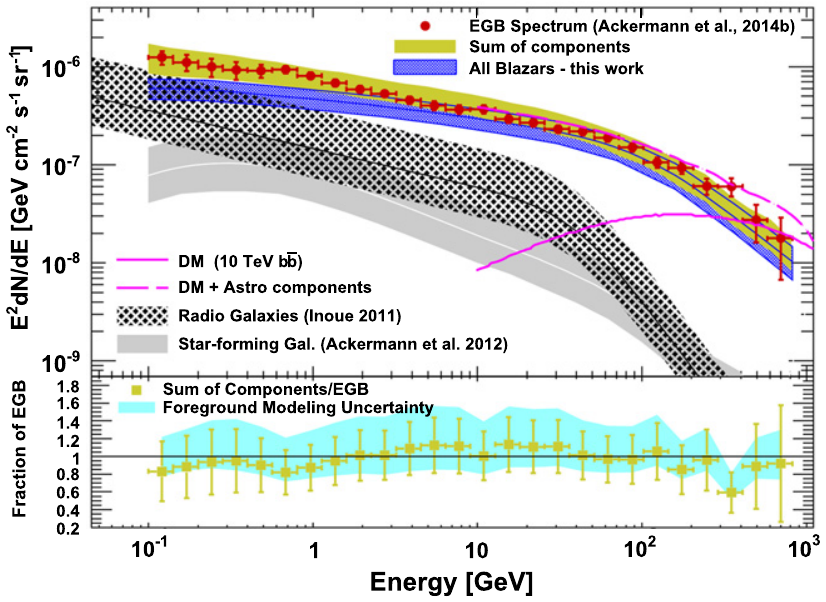}
  \end{center}
\caption{EGB (i.e. IGRB plus resolved extragalactic
  sources) as measured by the LAT in \cite{2015ApJ...799...86A} (red 
  points with error bars). The yellow band indicates the sum of
  contributions for unresolved sources of different types (blazars, 
  star-forming galaxies and radio galaxies). The lower inset shows 
  the ratio of this summed contribution to the EGB measurement, 
  as well as the uncertainty due to the foreground emission models. 
  This figure appeared as Fig.~3 of Ref.~\cite{Ajello:2015mfa}; additional details 
  are available there; reproduced by permission of the AAS.
  \label{fig:igrb_contributions}}
\end{figure}

Population studies of the source classes contributing to the 
EGB have shown that known \gray emitting source classes can 
account for most of it~\cite{Ajello:2015mfa}.  This can be seen in 
Fig.~\ref{fig:igrb_contributions} (see also \S\ref{sec:dm_cosmo}).

\begin{figure}[!htbp]
  \begin{center}
    \includegraphics[width=0.5\columnwidth]{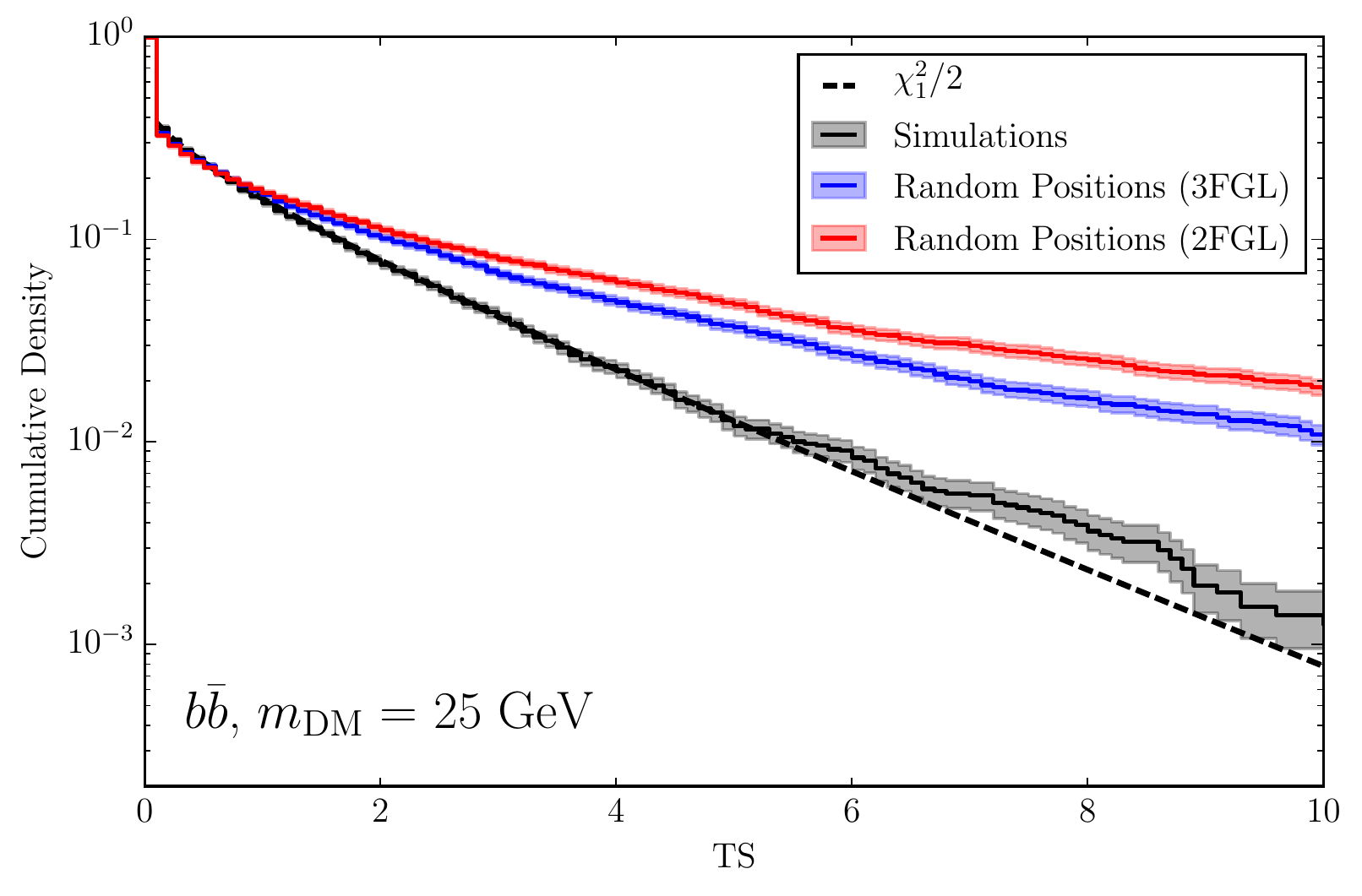}
  \end{center}
\caption{Distribution of $TS$ values obtained when testing for the existence of
  a point-like source with a spectrum characteristic of DM annihilation. The  dotted line shows the expected asymptotic
  distribution for the null hypothesis (a $\chi^2$-distribution with 1 degree
  of freedom). The gray band shows the distribution obtained
  from simulations without unresolved sources, while the red and blue 
  bands show the distributions obtained from random locations in the
  sky, when the background model used in the analysis included the
  2FGL (red) and 3FGL (blue) catalog sources.  This figure appeared as
  Fig.~5 of Ref.~\cite{2015PhRvL.115w1301A},
  additional details are available there, reproduced by permission of
  the APS and the authors.
  \label{fig:null_compare}}
\end{figure}

The distinction between a truly isotropic background and one composed primarily of 
unresolved sources is potentially very important for projecting search
sensitivities. This is because the nature of the statistical
fluctuations is quite different in the two cases.  In the former case
the background is characterized by Poisson fluctuations in the \gray
counts, while in the latter it is characterized by Poisson fluctuations in the 
number of sub-threshold sources overlapping the search target.  Since
the sub-threshold sources are generally not modeled, the distribution in the null-hypothesis case
will have more high test-statistic ($TS$, defined as twice the
different in log-likelihood between the null hypothesis and the
best-fit model) values than would be expected for truly isotropic
backgrounds.  An example of this can be seen in studies of the $TS$
distribution of random fields in flight data as compared to {\em
  all-sky photon simulations} shown in Fig.~\ref{fig:null_compare}.
Interestingly, although the fraction of high-$TS$ trials does increase
because of the unresolved source background, the median expected upper limits 
are not affected.  This is because the large majority of randomly
selected positions do not overlap with any \grs from unresolved sources.

\subsection{Residual Charged-Particle Backgrounds}
\label{sec:bkg_instrumental}

A final component of the background is the residual charged particles that contaminate
the LAT \gray sample. 
In this context we define \emph{particle backgrounds} as all events that are
classified as \grs in a given LAT event class but originate from
cosmic rays or the interactions of cosmic rays in the Earth's atmosphere. Therefore, the
particle backgrounds include both charged and neutral particles---including
secondary \grs.

Models of the particle background fluxes incident on the LAT are described in detail in 
Ref.~\cite{2004ApJ...614.1113M}.   Discussion of the calibration of the particle background 
contamination in the high-purity event classes, i.e., the event selections used for single source, source
population and diffuse emission analyses (\evtclass{SOURCE} class and cleaner) can be found 
in Ref.~\cite{2012ApJS..203....4A}.   Details of the residual charged particle contamination for 
Pass 8 event classes can be found in Ref.~\cite{2013arXiv1303.3514A}.    

For the studies described in this paper it is sufficient to note that the
residual charged particle backgrounds are treated as part of the isotropic \gray template, and that
the isotropic templates used for data analysis were derived from fits to the flight data.  

\begin{figure}[!htbp]
\begin{center}
\includegraphics[width=0.5\columnwidth]{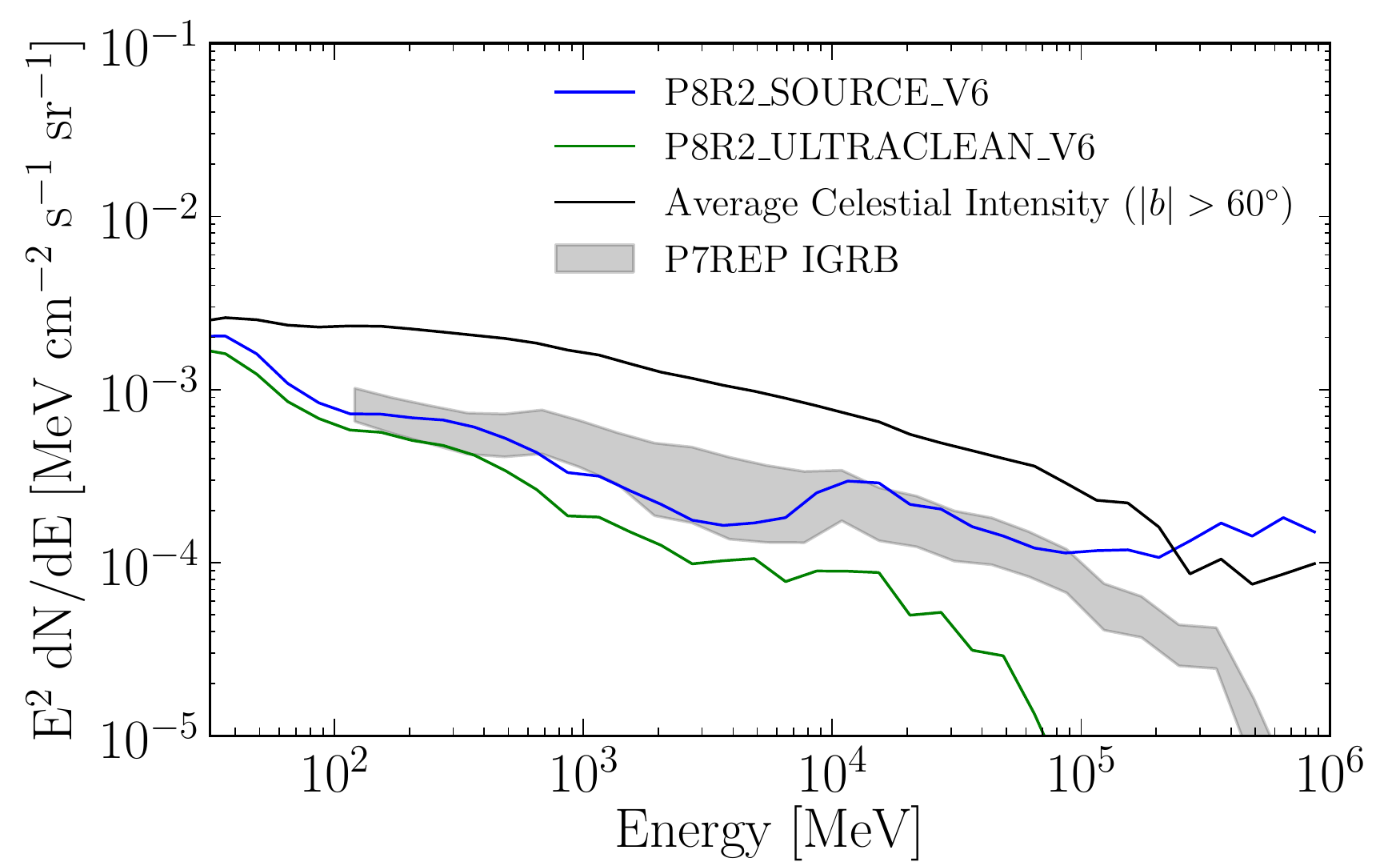}
\end{center}
\caption{Comparison of the residual charged-particle background contamination 
  for the \evtclass{P8R2\_SOURCE}  and
  \evtclass{P8R2\_ULTRACLEAN}  event classes with the average
  brightness of the \gray sky at intermediate and high
  latitudes and the estimated IGRB.
  \label{fig:bkg_levels_pass_8}}
\end{figure}

Estimates of the residual charged particle contamination for the various
Pass~8 event classes are shown in Fig.~\ref{fig:bkg_levels_pass_8}. The
contamination of the \evtclass{P8R2\_ULTRACLEAN} event classes is at or 
below the IGRB level for all energy $> 100\MeV$.

The simplest approach for dealing with residual charged particle backgrounds is to choose
an event class with low enough levels of residual charged particles that they
may safely be neglected.  This is possible for most of the analyses described in 
this paper, the exceptions being those targeting large fractions (or all)
of the sky, as even for the cleanest event class, the integrated contributions
from residual charged particle backgrounds can become important.  Specifically, searches for 
spectral lines (\S\ref{sec:dm_lines}) and large-scale, cosmological DM 
signatures (\S\ref{sec:dm_cosmo}) must consider this background component.

% LocalWords:  DM P8R2 V6 Aitoff 3FGL multi ROI anistropy extragalactic 3FGL
% LocalWords:  unassociated thresholding 3FGL 3FGL 1FGL 2FGL subhalos mis WIMPs
% LocalWords:  normalizations 4FGL efficiencies EGB IGRB blazars ULTRACLEAN AAS

% LocalWords:  3FGL 1FGL 2FGL 4FGL APS

\section{Search Strategies, Status, and Projections for Dark Matter
  Detection with the LAT}
\label{sec:dm_targets}

In this section we describe astrophysical objects that are the primary
targets for searching for signals from DM annihilation.  We first present 
an overview of the various WIMP search targets and results; then for
each target we summarize the status of current searches, and project 
how the sensitivity will improve with continued LAT data taking.
Finally we discuss searches for axion or ALP DM and how the sensitivity of
those searches will improve with additional LAT data.

\subsection{Dark Matter WIMP Search Targets}
\label{sec:dm_summary}

Tab.~\ref{tab:dm_targets} summarizes the targets for WIMP searches. 
The differences between the targets are 
pronounced enough to warrant significant modifications in the search 
techniques, as discussed earlier in this section.   For example,
searches targeting known dark-matter dominated Milky Way satellites
(\S\ref{sec:dm_satellites_status}) are very similar to blind searches
for point-like emission performed when creating catalogs of \gray
point sources such as the 3FGL; on the other hand, extracting a
isotropic signal from DM halos of galaxies at cosmological distances  (\S\ref{sec:dm_cosmo})
requires very detailed modeling of both the Galactic foreground
emission and the contributions of unresolved sources.

\begin{table}[!htbp]
\begin{tabular}{ccccc}
\hline \hline
Target & Distance (\kpc) & $J$~factor~($\GeV^2\cm^{-5}$) & Angular Extent $(^\circ)$ \\ \hline
Galactic center / halo (\S\ref{sec:dm_milky_way}) & 8.5 & $3\times10^{22}$ to $5\times10^{23}$ & $>10$ \\ 
Known Milky Way satellites  (\S\ref{sec:dm_satellites})  & $25$ to  $300$ & $3\times10^{17}$ to $3\times10^{19}$ & $<0.5$ \\ 
Dark satellites  (\S\ref{sec:dm_unid})  & \newText{up to $300$} & up to $3\times10^{19}$ & $<0.5$ \\ 
Galaxy Clusters  (\S\ref{sec:dm_clusters})  & $ > 5\times10^{4}$ & up
to $1 \times 10^{18}$ & up to $\sim 3$ \\ 
Cosmological DM (\S\ref{sec:dm_cosmo})   & $> 10^{6}$ & - & Isotropic \\ 
\hline \hline
\end{tabular}
\caption{Summary table of DM search targets discussed in this paper.\label{tab:dm_targets}}
\end{table}

\subsection{Current WIMP Search Sensitivity }
\label{sec:dm_current_summary}

\begin{figure}[!htbp]
  \begin{center}
    \includegraphics[width=0.80\columnwidth]{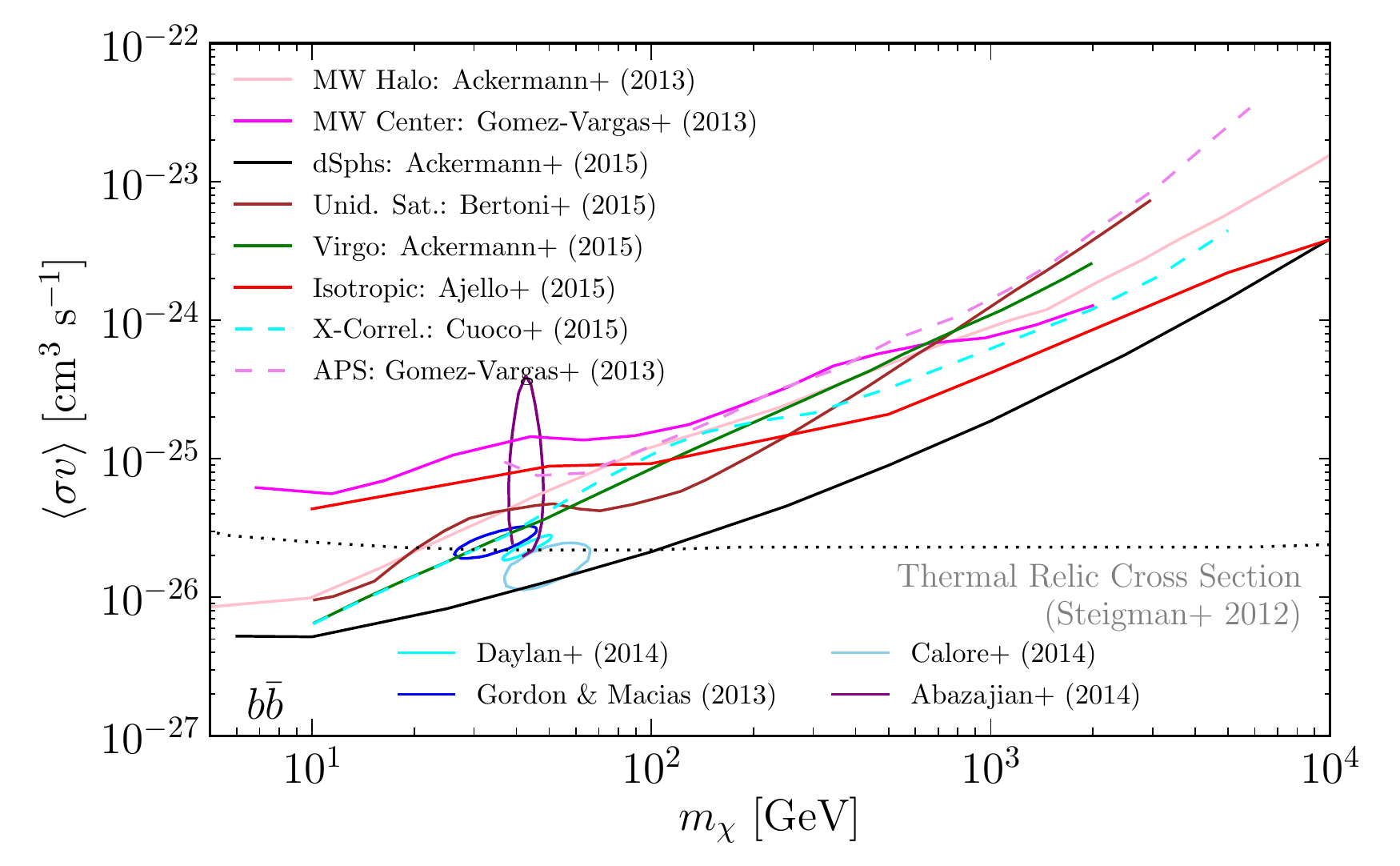}
  \end{center}
\caption{Comparison of representative published limits (curves) and best-fit 
  regions (ellipses) for the \bb channel found using LAT data for several DM
  targets.  References and details about the scenario selected as the
  representative limit or best-fit values for each DM target are 
  provided in Tab.~\ref{tab:target_summary}.  
  \label{fig:limit_summary}}
\end{figure}

We show a subset of published results for various DM targets 
for the \bb channel in Fig.~\ref{fig:limit_summary}. For each target,
we selected recent results that used moderate assumptions, i.e., neither
the most conservative nor the most optimistic cases.   
Because of differences in the datasets, DM profiles, and background 
modeling, these results should be taken as representative and absolute 
comparisons should be interpreted with caution.   Details about the 
scenarios considered (e.g., the DM distribution) for each of the 
targets are provided in Tab.~\ref{tab:target_summary}.

\begin{table}[!htbp]
\begin{tabular}{ccccc}
\hline \hline
Target & Ref. & Scenario & Other Refs. \\ \hline
Galactic halo~\S\ref{sec:dm_milky_way} & \cite{2012ApJ...761...91A} &
NFW profile ``constrained free source fits''  3~$\sigma$ ULs & - \\
Galactic center (limits)~\S\ref{sec:dm_milky_way} & \cite{2013JCAP...10..029G} & NFW profile, 3~$\sigma$ ULs & - \\
Galactic center (best-fits)~\S\ref{sec:dm_milky_way} &
\cite{2014arXiv1402.6703D,2015PhRvD..91f3003C,2014PhRvD..90b3526A,2013PhRvD..88h3521G,2016PhRvD..93h3514A,2016JCAP...04..030H}
& gNFW profile with $\gamma \sim 1.2$ & \cite{2011PhLB..697..412H,2015PhRvD..91l3010Z}  \\
dSphs~\S\ref{sec:dm_satellites} & \cite{2015PhRvL.115w1301A} & NFW profile & \cite{2015ApJ...809L...4D, 2015PhRvD..91h3535G, 2012ApJ...747..121A, 2012APh....37...26M, 2011PhRvL.107x1303G, 2011PhRvL.107x1302A, 2010ApJ...712..147A,2016PhRvD..93d3518L,2013JCAP...03..018S} \\
Unid. Satellites~\S\ref{sec:dm_unid} & \cite{2015JCAP...12..035B} &
95\% CL ULs & \cite{2010ApJ...718..899A,2011PhRvD..83b3518P,2012PhRvD..86d3504B,2012ApJ...747..121A,2014PhRvD..89a6014B} \\
Galaxy clusters~\S\ref{sec:dm_clusters} & \cite{Ackermann:2015aa} &
Virgo, ``DM-I'' conservative boost model & 
\cite{2010JCAP...05..025A,Huang2011,2012ApJ...757..123A,2012JCAP...07..017A,2012JCAP...01..042H,2014MNRAS.441.2309P,Ackermann:2015ab,2016arXiv160206527L} \\
Isotropic~\S\ref{sec:dm_cosmo} & \cite{Ajello:2015mfa} & 2~$\sigma$ ULs & 
\cite{Ando:2015qda,2015PhRvD..91l3001D,2015JCAP...09..008T} \\
Cross-correlation~\S\ref{sec:dm_cosmo} & \cite{2015ApJS..221...29C} &
``annLOW, ALLGeV'', 95\% CL ULs  &
\cite{Xia:2011ax,
  Cuoco:2007sh, Ando:2009nk, Ando:2014aoa, Fornengo:2014cya,
  Camera:2014rja, Xia:2015wka, Regis:2015zka,
  Ando:2013xwa,  Shirasaki:2014noa, Camera:2012cj}  \\
Ang. Power Spectrum~\S\ref{sec:dm_cosmo} & \cite{Gomez-Vargas:2014yla}
&  ``Galactic + Extragal HIGH DM'' & 
\cite{DiMauro:2014wha, Campbell:2014mpa,
  Calore:2014hna, Fornengo:2013rga,  Campbell:2013rua, Inoue:2013vza,
  Chang:2013ada, Gomez-Vargas:2014yla, Ando:2013ff, Ripken:2012db,
  Harding:2012gk, Cuoco:2012yf, Ando:2005xg,
  Ando:2006cr,Fornasa:2012gu} \\
\hline \hline

\end{tabular}
\caption{The works referenced here for the various DM targets are the
  representative results shown in Fig.~\ref{fig:limit_summary}.  
  The scenarios given in quotation marks appear as they were named in the original 
  reference to distinguish them from other scenarios presented in the same papers.
  For details about the exact parametrization of the various DM signals as well 
  as the modeling of astrophysical backgrounds the reader is referred to the 
  original references.
  \label{tab:target_summary}}
\end{table}

\subsection{Limiting Factors in Search Sensitivity }
\label{sec:dm_limiting_factors}
 
To understand the benefits of additional data taking for DM searches 
we consider three cases.   Because 
of the rapidly falling power-law backgrounds, individual
search targets might fall under different cases at low and high energies.

\begin{enumerate}
\item{{\em Searches that are or will be systematics limited:}  for these we
    cannot expect to improve the sensitivity dramatically.   We do
    expect that the sensitivity will improve as our knowledge of the
    \gray sky and the astrophysical backgrounds improve, but these
    improvements are likely to be incremental.}
\item{{\em Searches that are background limited:} here we can expect
    the sensitivity to improve as $\sqrt{t}$.    As the
    mission continues, the relative gain in sensitivity from these
    searches is moderate.   Doubling the current data set would result in a
    good, but not overwhelming, $\sim 40\%$ improvement in sensitivity
    for these searches.}
\item{{\em Searches that are signal limited:} here we can expect the
    the sensitivity to improve proportionately with time.   These are
    the searches that constitute the strongest case for continued data 
    taking.}
\end{enumerate}

For many of the targets that we will discuss, the searches are for a small
signal against a large background.   For those cases the largest 
and most problematic systematic uncertainties come for mis-modeling the 
background in a way that would induce a fake signal or mask a real
signal.    Typically such uncertainties will scale roughly linearly
with the background.  We have developed and applied a technique
for DM searches with significant systematic uncertainties, 
namely calculating the ``effective background'' \beff (i.e., the 
background weighted by how strongly it overlaps with the signal) for
the search~\cite{2013PhRvD..88h2002A,2015PhRvD..91j2001B,2014JCAP...10..023A}.
We estimate systematic uncertainties by positing that they can be 
expressed as a fraction of $b_{\rm eff}$, i.e., $f = n_{\rm sig} / b_{\rm eff}$
and measuring the observed signal in control regions, where no DM signal should be
present.  This effective background methodology is discussed in 
more detail in App.~\ref{app:method_effective_background} and in 
Refs~\cite{1995ApJ...447L..25B,1996ApJ...462..563N,1998ApJ...502...48K,2010MNRAS.402...21N}.

The statistical uncertainties on the number of counts assigned to a
signal arising from fluctuation in the background are expected to be 
roughly $\delta \nsig = \beff^{1/2}$.    To include the
effect of systematic uncertainties we estimate the total uncertainty
of the signal as a fraction of the effective background added in
quadrature with the statistical uncertainty: $\delta \nsig= \fsyst
\beff \oplus \beff^{1/2}$.    The value of \fsyst depends on the
analysis (and ranges from 0.015~\cite{2014JCAP...10..023A} to
0.05~\cite{2015PhRvD..91j2001B}) 
and for each analysis we estimate it with the judicious use of control 
samples in the flight data.

\begin{figure}[!htbp]
  \begin{center}
    \includegraphics[width=0.50\columnwidth]{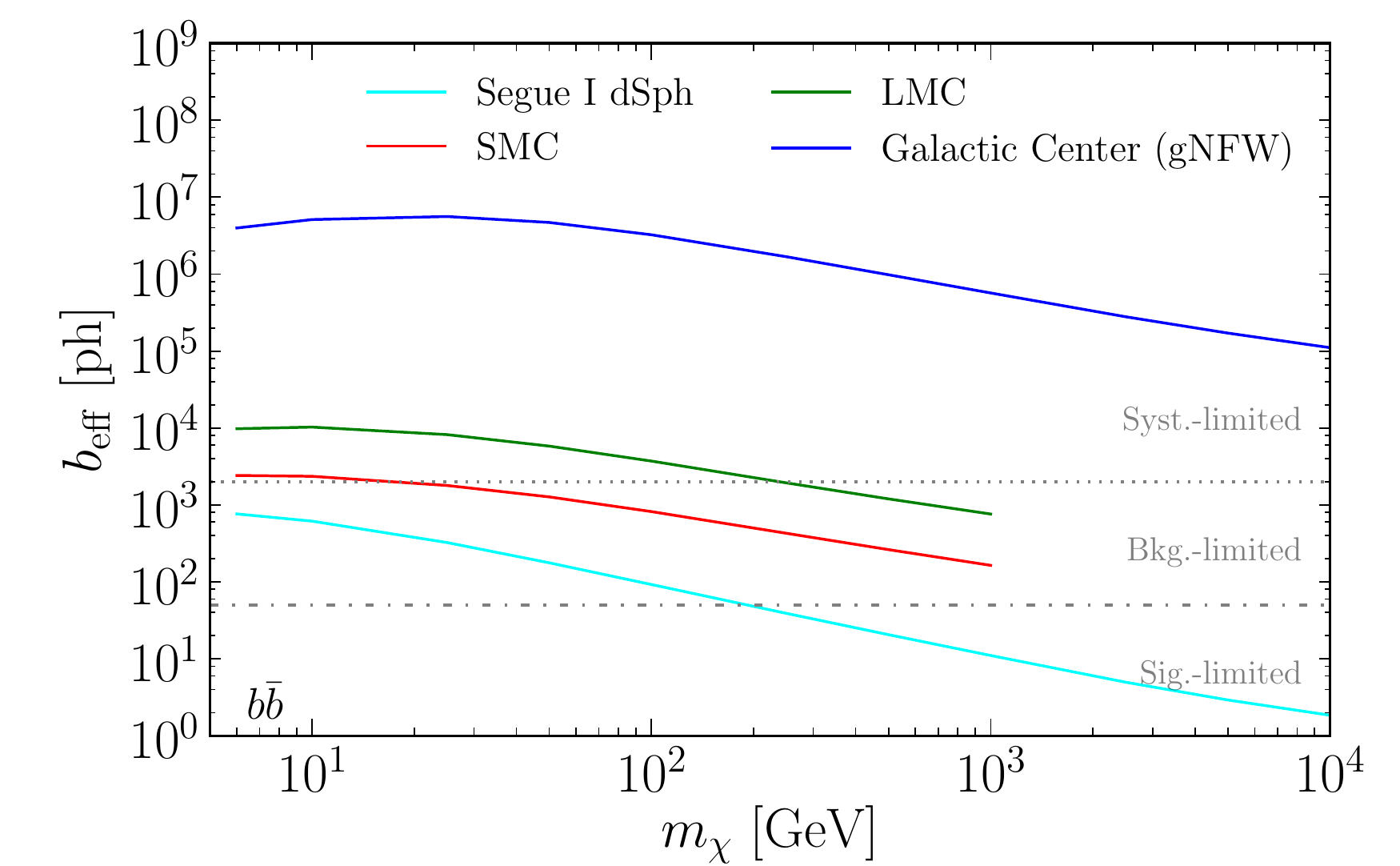}
  \end{center}
\caption{Effective background for six years of  \evtclass{P8R2\_SOURCE} 
  data, as a function of \mchi for the \bb channel, for four DM search targets with different
  angular scales and astrophysical backgrounds: the Galactic center 
  (\S\ref{sec:dm_milky_way}) the Large and Small Magellanic clouds
  (\S\ref{sec:dm_satellites}) and the Segue~I dwarf galaxy.  
  Also shown are the approximate cross-over levels between 
  different search sensitivity limiting factors.
  \label{fig:beff_targets}}
\end{figure}

Given the observed range of \fsyst, we can roughly state that
we expect searches to be systematics limited when $\beff \gtrsim 2000
\photons$, background limited when $2000 \photons  \gtrsim  \beff
\gtrsim 50 \photons$ and signal limited when  $\beff \lesssim 50 \photons$.
The ranges are shown in comparison with values of \beff as a 
function of \mchi for the \bb channel in Fig.~\ref{fig:beff_targets} 

\newText{The main difference when considering DM decay, as opposed to annihilation, 
is that \gray emission is less concentrated spatially since it scales as the 
DM density $\rho$ instead of $\rho^2$.   This will change the relative
powers of different search targets, reducing the impact of very high densities
in the cores of DM halos.  In terms of the scaling behavior
with additional data, targets that are larger than
the LAT angular resolution (such as the Galactic Halo or local clusters of galaxies) 
will have larger \beff.   Conversely, targets smaller 
than the LAT angular resolution, such as the dSphs or unidentified satellites
will have the same \beff, and hence the same scaling behavior.}

\subsection{The Milky Way Galactic Halo}
\label{sec:dm_milky_way}

The halo of our own Milky Way is
by far the brightest potential source of LAT-detectable \grs
from DM particle interactions.  However, as we discussed in 
\S\ref{sec:backgrounds}, DM searches targeting the Milky Way
halo must contend with large backgrounds from 
standard astrophysical processes in the Milky Way.

Although N-body simulations lead to the expectation that DM halos of galaxies
exhibit some degree of triaxiality and sub-structure
(see~\cite{2010MNRAS.402...21N} and references therein), the
Milky Way's DM distribution is generally assumed to be approximately 
spherically symmetric, with the density a
function of only the distance from the Galactic center, $r$.  As
discussed above, in this paper we consider
a generalized NFW profile for the radial density profile of the DM (Eq.~\ref{eq:gNFW}).
When considering the Milky Way, we adopt a
scale radius of $r_{\rm rms} = 20\kpc$, and select $\rho_{0}$ such that
the local DM density (at $8.5\kpc$ from the Galactic center)
is $0.4\GeV\cm^{-3}$, consistent with dynamical 
constraints~\cite[e.g.,][]{2013JCAP...07..016N,2012PASJ...64...75S,2010A&A...509A..25W}.
Note, however, that the actual DM density profile in the
inner galaxy~\cite[e.g.,][]{2015NatPh..11..245I,2015arXiv150307501D,2015JCAP...12..001P}, 
and the local density~\cite[e.g.,][]{2012MNRAS.425.1445G,2011BAAA...54..289M,2012ApJ...751...30M,2012ApJ...756...89B,2015RMxAC..46..105M}
are the subjects of vigorous debate.

DM-only simulations generally favor inner slopes
near the canonical NFW value ($\gamma =
1$)~\cite{2009MNRAS.398L..21S,2008MNRAS.391.1685S}, 
although baryonic effects are expected to have a non-negligible 
impact on the DM distribution within the inner $\sim 10\kpc$ 
of the Milky Way~\cite{2006ApJ...644..687C,2006PhRvD..74l3522G,2010MNRAS.406..922T,2011arXiv1108.5736G}. 
The magnitude of such effects, however, is currently uncertain.
With this in mind, we consider different DM radial profiles when estimating
the sensitivity of searches for DM signals from the Galactic halo. 
The radial profiles of both the density and the integrated $J$~factors 
for several density profiles were shown in
Fig.~\ref{fig:gc_radial} in \S\ref{sec:dm_signal}.

\subsubsection{Galactic Halo: Status of Current Searches}
\label{sec:dm_milky_way_current}

Several groups have reported an excess of \grs in the 
1--10\GeV energy range from the region of the Galactic center 
with respect to expectations based on cataloged point sources 
and detailed models predicting the Galactic diffuse \gray
emission (see, e.g.,~\cite{2011PhLB..697..412H,2014PhRvD..90b3526A,2014arXiv1402.6703D}).

\begin{figure}[!htbp]
  \begin{center}
    \includegraphics[width=0.50\columnwidth]{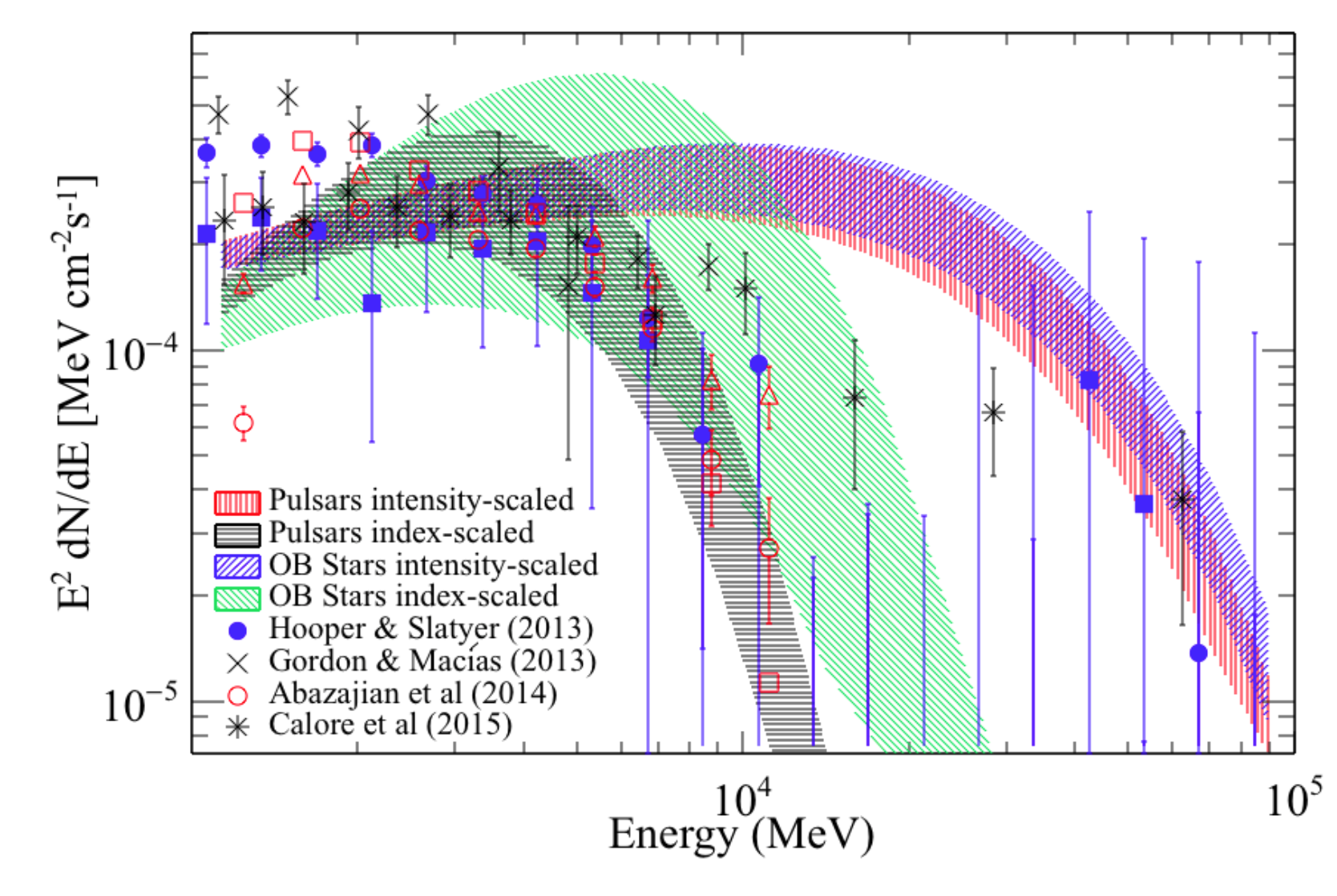}
  \end{center}
\caption{Spectrum of the Galactic center excess in the inner
  $15\degree\times15\degree$ region obtained with four different models
  of the Galactic diffuse emission compared with spectra obtained 
  from other published analyses based on still-different models of the Galactic diffuse emission.  This figure is from Ref.~\cite{2015arXiv151102938T} (reproduced by permission of the AAS); see that reference for
  additional details about the Galactic emission diffuse models and
  fitting.
  \label{fig:lat_gc_spectra}}
\end{figure}

Although the existence and spatial extension of the excess have 
been found to be robust against uncertainties of modeling the 
Galactic diffuse \gray emission, the measured spectrum 
of the excess has been shown to depend strongly on the Galactic diffuse 
emission model~\cite{2015PhRvD..91f3003C,2015arXiv151102938T,2015arXiv151102938T,2015PhRvD..91l3010Z,2016JCAP...04..030H}.  Fig.~\ref{fig:lat_gc_spectra} shows
four different fitted spectra that are extracted with for different
representative Galactic diffuse emission models.

The bright, structured and relatively uncertain astrophysical
fore/backgrounds toward the Galactic center complicate the
interpretation of the excess.   Some authors have
argued that the excess is consistent with DM, and that other interpretations
are disfavored~\cite[e.g.,][]{2013PhRvD..88h3009H,2015JCAP...06..043C,2014PhRvD..90b3526A}.
Others have pointed out that the spectrum of the excess is very similar
to spectra observed from pulsars and argued that the excess is more
likely attributable to a large population of either young
pulsars~\cite{2015arXiv150402477O} or older millisecond 
pulsars~\cite{2012PhRvD..86h3511A,2013MNRAS.436.2461M,2014JHEAp...3....1Y,2015JCAP...02..023P,Bartels:2015aea,Lee:2015fea}
near the Galactic center.   In particular, several analyses of the
spatial fluctuations of the \gray emission within the Galactic center excess found that they are more consistent 
with arising from a population of unresolved sources (such as
pulsars) than from a smooth distribution (as might be
expected for DM)~\cite{2015JCAP...02..023P,Bartels:2015aea,Lee:2015fea}.
To summarize, in the LAT energy band, the \gray spectrum 
expected from a population of pulsars is very similar to the \gray 
spectrum expected from 50\GeV DM annihilating to \bb,
and additional data, either relating to the spatial distribution 
of the signal, observations in other energy bands, or \gray observations
of other targets are required to clarify the interpretation of the
excess.

\begin{figure}[!htbp]
  \begin{center}
    \includegraphics[width=0.8\columnwidth]{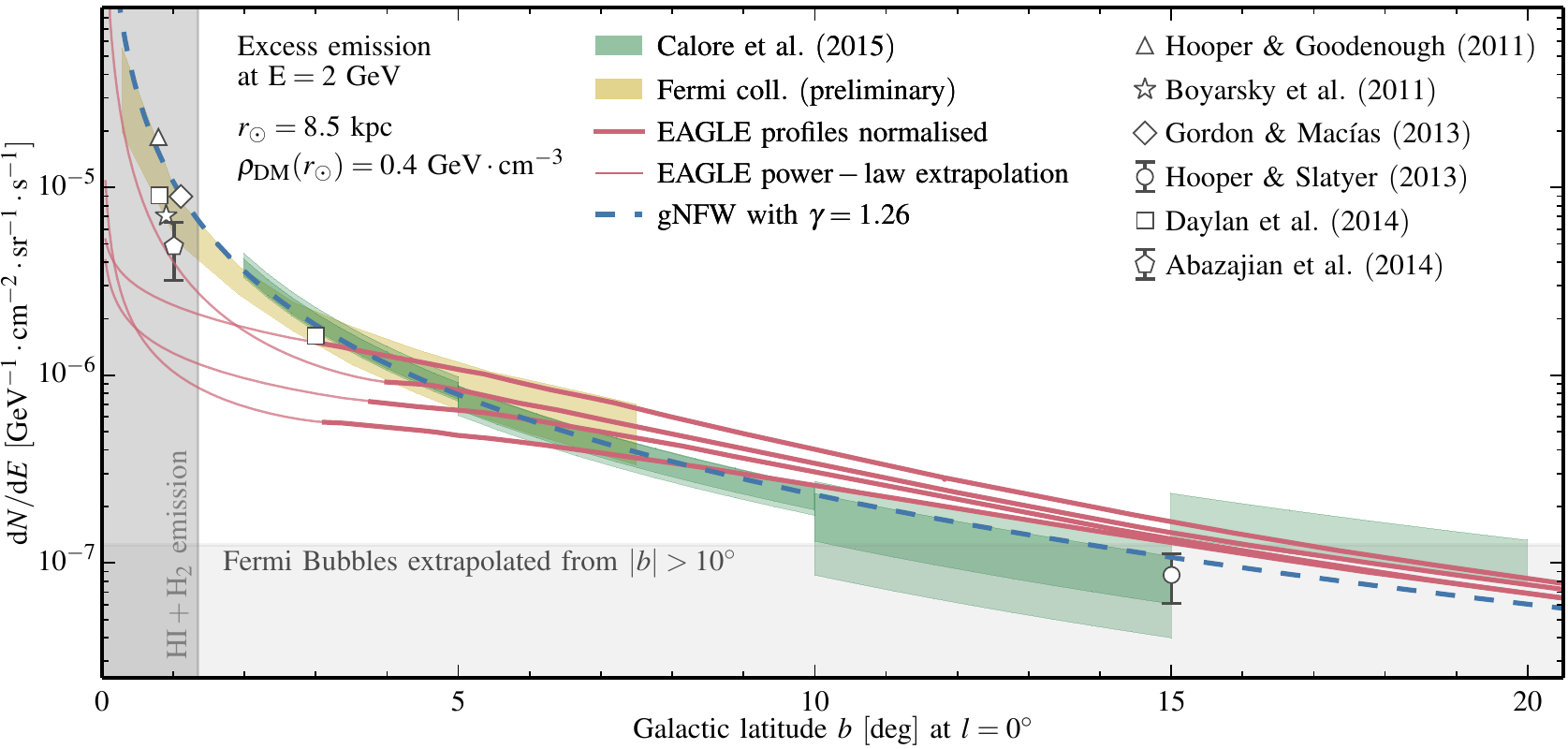}
   \end{center}
\caption{Measurements of the radial profile of the Galactic center
  excess (markers and bands) compared with predictions of
  hydrodynamical and N-body simulations
  of Milky Way-like structures (red lines).  This figure is from 
  Ref.~\cite{2016MNRAS.455.4442S} (reproduced by permission of the AAS); see that reference for additional details.   
  \label{fig:lat_gc_radial}}
\end{figure}

The morphology of the excess has also been studied
in detail and found to be consistent with spherical symmetry,
in particular an NFW profile with an inner slope of $\gamma = 1.2$~\cite{2014arXiv1402.6703D,2015PhRvD..91f3003C,2014PhRvD..90b3526A}.
Refs.~\cite{2016MNRAS.455.4442S,2015JCAP...12..053C} compared this to N-body simulations
of Milky Way-like structures and found it to be broadly consistent
with expectations, though somewhat more peaked than the simulations in 
the innermost \roughly 2\degree, see Fig.~\ref{fig:lat_gc_radial}.

Finally, the best-fit cross-section and mass range for a DM interpretation of the Galactic center 
excess is in mild tension with the current limits from a joint
analysis of several Milky Way satellite dwarf galaxies (see 
\S\ref{sec:dm_satellites}).

\subsubsection{Galactic Halo: Sensitivity Projections}
\label{sec:dm_milky_way_projections}

Although the DM signal from the Galactic center would be much larger
than for any other target, the astrophysical backgrounds are also
very large and imperfectly known toward the Galactic center.
Furthermore, it is important to distinguish between measuring an
excess with respect to models of \gray emission from predicted
cosmic-ray populations interacting with estimated dust, gas and radiation
field, and being able to interpret that excess as a clear signal of DM.
Accordingly, we can expect systematic uncertainties in modeling the Galactic fore/background to significantly limit the sensitivity of searches for DM signals 
from the Galactic center.    Furthermore, as described above, a
population of unresolved pulsars in the inner Galaxy would be a
difficult-to-reduce background for the best-fit DM models.  

\begin{figure}[!htbp]
  \begin{center}
    \includegraphics[width=0.49\columnwidth]{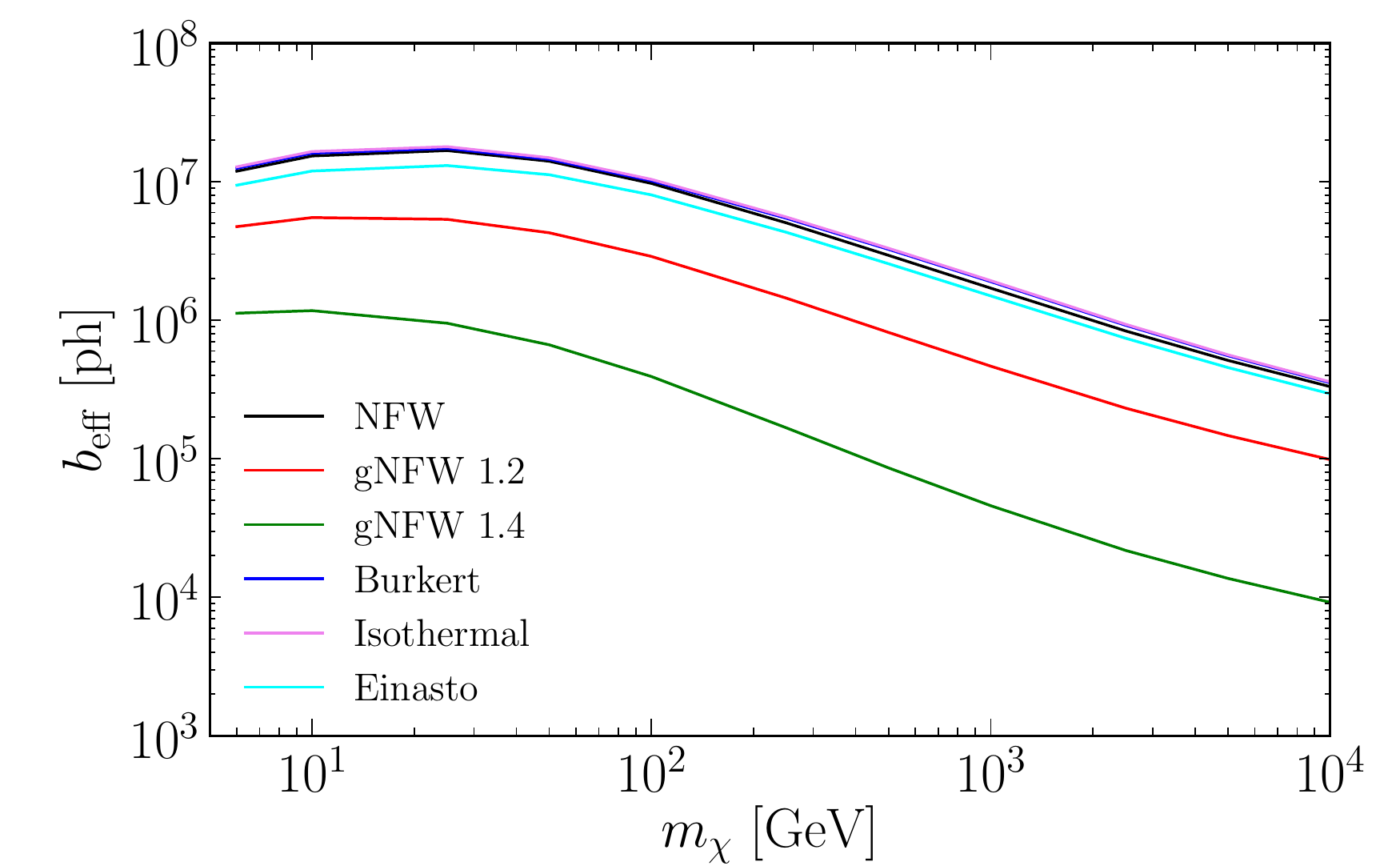}
    \includegraphics[width=0.49\columnwidth]{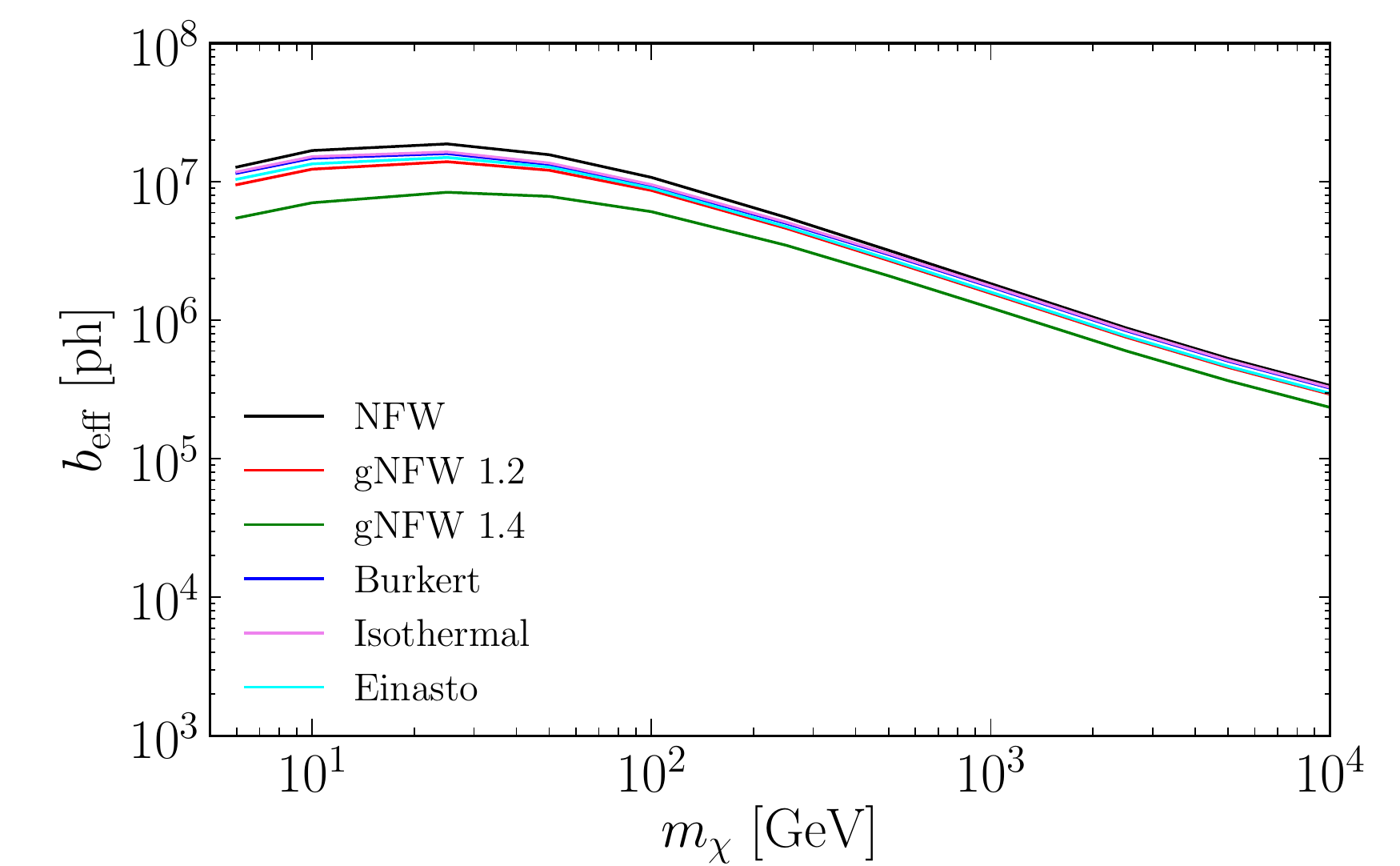}
  \end{center}
\caption{Estimated \beff for several DM radial profiles, 
  for a $60\degree \times 60\degree$ area centered on the
  Galactic center for 15~years of \evtclass{P8R2\_SOURCE} data.
  The plot shows the total integrated \beff for annihilations to \bb as a 
  function of the WIMP mass, \mchi.   
  The left-hand plot includes all Galactic latitudes $|b| < 30\degree$, the right-hand plot excludes
  the Galactic plane ($|b| < 2\degree$).
  \label{fig:gc_bkg_eff}}
\end{figure}

Therefore, in projecting the search sensitivity we account for such
systematic limitations.  The \beff (in counts)
for several radial profiles are shown in
Fig.~\ref{fig:gc_bkg_eff}.\footnote{Fig.~\ref{fig:gc_bkg_eff} was made
  using the ``binned model map simulations'' for the
  diffuse Galactic and isotropic background components, 
  together with the ``all-sky photon simulations'' of the 
  cataloged point sources as described in App.~\ref{app:method_monte_carlo}.}

\begin{figure}[!htbp]
  \begin{center}
    \includegraphics[width=0.49\columnwidth]{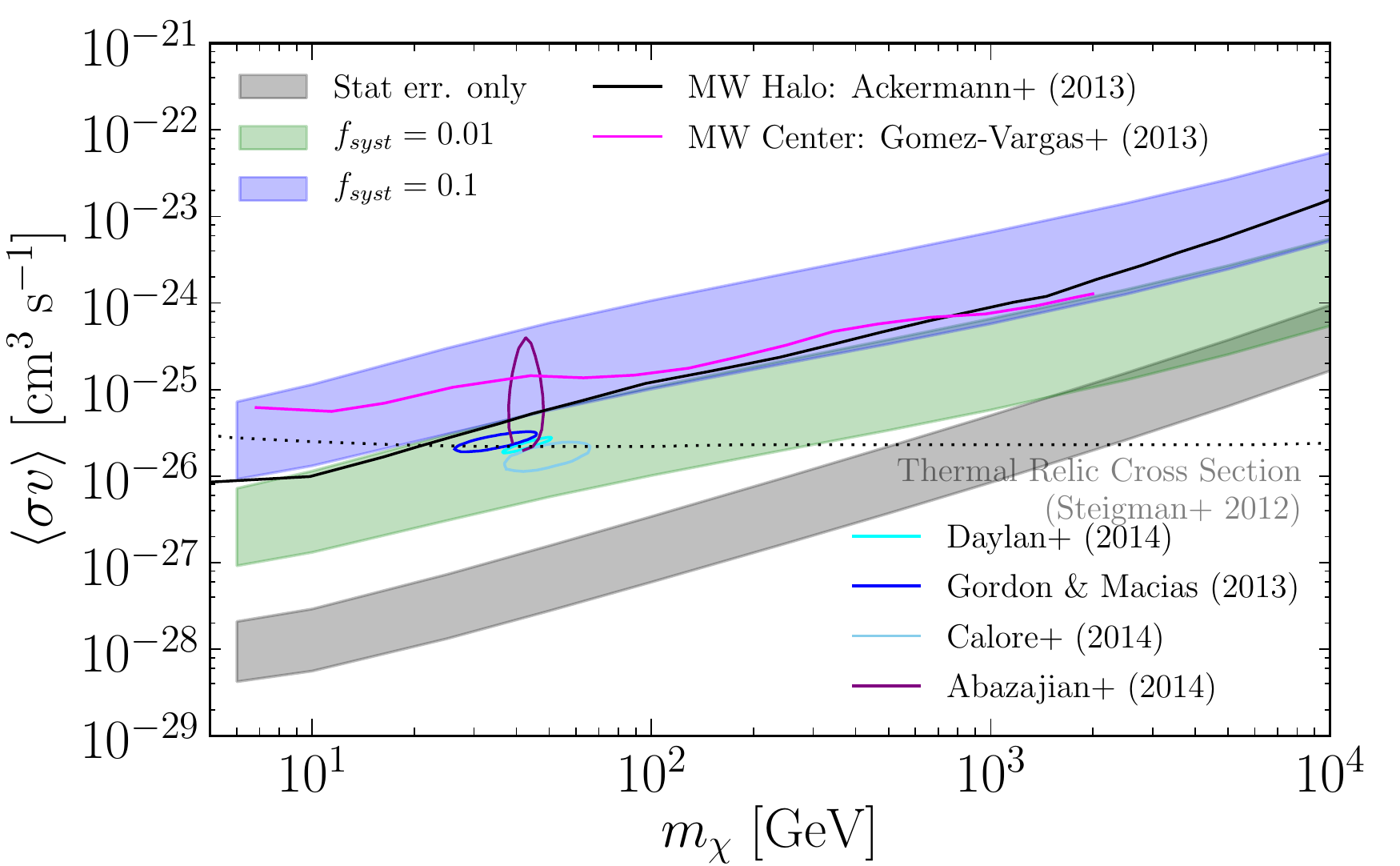}
    \includegraphics[width=0.49\columnwidth]{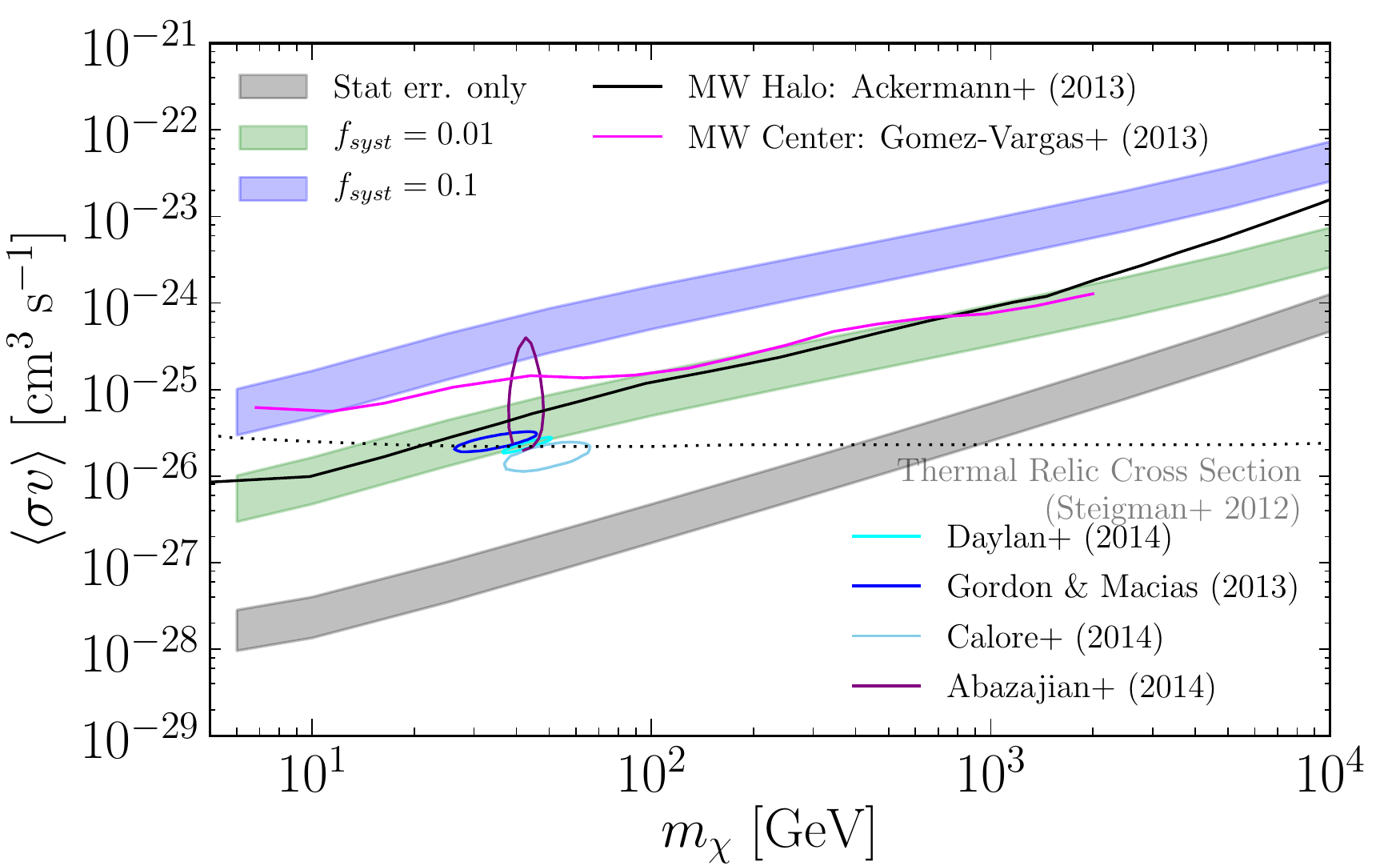}
  \end{center}
\caption{Projected limits for 15~years of \evtclass{P8R2\_SOURCE} 
  data, for annihilations to \bb as a function of 
  the WIMP mass, \mchi, when various levels of systematic 
  uncertainty are included.   The extrema of the 
  color bands give the projected limits for the NFW and gNFW 
  with $\gamma = 1.2$ radial profiles for a given level of systematic
  uncertainty.  The left plot includes all Galactic latitudes $|b| <
  30\degree$, the right plot excludes the Galactic plane ($|b| < 2\degree$).
  The ellipses and cross indicate the best-fit region obtained
  if the Galactic center excess is interpreted as a DM signal.
  \label{fig:gc_limits_nfw}}
\end{figure}

Fig.~\ref{fig:gc_limits_nfw} shows the expected upper-limit bands
for the statistical errors-only case as well as for indicative values 
of \fsyst (0.01 and 0.1).  

Fits for DM-like excesses in a scan of positions along the Galactic plane but away from 
the Galactic center (i.e., where one does not expect signals from DM)
find several other locations along the plane with putative signals
similar in size to the Galactic center excess (see, e.g., Fig. 11 of
~\cite{2015JCAP...03..038C}).    From a scan along the Galactic plane in 
$10\degree$ steps with 6 years of \evtclass{P8R2\_CLEAN} data we estimate
that the systematic uncertainties are of the order of $\fsyst =
0.02$, though they depend somewhat on energy and can reach $\fsyst
\sim 0.06$.  The fore/backgrounds are much brighter
within a few degrees of the Galactic plane, and some authors
(e.g.,~\cite{2014arXiv1402.6703D,2015JCAP...03..038C,2015PhRvD..91f3003C}) 
have chosen to mask the plane and confine the analysis to $|b| > 2\degree$, reducing
both the signal and the systematic uncertainties of the
fore/background modeling.

It is difficult to estimate how our understanding and modeling of the
fore/backgrounds will evolve, or our ability to rule out other
contributors such as pulsars.  However, as shown in Fig.~\ref{fig:gc_limits_nfw}, even
if the modeling uncertainties can be reduced to \newText{$\fsyst = 0.01$}
the interpretation of the nature of the Galactic center excess will be 
limited by systematic uncertainties of the background modeling.  
Thus, we believe that definitively confirming or ruling out a DM
interpretation of the Galactic center excess will require inputs from
searches of other DM targets \newText{or from other multi-wavelength data.
In particular, if the Galactic center excess is due to unresolved emission from 
pulsars, then additional LAT data should resolve some of those 
sources robustly and provide targets for dedicated radio pulsation searches.
If fact, arguments against pulsar interpretations
have relied on the non-observation
of many pulsars in the Galactic center region in the LAT data~\cite{2015JCAP...06..043C,2013PhRvD..88h3009H,2014arXiv1407.5583C}.}

\subsection{Known Satellites of the Milky Way}
\label{sec:dm_satellites}

Many DM subhalos have been %observationally 
discovered from optical surveys and follow-up spectroscopy as dark-matter-dominated dwarf 
spheroidal galaxies (dSphs) residing within the DM halo of our own Galaxy. 
The known dSphs have dynamical masses ranging from $\sim 10^{5}$ to $10^{8}\Msun$ 
and stellar half-light radii between 0.02 and 0.7\kpc~\cite{2012AJ....144....4M}.

The known dSphs are identified as collections of old, metal-poor 
stars, kinematic analyses of which reveals a binding DM halo up 
to a thousand times more massive than the stars themselves.  Multi-wavelength 
observations show that the objects contain little besides stars and
DM~\cite{Mateo:1998wg,Grcevich:2009gt,Spekkens:2014}. 
Without the requisite gas or cosmic-ray content, dSphs have no conventional 
mechanism for producing \grs.  This means that any DM signal
originating from dSphs must contend only with the diffuse \gray 
background.  As dSphs are primarily found far from the Galactic plane, the 
background intensity is often low and spectrally featureless.

Fitting the NFW density profile to the stellar velocity dispersions observed
in the nearest dSphs yields $J$~factors on the order of $10^{19}
\GeV^{2}\cm^{-5}$.  These values are \newText{fairly} robust; they are insensitive
to both the shape of the inner density profile and the level of
substructure within the dSph.  \newText{However, for particular dSphs
different authors 
have found a range of estimated $J$~factors 
that are larger than the quoted uncertainties of the individual
studies, suggesting that the measurements are potentially impacted by 
systematic biases (see,
e.g.,~\cite{2004PhRvD..69l3501E,2012PhRvD..86b3528C,2015ApJ...808L..36B,2016arXiv160307721U}).}

Although each individual $J$~factor is smaller than that of the
Galactic center, kinematically determined $J$~factors are known for 19 dSphs, so far.  
The predicted annihilation signal from the population of dSphs is
commensurately higher, and analyzing dSphs as a group results in
sensitivity competitive with other targets.  \newText{Furthermore, 
systematic disagreements of uncertainties of the $J$~factors are peculiar
to individual dSphs, rather than systematic across the set of dSphs.
Decreasing the $J$~factor of a single dSph, or increasing the uncertainty
of that $J$~factor, will lessen the impact of that dSph on the joint 
analysis, but will not greatly affect the overall result.  The effect
of the uncertainties in $J$~factors in a joint likelihood analysis 
was studied in detail in Ref.~\cite{2015PhRvL.115w1301A} (see in
particular Fig.~7), and the overall limits changed by less that 40\%
for all of the scenarios considered.}

\subsubsection{Known Satellites: Current Status}
\label{sec:dm_satellites_status}

Since early in the \Fermi mission, dSph analyses have provided cutting-edge 
constraints on \sigmav~\cite{2010JCAP...01..031S,2010ApJ...712..147A,
2011PhRvL.107x1302A,2011PhRvL.107x1303G,2012APh....37...26M,2013JCAP...03..018S}.  Upper 
limits using the new Pass~8 data set are some of the most constraining to date, 
ruling out WIMPs with masses below 100\GeV that annihilate through quark or $\tau$-lepton 
channels at the thermal relic cross section~\cite{2015PhRvL.115w1301A}.  
These limits are in mild tension with the masses and cross-sections best-fit for DM
interpretations of the Galactic center excess (\S\ref{sec:dm_milky_way}), 
and provide an essential cross-check for those claims.  

In addition to the accumulation of \gray data, dSph searches for DM are now
benefiting from a rapid increase in the number of known dSphs.  
Until recently, ultra-faint dSphs (those with luminosities $\lesssim 5\times 10^{4} \Lsun$) 
had only been discovered by the Sloan Digital Sky Survey (SDSS), which
primarily covered the northern hemisphere~\cite{York:2000}.  
Over the past two years, the Dark Energy Survey
(DES)~\cite{Abbott:2005bi} has begun to explore the 
southern hemisphere, discovering 17 new dSph candidates~\cite{Bechtol:2015wya,Koposov:2015cua,Kim:2015c,2015ApJ...813..109D}.  
Over roughly the same time period, the PanSTARRS survey
\cite{Kaiser:2002zz} contributed 
an additional 3 dSph candidates~\cite{Laevens:2015a,2015ApJ...813...44L}.
Two more candidates were also discovered in other optical surveys. 
If all of these are confirmed as dSphs with spectroscopic data, these surveys will have
more than doubled the size of the ultra-faint dSph population.
Indeed, Ref.~{\cite{DESY2_DSPHS} consider 28 kinematically confirmed dSphs
galaxies, 13 candidates that are likely to be dSphs (based on their 
optical properties), and 4 systems that may be either dSphs or globular
clusters.  The initial DES discoveries were promptly
investigated for \gray emission~\cite{2015ApJ...808L..36B,2015ApJ...809L...4D,2016PhRvD..93d3518L}, 
and yield results consistent with the previous studies of dSphs.  A projection of 
combined dSph sensitivity, including the future accumulation of 
both data and targets is discussed in \S\ref{sec:dm_satellites_projections}.

Finally, searches have also targeted other DM-rich
Milky Way satellites and structures, including the Large Magellanic
Cloud (LMC,~\cite{2015PhRvD..91j2001B}), the Small Magellanic Cloud
(SMC,~\cite{2016PhRvD..93f2004C}) and the Smith high-velocity cloud~\cite{2014ApJ...790...24D}).   
However, these targets are spatially
extended, and in the cases of the LMC and SMC have significant astrophysical \gray
emission.  Therefore %we expect that 
these searches %will be 
are
limited by the systematic uncertainties of the background modeling.

\subsubsection{Known Satellites: Sensitivity Projections}
\label{sec:dm_satellites_projections}

In the light of the active DES and PanSTARRS large-area optical
surveys and the upcoming LSST survey \cite{Ivezic:2008fe}, 
any projection of future dSph DM search sensitivity must include an
estimate of an expanded set of targets.  
After two years (out of five) of operation, 
DES has contributed several new likely dSph candidates~\cite{Bechtol:2015wya,Koposov:2015cua,Kim:2015c,2015ApJ...813..109D},
including a few that have already been 
confirmed as dSphs with spectroscopic followups~\cite{2015ApJ...808L..36B,2015ApJ...809L...4D}.

\newText{Predictions about the number and 
$J$~factor distribution of undiscovered dSphs
are very uncertain.  In particular, the faint end of the dwarf
galaxy luminosity function, the structural properties (and DM distributions) 
of the smallest satellites, and the radial distribution of 
subhalos that would host dSphs are not well known.}

\newText{The SDSS survey covered roughly 1/3 of the sky and 
discovered 15 ultra-faint dSphs; DES, PanSTARRS, and in particular
LSST, will cover complementary regions of the sky to 
significantly great depth.  Combining the distribution of optical luminosities of 
known dSphs with} N-body DM simulations and the expanded depth 
and sky coverage of the new surveys, we can anticipate 25 to 40 total dSphs to be
discovered by DES, and possibly hundreds by
LSST~\cite{2008ApJ...688..277T,2014ApJ...795L..13H}, 
\newText{however many of these dSphs would be more distant and 
have correspondingly smaller $J$~factors.   Even so, LSST is still
likely to contribute many dSphs with
$J$~factors above $10^{19}$ GeV$^{-2}$ cm$^{-5}$, and is also 
likely to contribute at least some dSphs with larger $J$~factors than
any discovered by DES~\cite{2015PhRvD..91f3515H}.}

\newText{In practice, the distribution of $J$ factors for the DES dSphs
has been similar to previously discovered dSphs, in spite of the
greater depth of the DES survey.   This could reflect that the
dwarf galaxy luminosity function continues below the faintest objects
discovered by SDSS, or it could simply be that the DES survey region 
has an excess of dSphs, because of the influence of the nearby 
Magellanic clouds.}

We will take 60 total dSphs as a conservative estimate of the 
total number of dSphs that can be used as targets for LAT 
searches, \newText{i.e., having $J$~factors that are large enough and
well determined enough to contribute the sensitivity of a joint analysis.}

As an all-sky monitor, the LAT has already, and will continue
to, observe the new targets for the duration of its lifetime.  All
that is required to incorporate them into a joint analysis are
locations and $J$~factors and their uncertainties.  To project the increased
sensitivity that will result, we simulated \newText{200} realizations of our entire search using the
``ROI-specific photon simulations'' and ``Binned model map
simulations'' described in App.~\ref{app:method_monte_carlo} but
duplicated our target set\footnote{The targets were placed at random locations and with
  $J$~factors (and $J$~factor uncertainties) sampled from the posterior distribution.}  up to three
times to reach 60 in total.
The effect of additional targets on the search sensitivity is illustrated in Fig.~\ref{fig:dsph_projected_targets}.

\begin{figure}[!htbp]
  \begin{center}
    \includegraphics[width=0.49\columnwidth]{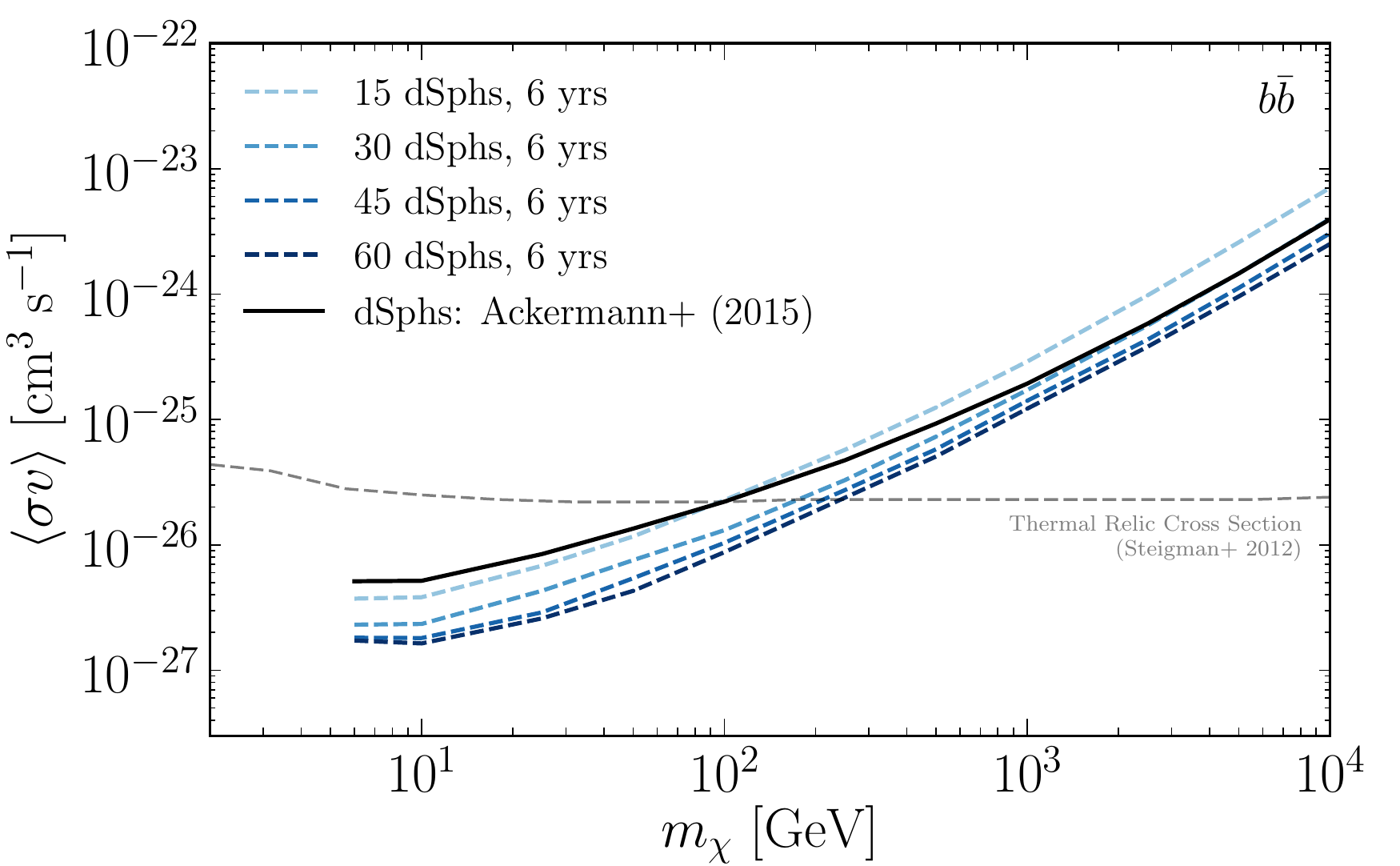}
    \includegraphics[width=0.49\columnwidth]{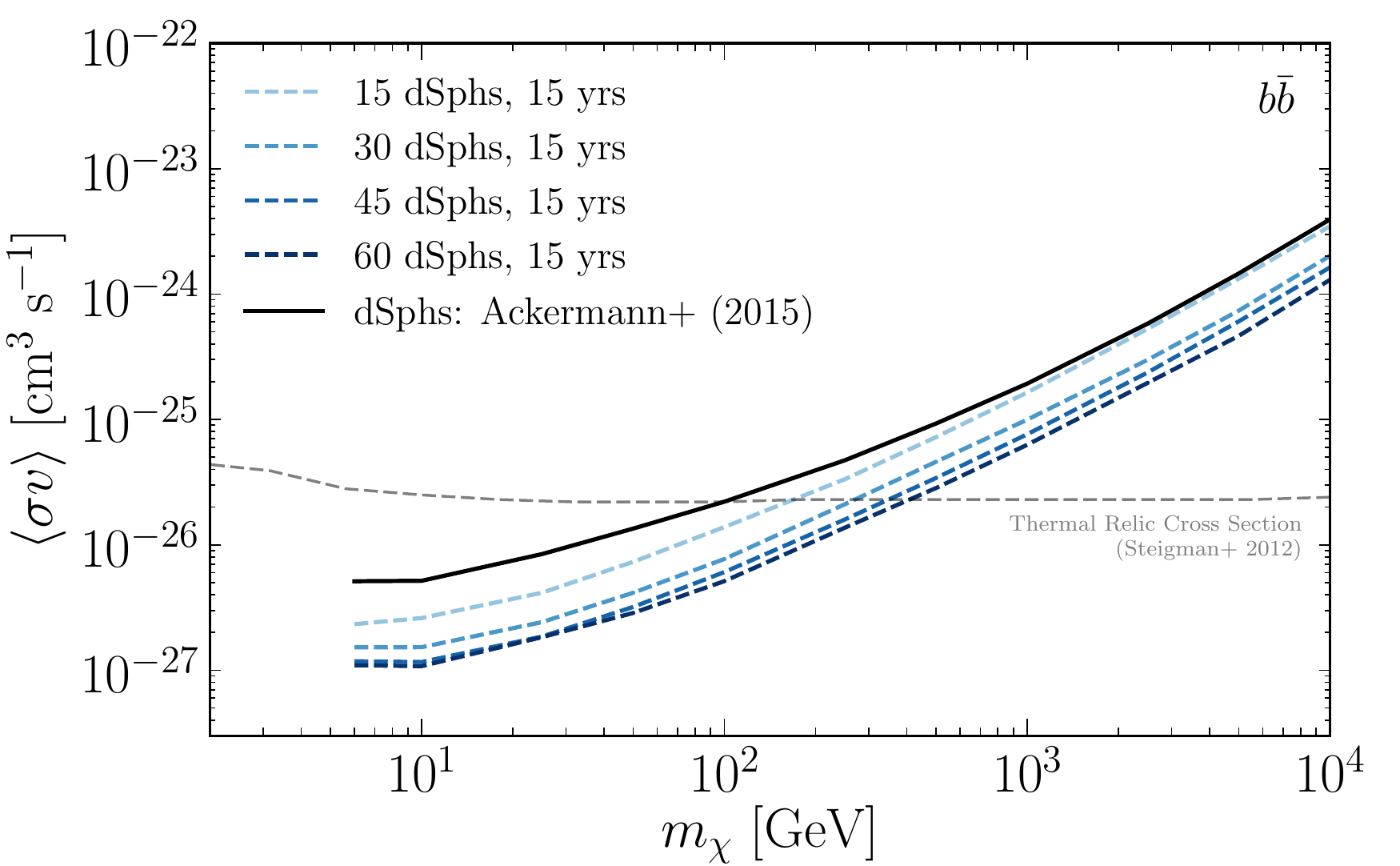}
  \end{center}
\caption{Projected upper limits on the WIMP annihilation cross section
  from the joint analysis of dSphs as a function of the size of the
  dSph sample on the assumption of 6- (left) and 15-year (right) data sets
  with \evtclass{P8R2\_SOURCE} data.  The solid black curve shows the observed 
  limit from the analysis of 15 known dSphs with 6 years of \evtclass{P8R2\_SOURCE} data~\cite{2015PhRvL.115w1301A}.
  Projections correspond to the median expected limit for the given number of dSphs and observation period
  \newText{from 200 \newText simulated realizations of the entire search (see text for details).}
  \label{fig:dsph_projected_targets}}
\end{figure}

We also examined how the expected sensitivity scales with time
for different masses and annihilation channels.
Fig.~\ref{fig:dsph_limit_scaling} shows the mean of the ratio of expected
limits for all of the simulated dSphs.  Because of the softer
spectrum in the \bb channel, the improvement in that channel 
is close to the expectation for a background-limited search (i.e., it 
scales as $\sqrt{t}$) for low masses, improves with increasing mass,
but does not reach the linear scaling we would expect for a purely
signal limited search.   On the other hand, in the harder
\tautau channel, the scaling behavior transitions from the 
background limited to signal limited cases around 100\GeV.

\begin{figure}[!htbp]
  \begin{center}
    \includegraphics[width=0.49\columnwidth]{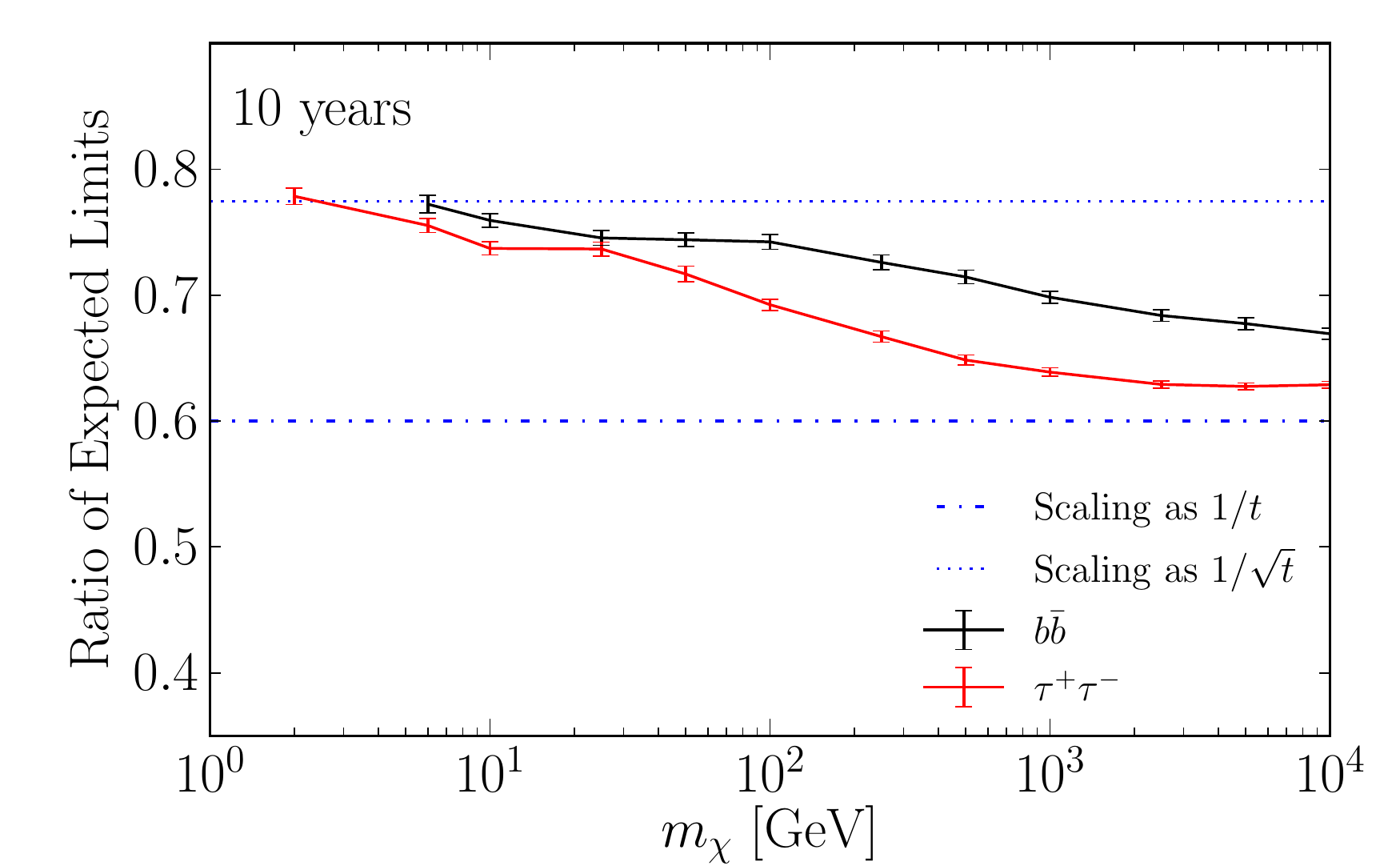}
    \includegraphics[width=0.49\columnwidth]{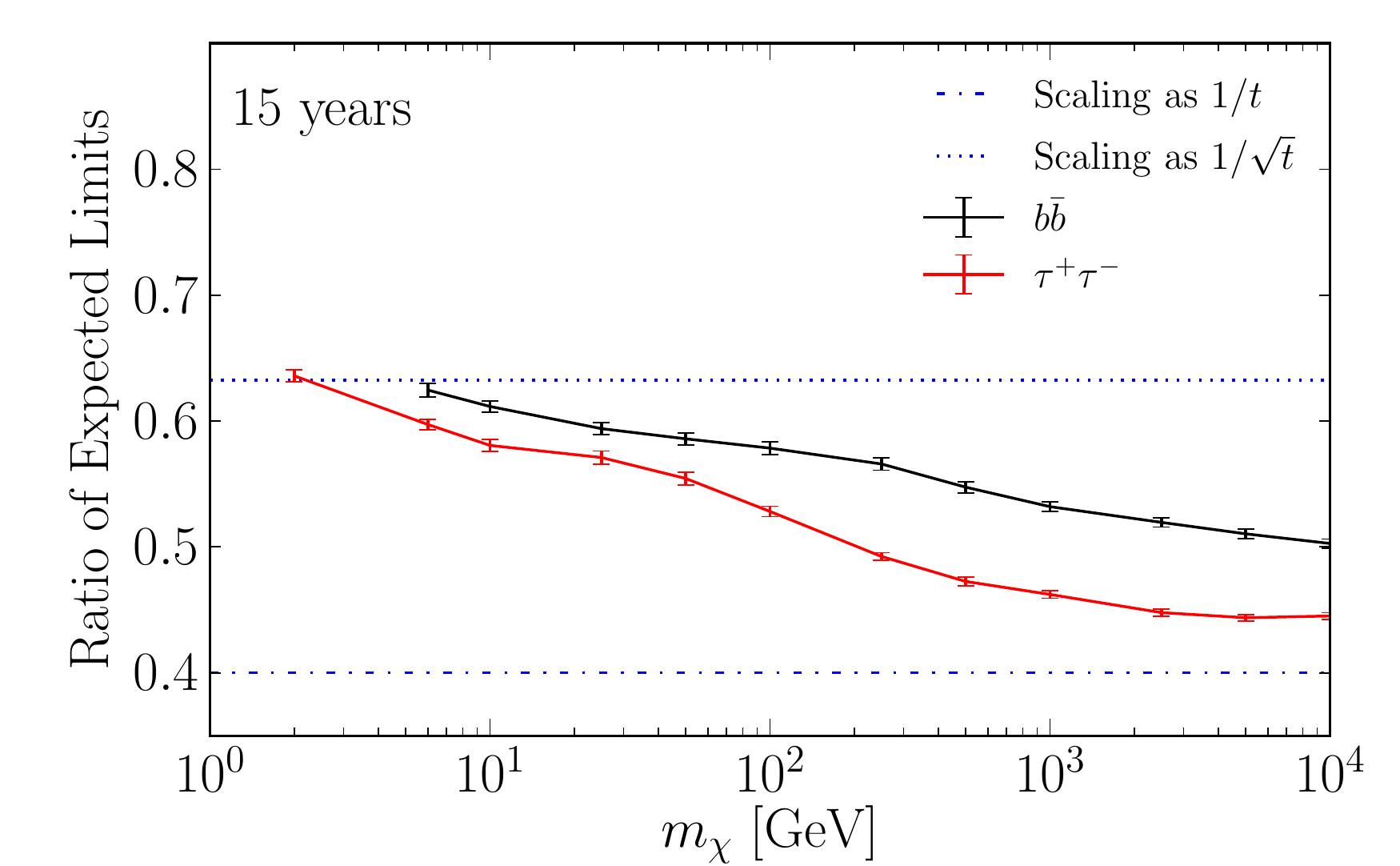}
  \end{center}
\caption{Ratio of the median expected upper limits on the WIMP annihilation cross section
  for 10~years of LAT data relative to 6~years (left) and 15~years relative to 6~years (right) for the
  \bb and \tautau annihilation channels.  The upper and
  lower horizontal lines are the expectations for the purely
  background limited (sensitivity scales as $\sqrt{t}$) and purely 
  signal limited (sensitivity scales linearly with $t$) cases.  
  We use the results for the projected sample of 60 dSphs (see
  Fig.~\ref{fig:dsph_projected_targets}) for these comparisons.
 \label{fig:dsph_limit_scaling}}
\end{figure}

The slope of the projected upper limit curve
near 100\GeV is close to one (\roughly 1.1 to 1.2 ) $\cm^3
\second^{-1} / \GeV$.   The mass 
for which the thermal relic cross section will be excluded
scales as the inverse of the slope times the improvement on the limits on
\sigmav.     This results in considerable extension of the mass range 
with limits at or below the 
thermal relic cross section with additional data, up to 
$> 400\GeV$ ($ > 200\GeV$) in the \bb (\tautau) 
channel with 60~dSphs and 15~years of data, see Fig.\ref{fig:dsph_limit_scaling_relic}.

\begin{figure}[!htbp]
  \begin{center}
    \includegraphics[width=0.49\columnwidth]{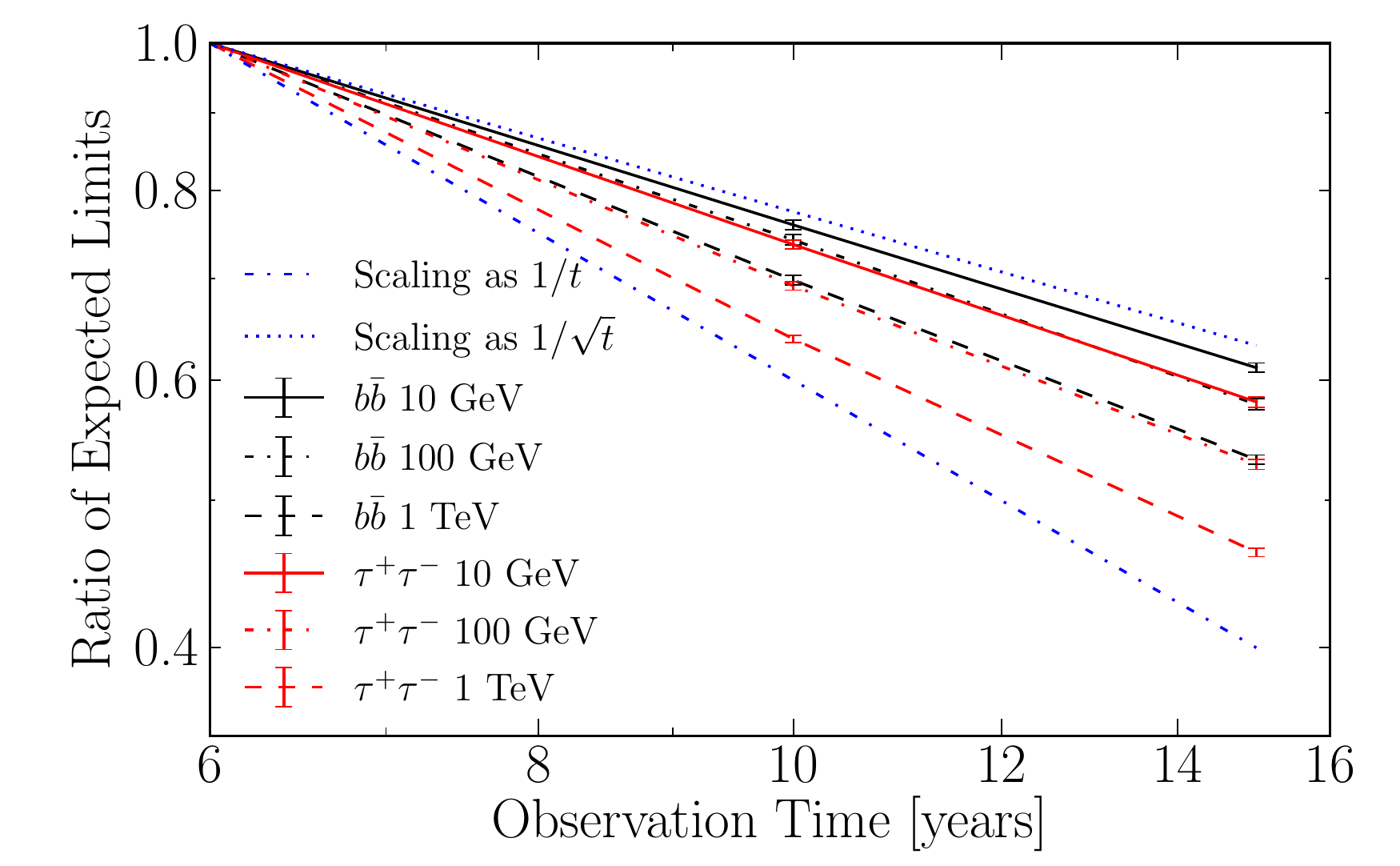}
    \includegraphics[width=0.49\columnwidth]{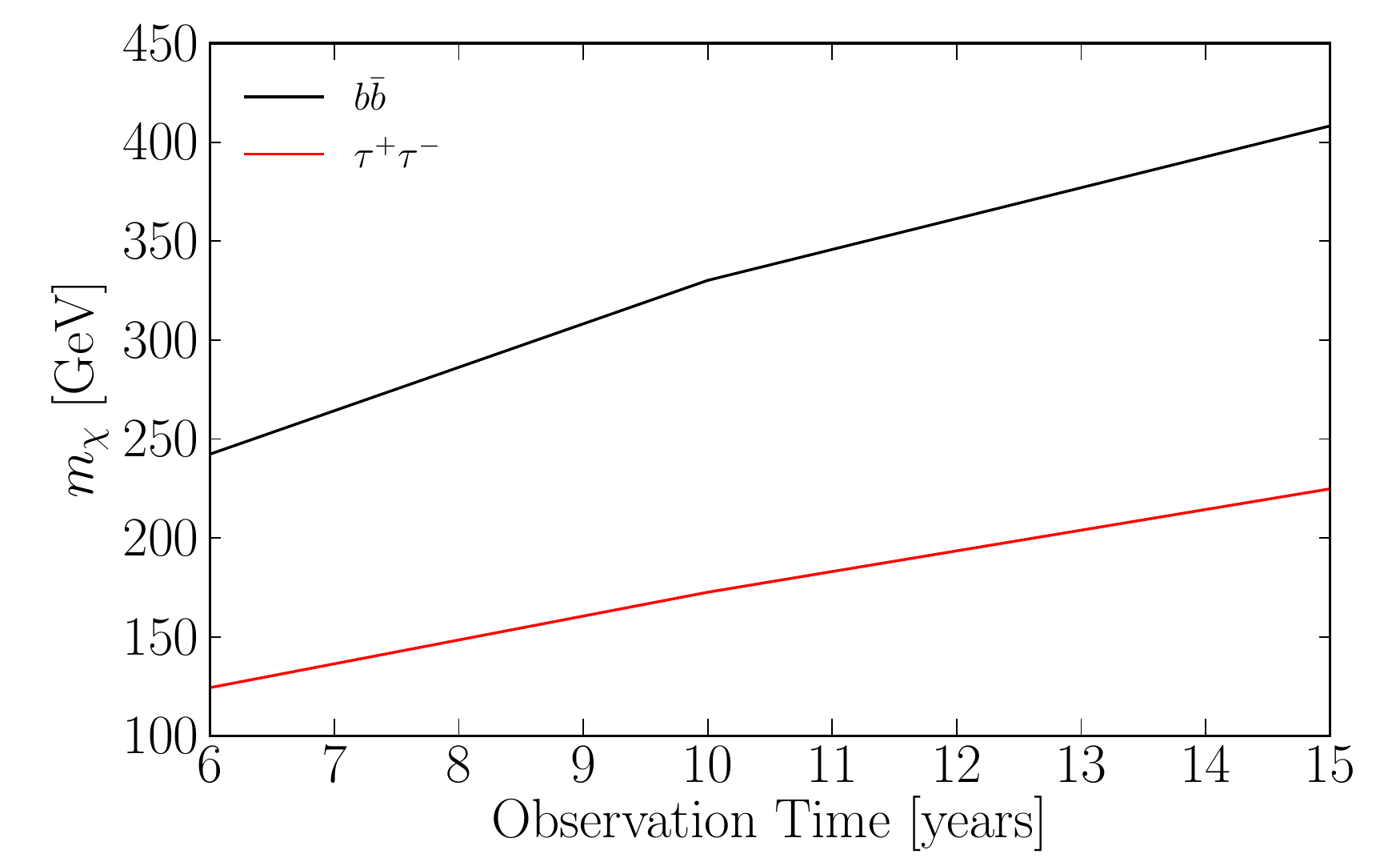}
  \end{center}
\caption{Left: ratio of the median expected upper limits on the WIMP annihilation cross section
  from 6 years of LAT data as a function of time for 10\GeV, 100\GeV and 1\TeV for the
  \bb and \tautau channels.  Right:  
  upper extent of the mass range over which the thermal relic cross section can be excluded, as a function of
  observation time, for the \bb and \tautau channels.  
  We use the results for the projected sample of 60 dSphs (see
  Fig.~\ref{fig:dsph_projected_targets}) for these comparisons.
 \label{fig:dsph_limit_scaling_relic}}
\end{figure}

Finally, given that optical surveys will be discovering new targets for
years to come, we consider the possibility that they
discover a dSph that is near enough and massive enough that
it would be clearly observable by the LAT for plausible DM annihilation cross-sections.

The $J$~factors of known dSphs scale approximately with the square 
of distance, i.e., the dSphs would have similar 
$J$ factors  (within about $0.4$~dex) if they were all at the same
distance: $J \sim 10^{18.3} \GeV^2\cm^{-5} \times ( d / 100\kpc )$.
For a 100\GeV DM particle annihilating to \bb at the 
thermal relic cross section we currently could expect a 5$\sigma$ significance
detection of any dSphs following that $J$~factor scaling relation
within \roughly 8\kpc and located away from the Galactic plane.   
That corresponds to a ``discovery'' volume of $\sim
2100\kpc^{3}$.   With 15 years of data that volume would at 
least double to $\sim 4200\kpc^{3}$.  For higher masses and in the 
harder \tautau channel the volume would increase
by a factor of four.   

\subsection{Undiscovered Satellites of the Milky Way}
\label{sec:dm_unid}

Current simulations of Milky Way-sized halos~\cite{2009Sci...325..970K,2008MNRAS.391.1685S} 
predict the existence of tens of thousands of Galactic DM subhalos. 
The most massive of these subhalos are expected to host the known
dSphs, while the stellar/baryonic component of smaller subhalos may 
be negligible. In fact, some of these subhalos may completely lack 
any astrophysical counterparts~\cite{2012ApJ...749...75S}.  Indeed, this leaves the 
exciting possibility that \grs from DM annihilation in these 
as {\it dark satellites} may be the only way to detect them. Given the 
typical masses and distances of dark satellites, they may represent 
excellent targets for DM searches in \grs, 
e.g.,~\cite{2005MNRAS.364..367D,2012JCAP...11..050Z,2014PhRvD..89a6014B,2012ApJ...747..121A,2010ApJ...718..899A,2011PhRvD..83b3518P,2012PhRvD..86d3504B}. 

Since the LAT surveys the entire \gray sky 
with unprecedented sensitivity, it is an ideal instrument for 
the discovery of new source classes.  In fact,
$\sim 1/3$ of the sources in the 3FGL catalog
lack firm associations with known counterparts at other 
wavelengths~\cite{2015ApJS..218...23A}. Some fraction of the 
unassociated \gray source population may be composed of dark
satellites. By comparing the predictions of N-body cosmological simulations 
with the observed number of unassociated sources, it is possible 
to place firm constraints on the DM annihilation cross section.
Additionally, if a population of candidate dark satellites is
found in \gray observations, the findings could be used to target deep
optical searches for hyper-faint dwarf galaxies.

\subsubsection{Undiscovered Satellites: Current Status}
\label{sec:dm_unid_status}

Several diagnostics have been explored to distinguish a
potential population of dark satellites from more conventional astrophysical sources in \grays:
\begin{enumerate}
\item{ \newText{{\it Multi-wavelength associations and identified sources:}  Dark satellites are
  not expected to emit in other wavelengths, so cataloged sources with 
  such associations are excluded.  Likewise, the detection of pulsations,
  typically either in \gray or radio data, can rule out sources as potential 
  dark satellite.}}
\item{ {\it Temporal variability:} Dark satellites are expected to be
  steady sources. This distinguishes them from background blazars,
  which can vary on minute-to-month time scales.}
\item{ {\it Spatial distribution:} Dark satellites are expected to be
  nearly isotropically distributed on the sky. Thus, selecting sources
  at high Galactic latitudes can increase the contrast with respect to
  a population of Galactic sources, such as young pulsars.}
\item{ {\it Spatial extension:} Nearby dark satellites may have
  spatial extensions that are measurable in \grs. This is unlikely for
  astrophysical sources at high Galactic latitudes, which are
  primarily either blazars or millisecond pulsars~\cite{2016arXiv160207303B}.
  \newText{When considering DM decay, as opposed to annihilation,
    the expected extension can be even larger, making spatial 
    extension a stronger discriminant.}}
\item{ {\it Spectral character:} In some cases, further restrictive
  cuts have been applied, motivated by some of the spectral properties
  expected from dark satellites (e.g., hard power-law indices,
  spectral features, detection above 10 GeV, etc.). A population of
  dark satellites should have a common spectral shape.}
\end{enumerate}

Current simulations of Milky Way-sized halos (e.g., Via
Lactea~II~\cite{2009Sci...325..970K}  or Aquarius~\cite{2008MNRAS.391.1685S})
are unable to simulate the subhalo hierarchy below a mass $\sim 10^5\msun$, 
while the theoretical minimum DM halo mass can be as low as 
$10^{-6}\msun$~\cite{Profumo:2004qt,Bringmann:2009vf,Cornell:2013rza}. Thus, 
extrapolations to lower masses are required to assess the contribution of nearby, low-mass subhalos 
to the potentially detectable dark satellites. 

Understanding the number and distribution of DM subhalos is an
important step toward their potential discovery. Previous work has begun to address 
this issue, e.g.,~\cite{2010ApJ...718..899A,2011PhRvD..83b3518P,2012PhRvD..86d3504B}. 
The authors of Refs.~\cite{2012JCAP...11..050Z,2014PhRvD..89a6014B} 
concluded from a search for DM subhalos in the 2FGL catalog~\cite{2012ApJS..199...31N}
that none of the unidentified sources in the catalog is
a clear DM candidate.  Ref.~\cite{2014PhRvD..89a6014B} found that
although many faint sources could not be ruled out as DM 
candidates, among sources with integral fluxes above 1\GeV,
$\Phi_\text{1~GeV} > 10^{-9}\cm^{-2}\second^{-1}$ (that is, the brightest ones) all but a handful 
could be ruled out as DM candidates for almost any
combination of annihilation channel and \mchi.   They used this
finding in combination with results from the Aquarius simulation on
subhalo abundance and internal structure to demonstrate that this
technique could be used to obtain competitive limits on \sigmav.

The systematic uncertainties associated with 
modeling the subhalo population (e.g., different internal structure
models, mass functions, radial distributions, sub-substructure) have
not been deeply explored. For example, an updated analysis using the 3FGL catalog
and a more conservative model to describe the dark satellites'
structural properties obtained a factor \roughly 4 weaker upper 
limits on \sigmav~\cite{2015JCAP...12..035B}, and in
fact identified a set of unassociated catalog sources that is
consistent with a DM-interpretation of the Galactic center excess
(e.g., \roughly50\GeV DM annihilation to \bb).

\newText{The identification and association of cataloged sources continues
after the production of the catalog.  In particular, many unassociated 
LAT sources have been identified as pulsars from radio observations
(see e.g.,~\cite{2013ApJS..208...17A} and references therein) and others have been associated with blazars 
using data from the {\it WISE} mission and optical spectroscopy~\cite{2015ApJS..217....2M}.}

\subsubsection{Undiscovered Satellites: Sensitivity Projections}
\label{sec:dm_unid_projections}

An extended LAT mission will benefit the search for dark satellites
 in the following ways:
\begin{enumerate}
\item{Setting a fainter threshold for source detection. A lower source
  detection threshold increases the number of faint dark satellites
  that could be detected as unassociated sources.}
\item{Providing better localization of sources, which effectively
  lowers the threshold for source association.  The source association
  procedure assesses the chance false association rate, which scales
  with the area of the source localization region.  Improving
  the localization reduces this area and hence the chance
  false association rate, allowing for more associations at a given
  probability threshold (see, e.g., \S~5 of
  \cite{{2015ApJS..218...23A}} for a discussion of the source
  association procedure).}
\item{Providing better spectral characterization of unassociated
  sources, thus setting a fainter threshold at which
  highly curved DM-like spectra and power-law
  spectra can be distinguished.}
\item{Providing greater sensitivity to spatial extension of
  unassociated sources.}
\item{Providing a longer baseline for detecting temporal
  variability. Temporal variability can be used to associate sources
  to background blazars, reducing the number of unassociated
  sources}
\end{enumerate} 

To project the sensitivity for dark satellite searches, we have
performed a study based on Refs.~\cite{2014PhRvD..89a6014B} and
\cite{2015JCAP...12..035B}, which examine the unassociated sources 
for potential DM candidates.  To do this we started with all of
the sources in each of the LAT published point source catalogs and
applied a few simple criteria:

\begin{enumerate}
\item{We excluded all associated sources.}
\item{We excluded all sources within $20\degree$ of the Galactic plane.  This 
    criterion was applied because the source detection threshold is much higher in 
    the plane than at higher Galactic latitudes.}
\item{We excluded variable sources; specifically we exclude sources with a variability 
    index $h > 80$, following~\cite{2014PhRvD..89a6014B}.}
\item{Finally, we excluded high-significance sources, i.e., 
    those that were detected with $TS > 100$; the better
    characterization of these sources makes it much easier to 
    identify multi-wavelength counterparts and to 
    distinguish astrophysical sources from potential 
    dark satellites.}
\end{enumerate}

In Fig.~\ref{fig:catalog_assoc} we show that each of the LAT source
catalogs has an ``exclusion threshold'' above which very few (i.e.,
$N_{\rm cand} \lesssim 20$) 
catalog sources pass the above criteria to be viable dark satellite
candidates.   The right panel of Fig.~\ref{fig:catalog_assoc} also indicates that
for a given $N_{\rm cand}$ the effective threshold scales 
roughly as $\sqrt{t}$.

\begin{figure}[!htbp]
  \begin{center}
    \includegraphics[width=0.49\columnwidth]{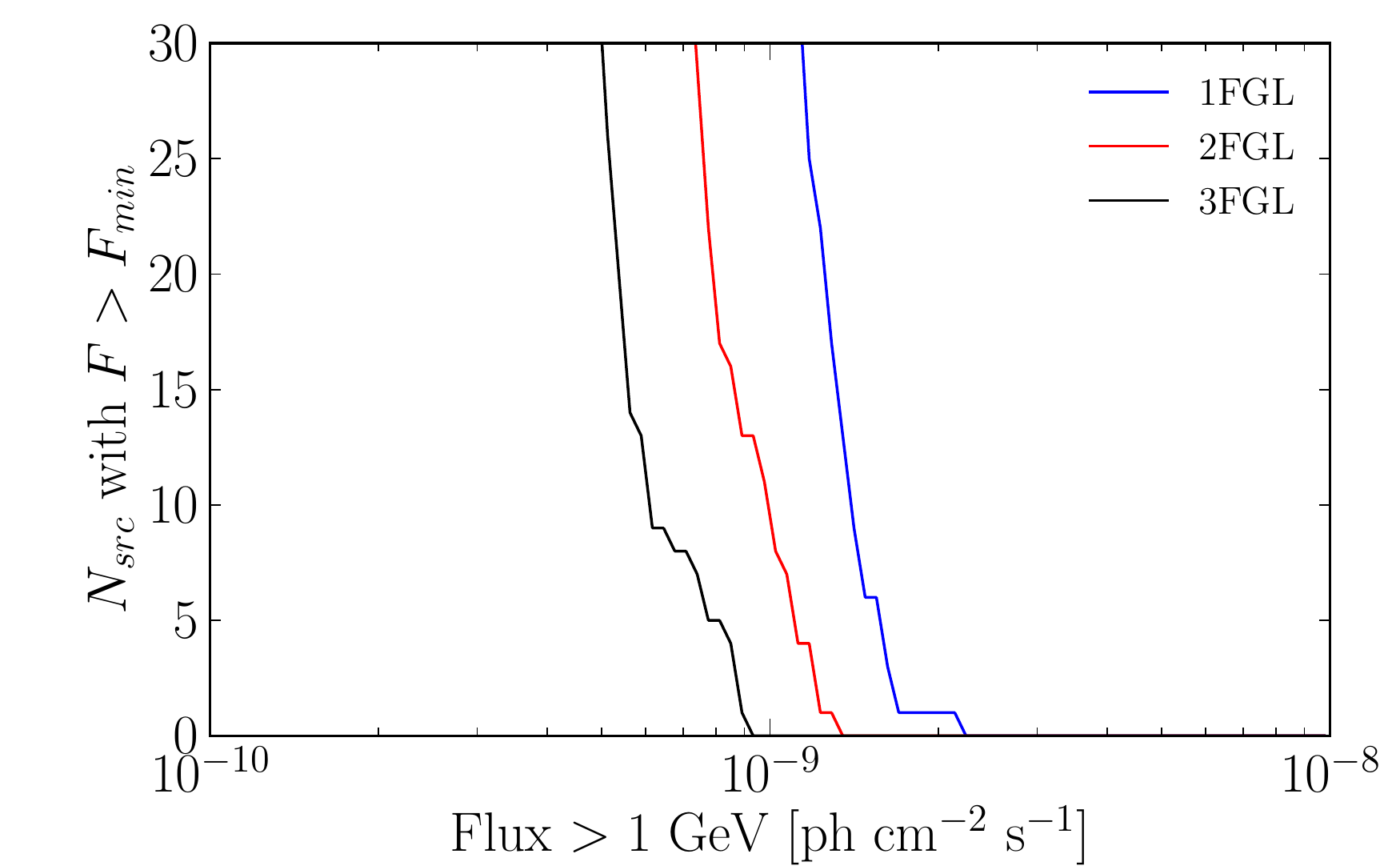}
    \includegraphics[width=0.49\columnwidth]{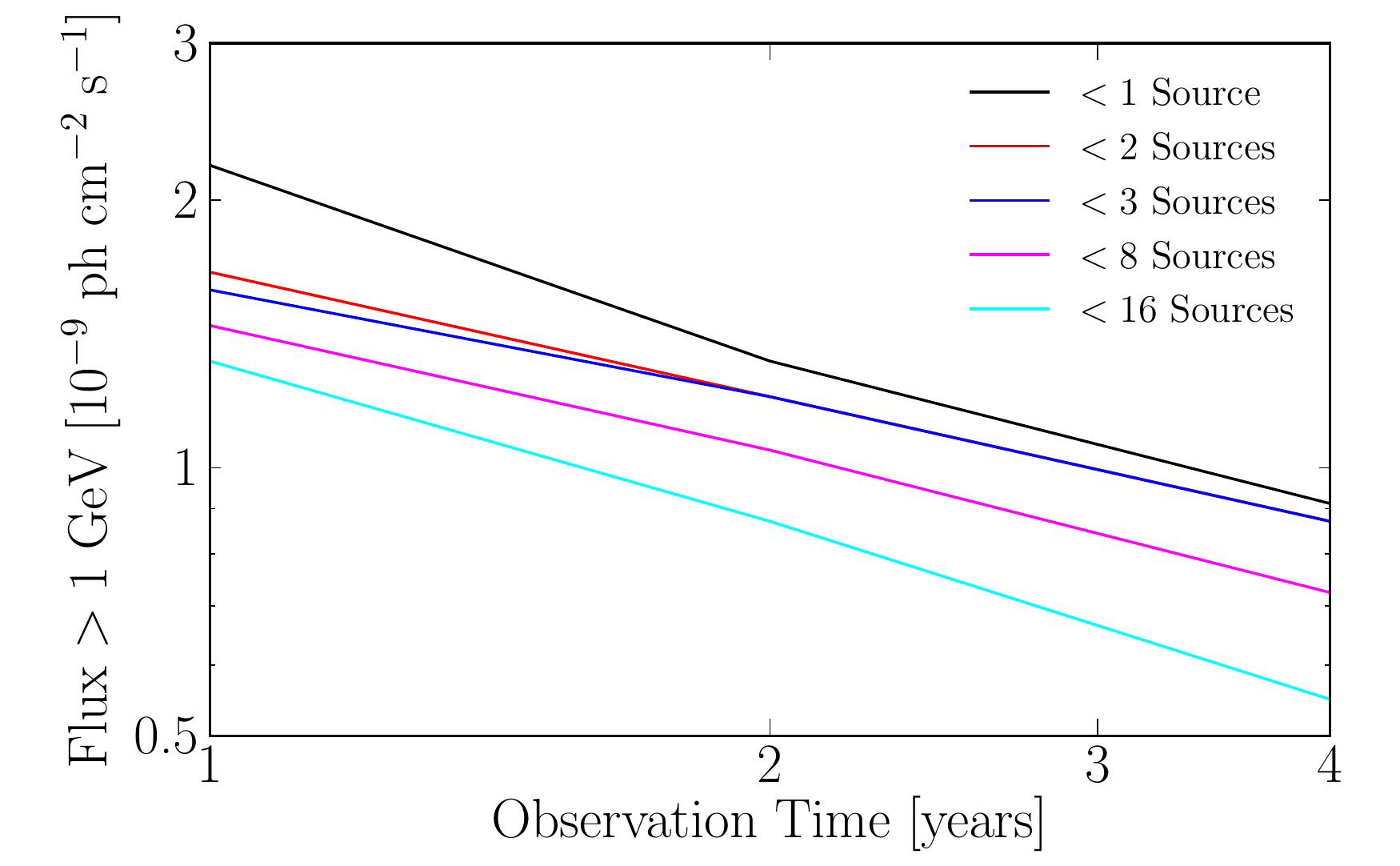}
  \end{center}
\caption{Left: number of potential DM candidate sources 
  in the 1FGL (1 year of LAT data~\cite{2010ApJS..188..405A}), 2FGL (2 years of data~\cite{2012ApJS..199...31N}) and 3FGL 
  (4 years of data~\cite{2015ApJS..218...23A}) catalogs.  Right: time-progression of the 
  threshold above which all but $N_{\rm cand}$ sources have been excluded as
  potential DM candidates.     
  \label{fig:catalog_assoc}}
\end{figure}

We can then use estimates of the expected flux distribution of dark 
satellites from N-body simulations of Galactic structure.   Following 
Refs.~\cite{2015JCAP...12..035B} we have adopted the relationship
for $N_{\rm cand}$ with integral flux above 1~GeV, $I_{\rm
  1 GeV}$, greater than a threshold sensitivity S: 

\begin{equation}
\label{eq:dNdS_unid}
N(S) = \left( \frac{S_0}{S}\right)^{1.5} \frac{\sigmav}{\sigmav_{0}},
\end{equation}

\noindent where the scaling factors are $S_0 = 9.83\times10^{13} \photon
\cm^{-2} \second^{-1}$ and $\sigmav_0 = 6.23\times10^{-27} \cm^3 \second^{-1}$.

Using the relationship from Eq.~\ref{eq:dNdS_unid} we can project the 
95\% CL upper limits by solving for the value of \sigmav for which 
$N_{\rm cand} > 3$ but no candidates are observed.   
The results are shown in Fig.~\ref{fig:unid_limits}.   We expect this search 
to be sensitive to annihilation at the thermal relic cross section 
for masses up to 100\GeV in the \bb channel.

\begin{figure}[!htbp]
  \begin{center}
    \includegraphics[width=0.49\columnwidth]{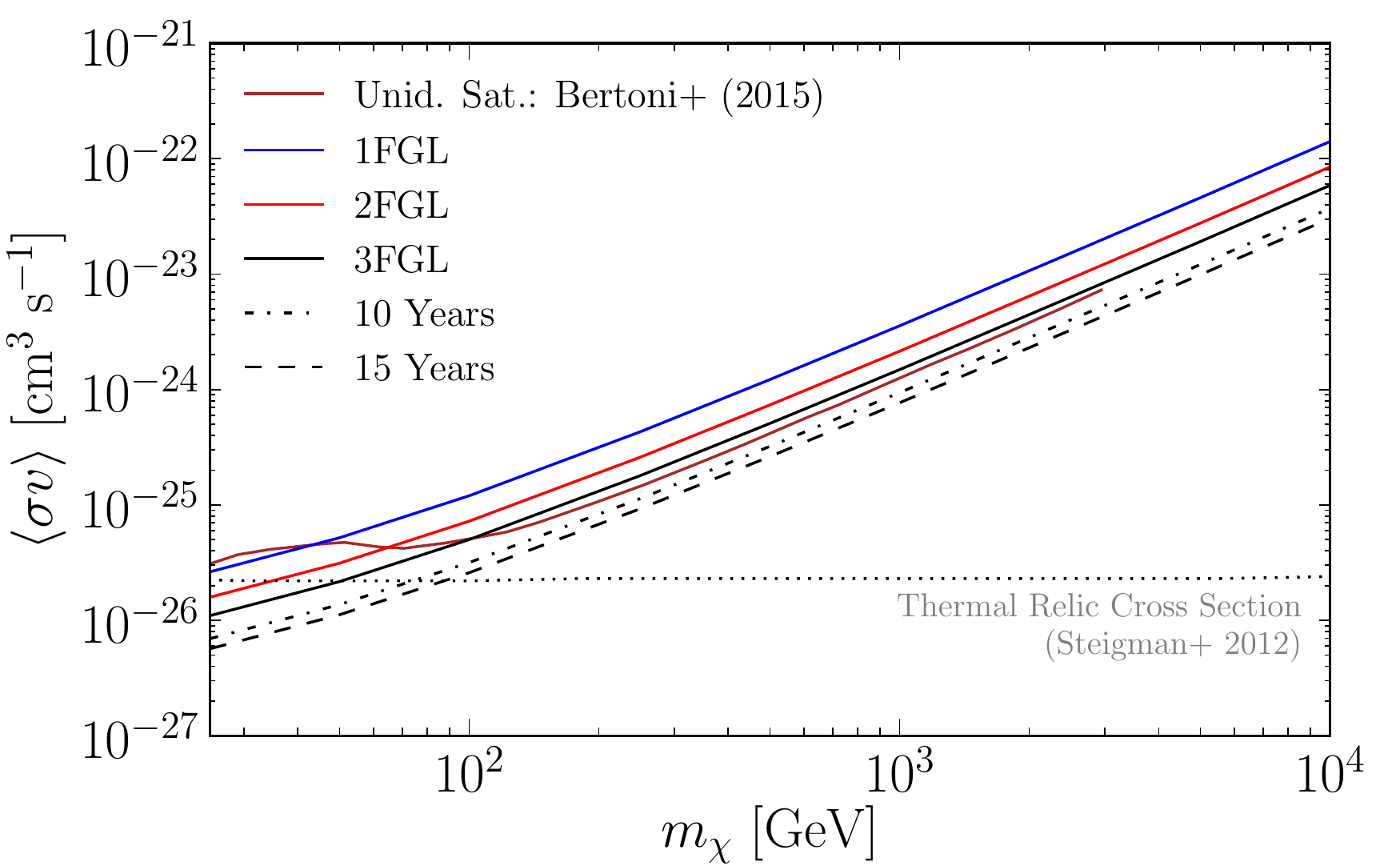}
  \end{center}
\caption{Projected upper limits derived from considering the expected
  number of dark satellite candidates $N_{\rm cand}$ in LAT sources
  catalogs as a function of \sigmav.   The extrapolations to 10 and
  15~years are based on an assumed $\sqrt{t}$ scaling of exclusion
  threshold for $N_{\rm cand} < 1$ of the 3FGL shown in Fig.~\ref{fig:catalog_assoc}.
 \label{fig:unid_limits}}
\end{figure}

\subsection{Galaxy Clusters}

\label{sec:dm_clusters}

The study of DM in galaxy clusters has a long history, dating
back to the first evidence of DM found by F. Zwicky \cite{Zwicky:1933gu}. The systematic uncertainty 
in the determination of DM density profiles and $J$~factors in 
clusters is similar in many ways to the case of dSphs discussed
above.  The dominant uncertainty in the $J$~factors for clusters arises 
from two orthogonal astrophysical considerations. First,
the empirically-measured cluster mass profiles, which 
are derived from a combination of X-ray temperature profiles and gas
kinematics, e.g.,~\cite{1986RvMP...58....1S}, are uncertain. 
Second, the predicted 
\gray luminosity that arises from the contribution of DM 
substructure in the clusters is also uncertain. The role of DM substructure (or
subhalos) for the total DM annihilation flux can be conveniently 
expressed in terms of a boost factor $b$. 
Determining values for $b$ requires 
assumptions on the relative abundance of subhalos and of
their structural properties. 
These assumptions are usually inspired by the results from
N-body cosmological simulations. However, even the highest-resolution 
current simulations fail to resolve the whole subhalo
hierarchy.  For cluster-sized simulations,
this mass resolution limit is
$\sim10^{8}\msun$~\cite{Gao:2012aa,2016MNRAS.tmp...66H},
as compared to the smallest subhalos, which may have masses as low as 
$10^{-6}\msun$ \cite{Profumo:2004qt,Bringmann:2009vf,Cornell:2013rza}, 
Estimates of the boost to the DM annihilation 
signal due to subhalos is thus dependent on extrapolations over 
several orders of magnitude \cite{Sanchez-Conde:2014aa}. 

Another challenge to searching for \grs from DM interactions 
in galaxy clusters is that clusters are believed to host 
relativistic particle populations, most notably cosmic-ray electrons, 
but possibly protons as well \cite{2014IJMPD..2330007B}. As such, 
\grs may arise from astrophysical processes in
the intra-cluster medium (ICM).

\subsubsection{Galaxy Clusters: Current Status}
\label{sec:dm_clusters_status}
Historically, X-ray observations have been pivotal for the selection
of cluster samples. More specifically, the HIFLUCGS
catalog~\cite{Reiprich:2002aa} has been the basis for several recent
analyses.  The catalog contains a complete flux-limited sample of the 
brightest X-ray clusters at low redshift $z\leq0.2$. 
In most cases, the \gray analyses have considered either a subset of 
the most nearby, most massive clusters (e.g., Coma, Fornax or the Virgo 
cluster~\cite{2010JCAP...05..025A,Huang2011,2012ApJ...757..123A,2012JCAP...07..017A,2012JCAP...01..042H,2014MNRAS.441.2309P,Ackermann:2015ab}) 
or targeted entire samples~\cite{Ackermann2010b,2016arXiv160206527L}. The Perseus cluster is
another well-motivated target but due to the presence of two
bright Active Galactic Nuclei (AGN) in the cluster along with its proximity to the Galactic
plane, this target is probably better suited for Cerenkov
telescopes~\cite{Aleksic:2010aa,Aleksic:2012aa}. 
At present no cluster-wide emission has been detected in 
\grs.\footnote{This is with the exception of a reported excess 
  from Virgo~\cite{Han2012,Macias-Ramirez2012,Ackermann:2015aa}. 
  However, the recent analysis in \cite{Ackermann:2015aa} indicated that
  this excess is likely associated with unmodeled Galactic diffuse 
  emission and/or a combination of weak point sources.}

In sample studies, a stacking analysis is often used to reach a
greater sensitivity~\cite{Reimer:2003aa,Huang2011,2012JCAP...07..017A,2013A&A...560A..64H,Ackermann:2014aa,Prokhorov:2014aa,2016arXiv160206527L}. 
For example, joint likelihood fitting of the sample may provide a factor of 2--3 improvement with respect to the case for which only the cluster with the
highest $J$~factor was used in the analysis~\cite{Huang2011,Zimmer:2011aa}.

One challenge when analyzing nearby galaxy clusters is
that both their predicted non-thermal and DM-induced 
\gray emissions are expected to be spatially extended (up to 
\roughly3\degree if the nearest cluster, Virgo, is excluded.\footnote{The Virgo cluster has a
  spatial extent of \roughly14\degree and analyzing the surrounding
  region in the sky presents a number of
  challenges~\cite{Ackermann:2015aa}.}) 
At the same time clusters host AGN. While
significant point-like flux can be expected from the central AGN in a
cluster, searches have only yielded upper
limits~\cite{2013MNRAS.429.2069D,Prokhorov:2014aa}. 
A sub-threshold AGN residing in a cluster would
manifest itself as diffuse \gr emission with characteristics similar 
to that expected from the ICM (e.g., a power-law spectrum with 
$\Gamma=2.3$ or softer). This would result in an additional foreground
emission, predominantly toward the center of the cluster, where the 
DM emission also peaks. However, the DM annihilation flux profile is 
expected to be shallower and more extended than the non-thermal 
emission profile once DM substructures are taken into 
account~\cite{2011JCAP...12..011S}. Hence, the cluster outskirts may 
be an attractive possibility for DM searches.

\subsubsection{Galaxy Clusters: Sensitivity Projections}
\label{sec:dm_clusters_projections}

For our sensitivity projections we consider a set of 32 clusters,
chosen from the HIFLUCGS catalog, maximizing the total $J$~factor of all 
targets and minimizing potential overlap between nearby 
clusters.\footnote{In our analysis we also consider each of the 106
  HIFLUGCS clusters as an isolated system.  Given the broad point-spread function 
  at the lowest energies, overlaps may occur between neighboring targets, which 
  would introduce a potential bias in a joint
  analysis~\cite{Ackermann:2014aa}.} 
Spatially, each cluster is modeled under the assumption of an NFW DM density
profile.  We include DM subhalos down to $10^{-6}\msun$. For the
description of the subhalos (and consequently of the subhalo boost,
$b$), we follow the recent work~\cite{Sanchez-Conde:2014aa}, which predicts a 
flattening of the halo concentration parameter for the smallest halo
masses, and consequently, moderate $b\sim$ 30--40 for galaxy clusters.  We adopt 
the same approach as for the dSphs (see Section
\ref{sec:dm_satellites_projections}), and use the ``ROI-specific
photon simulations'' described in App.~\ref{app:method_monte_carlo}.
Fig.~\ref{fig:cluster_predictions} shows the effect of continued data 
taking on the upper limits to the DM annihilation cross section.
These projections were made by rescaling the projected sensitivity
calculated in Ref.~\cite{Ackermann:2014aa} by the $\sqrt{t}$ 
and accounting for the improved sensitivity with Pass~8 data.
Overall, the limits improve by a factor $\roughly 2$ over
the whole WIMP mass range considered.

\begin{figure}[!htbp]
  \begin{center}
    \includegraphics[width=.50\columnwidth]{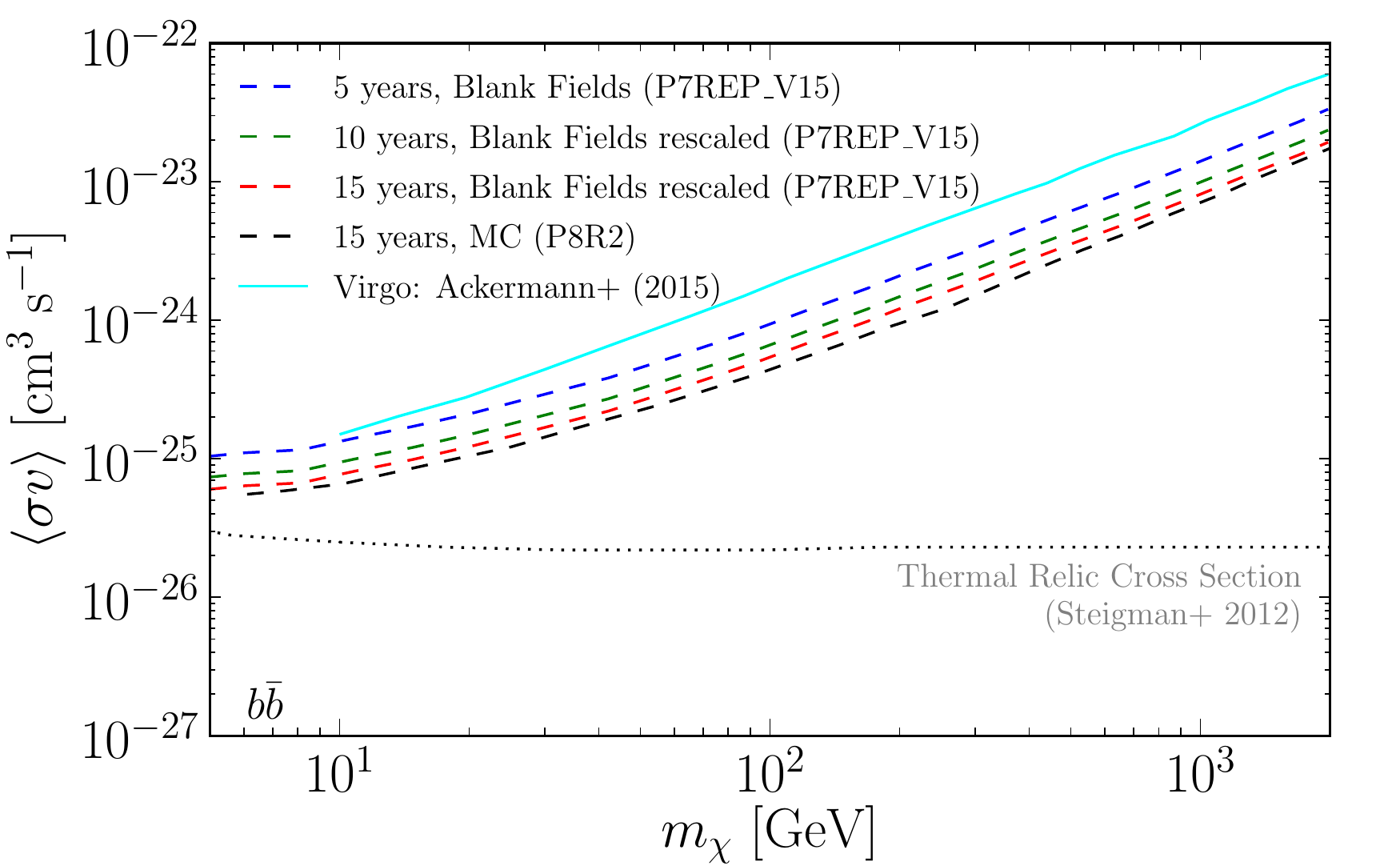}
  \end{center}
\caption{Projected upper limits on the WIMP annihilation cross 
section from a joint analysis of 32 nearby galaxy clusters as a function
of DM mass for DM annihilating into \bb. The colored dashed lines
correspond to rescaling of the obtained median sensitivity from an 
analysis of blank fields using Pass~7REP data~\cite{Zimmer:2015aa}, 
while the black dashed line corresponds to the expected median sensitivity 
based on simulations of the \gray sky. The limits obtained from the Virgo galaxy cluster in \cite{Ackermann:2015aa} 
are also shown for comparison.
\label{fig:cluster_predictions}}
\end{figure}

\subsection{Cosmological WIMP Searches}
\label{sec:dm_cosmo}

The discussion above has focused on DM annihilation in the Milky Way, 
Milky Way satellites and galaxy clusters.  Searches for DM signals in \gray data are 
also sensitive to the accumulated emission from DM
annihilation in all the DM halos that have formed in the
Universe~\cite{Fornasa:2015qua}. Though this emission is
``unresolved'', it can be modeled given a mass function for
DM halos and a description of the DM halo structural properties. The 
latter is usually made by assuming a universal form for the DM
density profile, e.g., NFW~\cite{1996ApJ...462..563N} or
Einasto~\cite{Einasto:1965} (see also App.~\ref{app:dark_matter_profiles}). As for the halo mass function, in
principle, it could be deduced from the observed
luminosity function of galaxies.  However, this method is hindered 
by the current uncertainties in the mapping of galaxy luminosity to 
DM halo mass.  A more robust understanding of the DM halo mass 
function and its redshift evolution comes from N-body cosmological 
simulations~\cite{Jenkins:2000bv,Springel:2005nw}. Two of the 
largest-volume cosmological simulations to date, e.g., provide 
a statistically-complete sample of DM halos down to a maximum circular 
velocity of about 50 km/s, or a mass of approximately
$10^9\Msun$~\cite{2009MNRAS.398.1150B,2011ApJ...740..102K}. 
Based on these and previous N-body simulation results, different 
functional forms for the halo mass function and its redshift
dependence have been proposed in recent years,
e.g.,~\cite{Sheth:2001dp,Tinker:2008ff}. 
The integrated DM annihilation signal from all halos at
all cosmic epochs can then be predicted by combining the information 
on the halo mass function and the DM density profile of individual halos.

\subsubsection{Searches for Contributions to the IGRB Spectrum: Current Status}
\label{sec:dm_isotropic}
\label{sec:dm_cosmo_status}

A possible way to find cosmological DM signatures is by searching for
features in the IGRB spectrum~\cite{2015JCAP...09..008T,2012PhRvD..85b3004C,2015PhRvD..91l3001D,2010JCAP...11..041A,2010JCAP...04..014A,Ando:2015qda}.
The statistics are good but this method is nonetheless limited by the
large systematic uncertainties on the reconstructed IGRB related to the
uncertainty of the Galactic diffuse emission (which is a strong foreground to
the IGRB).  At present these dominate the statistical uncertainties
over the entire energy range (see Fig.7 in Ref.~\cite{2015ApJ...799...86A} and
Appendix A in Ref.~\cite{2015JCAP...09..008T}).   
Known astrophysical source populations contributing to the IGRB typically can be modeled 
starting from the resolved component of the population present in the
EGB, or from observations at other wavelengths (radio, IR, X-ray).  From these models, blazars,
star-forming galaxies, and radio galaxies are found to be the main 
contributors, and altogether can make up the total of the measured
IGRB intensity although with large
uncertainties~\cite{Ajello:2015mfa}.   DM constraints then can be obtained 
by comparing the expected cosmological DM signal with the IGRB 
residual spectrum after subtracting the expected astrophysical 
contribution, e.g.~\cite{Ajello:2015mfa, 2015JCAP...09..008T, Ando:2015qda, 2015PhRvD..91l3001D}. 

Interestingly, studying the contributions to the IGRB can also shed
light on an entirely different class of DM candidate: evaporating 
primordial black-holes (PBHs, see~\cite{2010PhRvD..81j4019C} for a review). 
The strongest limits on the density of PBHs, in the mass range from $10^{14}$ to $10^{16}$\,g, come from studies of 
diffuse \gray emission~\cite{1998PhR...307..141C,2003PhLB..551..218B,2010PhRvD..81j4019C}.

\subsubsection{Searches for Contributions to the IGRB Spectrum: Sensitivity Projections}
\label{sec:dm_cosmo_projections}

Currently, the dominant uncertainties in the derivation of DM limits 
from the IGRB energy spectrum are from the large uncertainties 
associated with the modeling of the Galactic
foregrounds~\cite{2015JCAP...09..008T}. 
Thus, %it is expected that 
any IGRB-based DM sensitivity projections 
will largely depend on how these foregrounds will be understood in 
the future.  However, it is difficult to predict how  knowledge of the 
Galactic diffuse emission will improve in the next few years.  For this 
reason,  we considered two scenarios, for which the systematic
uncertainties from modeling the Galactic diffuse emission
are reduced to 50\% and 10\% of their current value.

On the other hand, as explained in \S\ref{sec:dm_cosmo_status}, the majority of the IGRB 
flux arises from the \gray emission of unresolved AGN and star-forming
galaxies. Thus, the projected sensitivity of the LAT to
extragalactic DM will also depend on the uncertainties in the
contribution of (unresolved) point sources to the IGRB. We discuss 
in detail the evolution of these uncertainties in App.~\ref{app:method_egb_sensitivity}. 

Fig.~\ref{fig:EGBlimts} shows the projected upper limits on the 
DM annihilation cross-section in the \bb annihilation channel,
evaluated using the same method that was employed
in~\cite{Ajello:2015mfa}.   The estimates and uncertainties of the contributions of different
source populations to the IGRB after 15 years of data (``optimized
astro'' in the figure) are described in
App.~\ref{app:method_egb_sensitivity}. 
We have assumed a reduction of a factor 2 
for the size of the statistical errors in the IGRB measurement, based 
on what is expected from both this larger data set and from the increased
LAT acceptance in Pass~8 (see App.~\ref{sec:lat} for details). In addition, two cases are shown 
in Fig.~\ref{fig:EGBlimts}, corresponding to different assumptions 
on the future evolution of the systematic errors: a first case
in which they will be reduced by a factor 2 with respect to their
current values (blue dashed line), and a second, more optimistic 
case in which a very good control on the systematics has been achieved 
down to a value of 10\% of the present level (red dot-dashed
line). For both cases we have the same level of theoretical
uncertainties on the strength of the DM annihilation signal
(represented by the red shaded region in Fig.~\ref{fig:EGBlimts}) as
the one derived in \cite{2015JCAP...09..008T} for the current DM
limits. This is conservative, as future N-body cosmological
simulations with higher particle resolution should help in lowering 
the main source of these uncertainties, namely the structural 
properties of low-mass DM halos and subhalos and the exact 
slope of the subhalo mass function (see, e.g., 
\cite{Ishiyama:2014uoa,2016MNRAS.tmp...66H} for recent 
and promising developments). It is however difficult to estimate the 
evolution of these uncertainties at the moment. 

In both cases, it can be seen in Fig.~\ref{fig:EGBlimts} that there is 
significant improvement of the current DM limits (solid line) 
over the whole WIMP mass range considered. The improvement is 
about a factor 2 for the case in which the systematics errors are 
half of the current ones, and ranges between a factor $\roughly 2$ to 5
for the case of having systematics reduced down to 10\% the present
values, depending on the WIMP mass.  Also, as expected, reducing
the systematics uncertainties has a larger impact on the DM sensitivity
at low WIMP masses, while increased statistics primarily improves
the sensitivity at higher masses.

\begin{figure}[!htbp]
  \begin{center}
    \includegraphics[width=0.50\columnwidth]{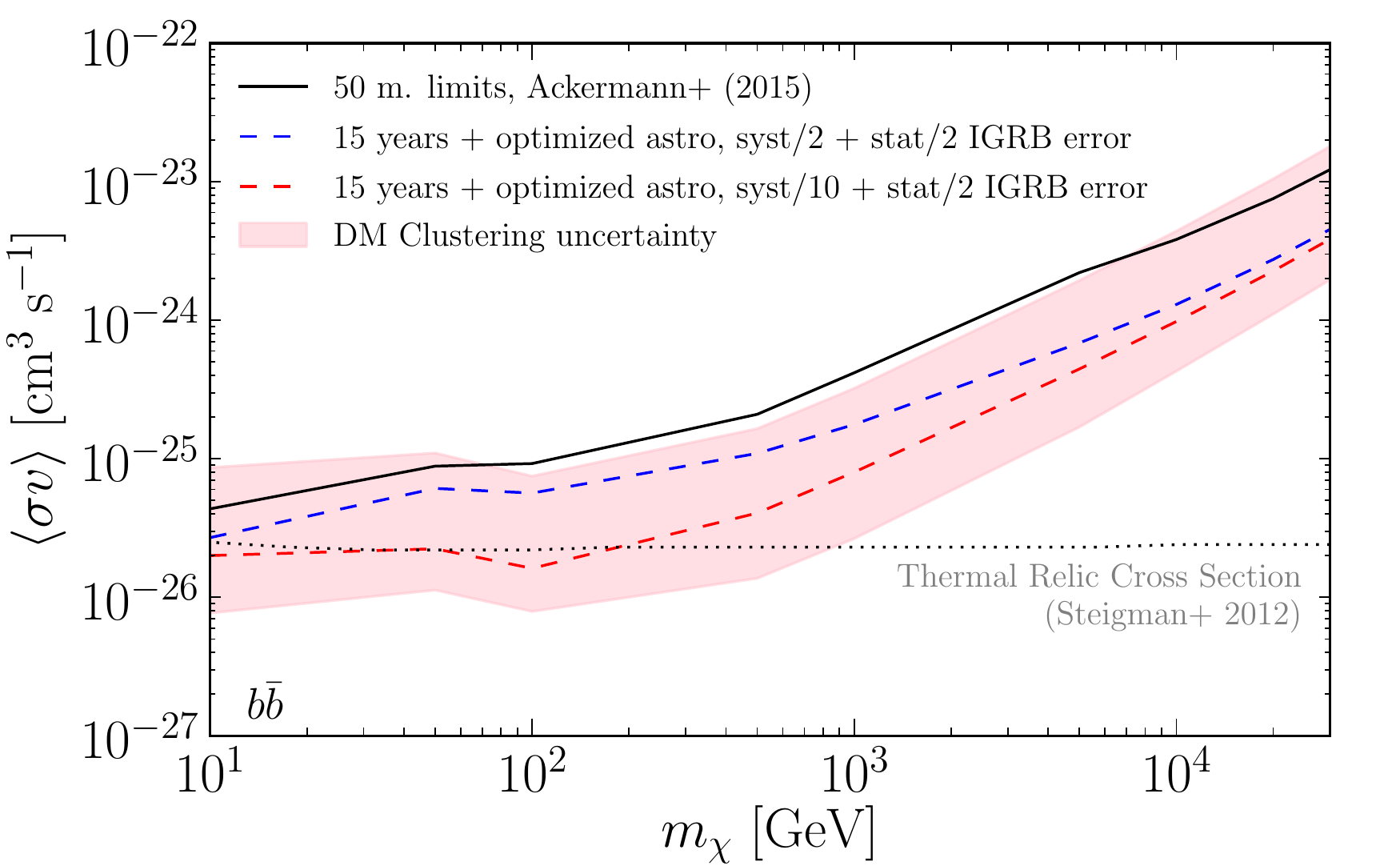}
  \end{center}
 \caption{Projected upper limits on the WIMP annihilation cross section 
    derived from the IGRB energy spectrum after 15 years of data taking, 
    compared to the most recent limits derived in~\cite{Ajello:2015mfa} 
    using 50 months of data (thin solid line). The dashed blue and red dotted curves
    correspond to cases in which the systematic uncertainties are
    reduced by factors of 2 and 10, respectively,
    relative to their values in the 50-month analysis. In both cases, 
    the statistical uncertainties have been reduced by a factor of 2 (see
    text for details). The shaded red area corresponds to the uncertainties from the modeling of the
    cosmological DM annihilation signal (see~\cite{2015JCAP...09..008T}).
    \label{fig:EGBlimts}}
\end{figure}

\subsubsection{IGRB anisotropies and Cross-Correlation Measurements: Current Status}
\label{sec:dm_anisotropy}
\label{sec:dm_cosmo_aps_status}

Integrating the IGRB over the entire sky results in a large
loss of information. More information can be recovered and exploited by
investigating the anisotropy properties of the measured IGRB.
Three measures of anisotropy currently are being studied: 
i) the auto-correlation or angular power spectrum (APS) of the IGRB~\cite{2012PhRvD..85h3007A,
  DiMauro:2014wha, Campbell:2014mpa, Gomez-Vargas:2014yla,
  Calore:2014hna, Fornengo:2013rga,  Campbell:2013rua, Inoue:2013vza,
  Chang:2013ada, Gomez-Vargas:2014yla, Ando:2013ff, Ripken:2012db,
  Harding:2012gk, Cuoco:2012yf, Ando:2005xg,
  Ando:2006cr,Fornasa:2012gu}, 
  ii) the cross-correlation with Large Scale Structure (LSS) tracers
(galaxy catalogs and lensing surveys)~\cite{Xia:2011ax,
  Cuoco:2007sh, Ando:2009nk, Ando:2014aoa, Fornengo:2014cya,
  Camera:2014rja, Xia:2015wka, Regis:2015zka,2015ApJS..221...29C,
  Ando:2013xwa,  Shirasaki:2014noa, Camera:2012cj},
and iii) pixel fluctuations (also referred to as one-point
statistics)~\cite{Malyshev:2011zi, Lee:2014mza, Feyereisen:2015cea,
  Lee:2015fea, Lee:2008fm, Dodelson:2009ih, Bartels:2015aea}. 

Relative to spectral analyses, auto-correlation analyses possess the
advantage that the background is dominated by astrophysical point sources just
below the detection threshold.
Thus, masking/removing newly detected point sources greatly 
reduces the astrophysical background and  enhances the DM 
sensitivity.  For example, the anisotropy of the high-latitude 
\gray sky is reduced by about \emph{four} orders of magnitude 
when the 3FGL sources are masked with respect to the case in 
which they are not.  In comparison, the intensity of the IGRB when 
masking the 3FGL sources is reduced by only 20--30\% over most 
of the energy range (it can be 50--80\% above 100 GeV). 
Clearly, using a deeper point-source catalog available with 
many years of data taking can substantially improve the sensitivity
to DM via auto-correlation. The auto-correlation of the IGRB was first measured
in~\cite{2012PhRvD..85h3007A} using 2 years of LAT data.
Constraints on DM were derived in~\cite{Gomez-Vargas:2014yla,
  Ando:2013ff,  Fornasa:2012gu}. 
As with the IGRB energy spectrum (\S\ref{sec:dm_cosmo_projections}), these constraints 
account for the expected contributions 
from astrophysical sources and their uncertainties.  
The expected astrophysical contribution to the 
APS is dominated by blazars at all energies.  In comparison, the expected contribution 
of blazars to the IGRB energy spectrum is sub-dominant below 
50~\GeV~\cite{Cuoco:2012yf, DiMauro:2014wha}.

The LAT has measured the APS of the diffuse emission at Galactic latitudes
$|b| > 30\degree$ using approximately 2 years of
data~\cite{2012PhRvD..85h3007A}. The APS,
denoted as $C_P(E)$, has been measured in four energy bins 
spanning 1 to 50 GeV and, at multipoles $l \geq 155$, an angular
power above the photon noise level is detected in all energy 
bins. Since the measured angular 
power is approximately constant at all multipoles, the origin of 
the APS is attributed to the contribution of one or more un-clustered 
point-source populations.  Ref.~\cite{DiMauro:2014wha} finds that 
measured values of $C_P$ are compatible with the anisotropy from 
Radio Loud AGN, namely blazars and mis-aligned AGN (MAGN).

Cross correlations with LSS provide a fundamental advance with
respect to the study of the IGRB spectrum only.  They, in fact, give 
access to the redshift dimension, which is integrated
out in the energy spectrum  and thus cannot be used in an 
analysis of LAT data by itself.
LSS tracers come with redshift information, and a positive correlation 
with a given tracer is equivalent to isolating the \gray 
emission coming from the redshift range spanned by the tracer itself in the IGRB. 
Moreover, since different tracers/catalogs are available,
each covering a different redshift range, using them in conjunction 
effectively allows \emph{redshift tomography} of the
IGRB, i.e., splitting the emission according to the different redshift
ranges where such emission originated. In particular, for DM
searches it is useful to isolate the low-redshift part of the IGRB ($z
\lesssim 0.5$) where the DM signal peaks, as opposed to standard astrophysical
components, which typically peak at $z\sim1$.
Constraints on DM from the cross-correlation are stronger than those
from the energy spectrum since the DM signal then can be
compared with the low-$z$ IGRB (where the DM signal peaks) as 
opposed to the whole $z$-integrated IGRB.

The cross-correlation signal is weak and has been detected only
slightly above $3\sigma$ very recently~\cite{Xia:2015wka} using 
5~years of LAT data. Thus, such cross-correlation studies 
will greatly benefit from additional data taking.  By performing 
accurate measurements of the cross correlation at different redshifts, 
distinguishing DM-induced emission from the other astrophysical 
contributors will be easier owing to the peculiar $z$-dependence of the LSS. 
The measured correlations have been used in~\cite{Regis:2015zka} 
to derive conservative constraints on DM neglecting any possible astrophysical
contribution, and in~\cite{2015ApJS..221...29C} to infer more aggressive 
constraints by fitting a multi-component model with both DM and
astrophysical sources.  When adopting the same DM annihilation boost 
factor from DM substructures, the constraints from cross-correlation 
studies are about a factor of 10 stronger than those derived from
the IGRB energy spectrum.

Finally,  one-point statistics measurements are also sensitive to DM. These DM signatures
in the IGRB have been investigated theoretically
in~\cite{Feyereisen:2015cea, Lee:2008fm, Dodelson:2009ih}  
but so far there have been no DM searches using 1-point
statistics with LAT data (see however~\cite{Malyshev:2011zi} for a non-DM oriented 
work using 11 months of LAT data between 1 and 300 GeV).
However, as discussed in \S\ref{sec:dm_milky_way_current}, this method has also been 
applied the Galactic center region to test whether the Galactic center 
excess can be better fitted by a smooth component or as a 
collection of confused/unresolved point sources~\cite{Lee:2015fea,Bartels:2015aea}. 

\newText{These studies have typically started with a low-background 
  event sample (e.g., \irf{P7REP\_CLEAN}) and then masked both the
  Galactic plane and some or all of the cataloged sources to increase the
  signal-to-noise ratio.  For example, Ref.~\cite{Xia:2015wka}
  used \irf{P7REP\_CLEAN} data for $|b| > 30^{\circ}$, then
  masked $1^{\circ}$ radius disks around all 3FGL catalog sources, and increased the disk radius
  to $2^{\circ}$ degrees below 1\GeV around the 500 brightest cataloged
  sources (see \S~3 of that paper for additional details on the
  data selection).}

\subsubsection{IGRB Anisotropies and Cross-Correlation Measurements: Sensitivity Projections}
\label{sec:dm_cosmo_aps_projections}

Rather than deriving projections for each of the methods described in 
\S\ref{sec:dm_cosmo_aps_status}, we will consider the specific case
of the APS analysis.   We expect the sensitivity of the pixel
fluctuations to scale similarly with time.   The sensitivity of the 
cross-correlation with LSS tracers depends critically on the
details of forthcoming catalogs of LSS tracers, which are difficult to
predict.

Following the work in \cite{Fornasa:2012gu,Gomez-Vargas:2013cna}, 
we consider in App.~\ref{app:method_aps_sensitivity} two cases for 
the properties of extragalactic DM halos and subhalos: the LOW case, 
where halos are relatively poor in subhalos, and the HIGH case, that 
imply large subhalo boosts to the DM annihilation cosmological signal.  
The constraints for these scenarios are shown in
Fig.~\ref{fig:Cp15years_limits}.
According to our sensitivity estimates, 15 years of data improve the 
current results (which use 2 years of data \irf{P7\_CLEAN}) by about a factor of 2 to
3 for the two DM substructure models considered.   This prediction
suggests that DM searches based on APS measurements are background 
limited and scale roughly as $\sqrt{t}$.  \newText{(The increased 
statistics from using Pass 8 data account for 10--20\%
of the improvement.)}

\begin{figure}[!htbp]
  \begin{center}
    \includegraphics[width=0.49\columnwidth]{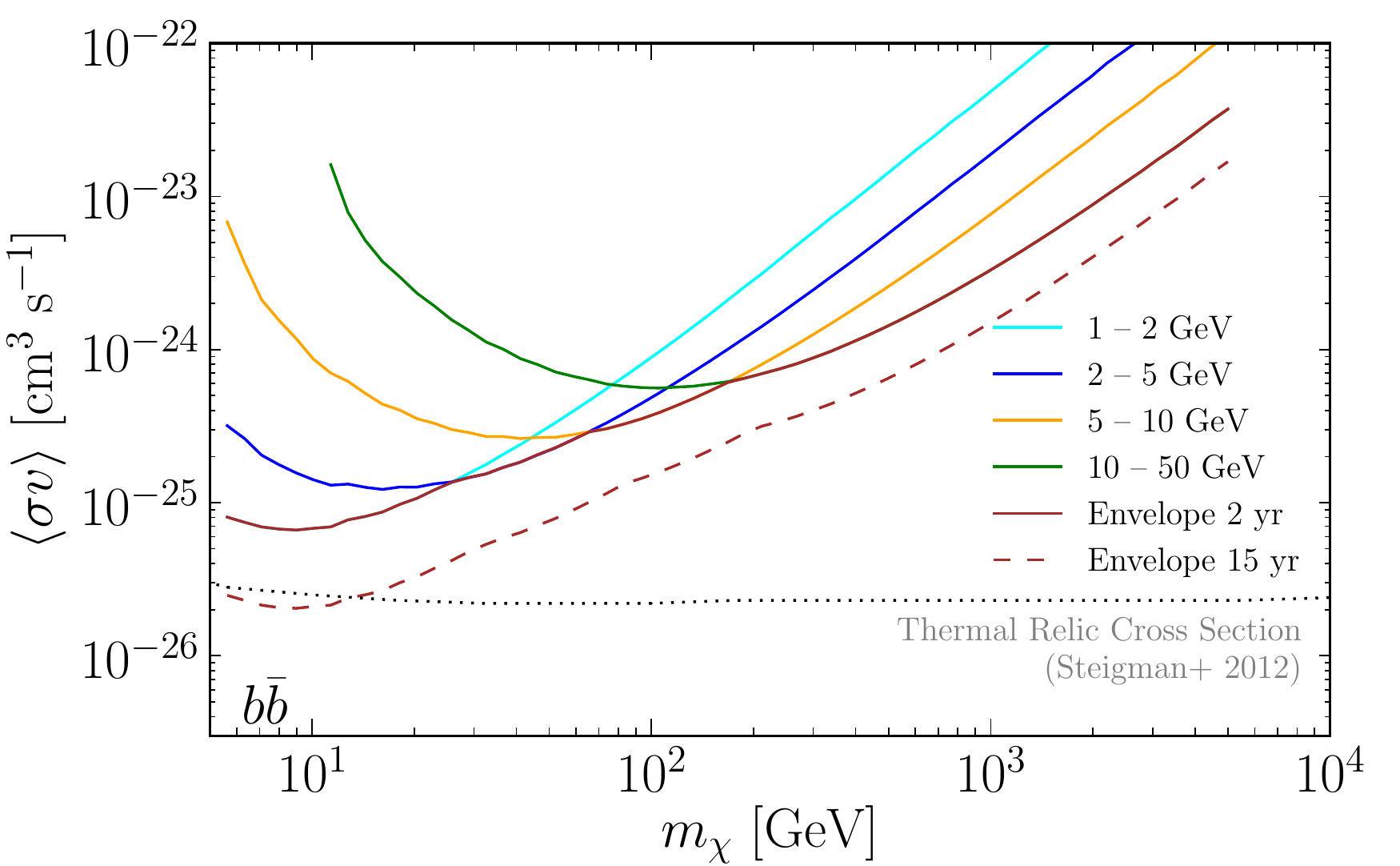}
    \includegraphics[width=0.49\columnwidth]{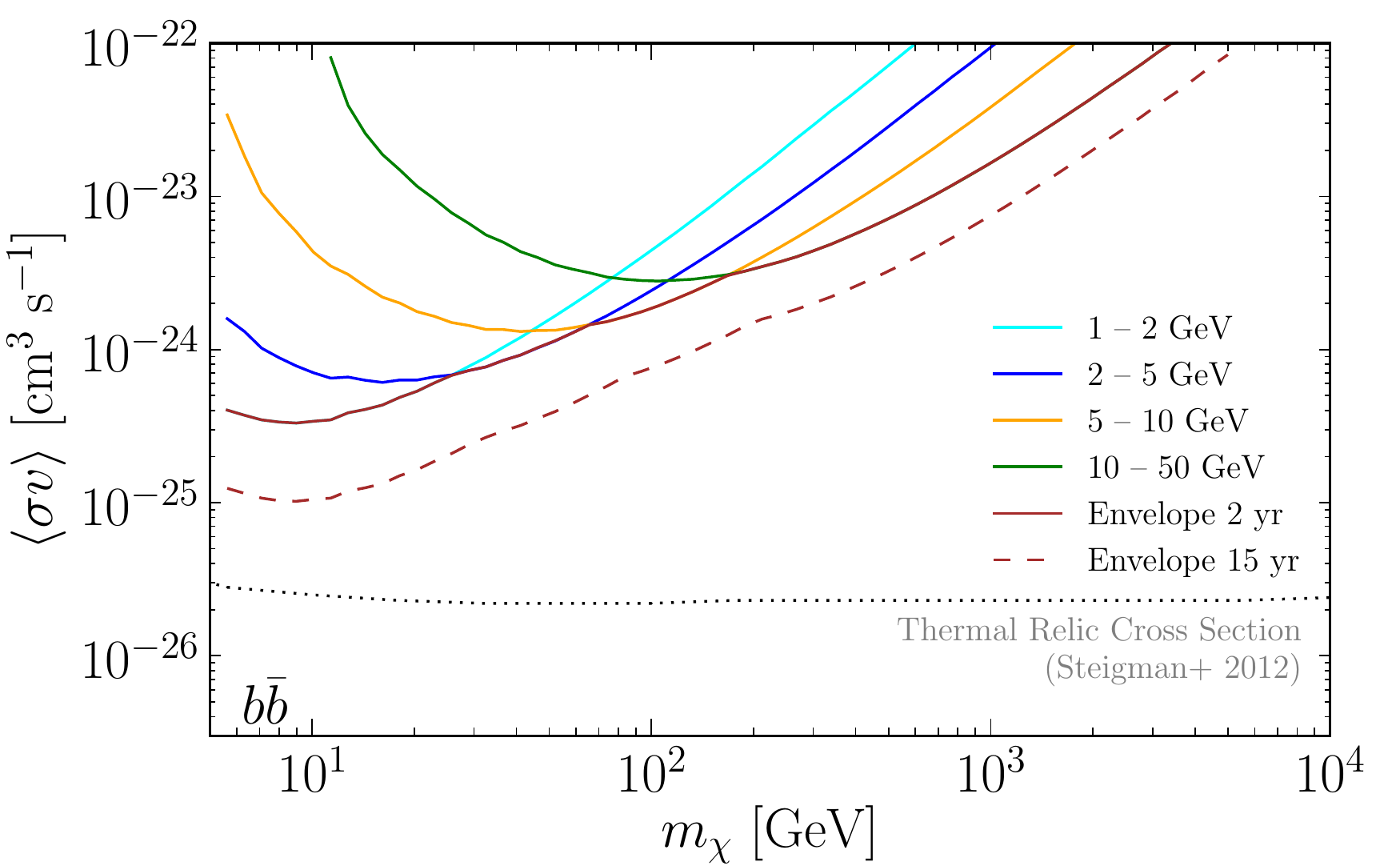}
  \end{center}
  \caption{95\% CL limits on the DM annihilation cross section
    considering two different scenarios for extragalactic DM halos 
    and their substructures (HIGH on left, LOW on right);
    see~\cite{Gomez-Vargas:2013cna} for further details. The limits 
    from 2 years of data are shown for the four different energy bins 
    adopted in the analysis of APS data (cyan, blue, orange, and
    green) and for a convolution of them (solid brown line). The 
    projections for the same energy bins and their convolution is 
    also shown for 15 years of data.}
  \label{fig:Cp15years_limits} 
\end{figure}

\subsection{Spectral Lines}
\label{sec:dm_lines}

In many DM models, pairs of DM particles can annihilate into a \gr
and a second particle ($X$), e.g., $\gamma \gamma$, 
$\gamma Z$, or $\gamma H$.   Since DM is strongly constrained 
to be electrically neutral, it has no direct coupling to photons.  
Thus the process $\chi \chi \rightarrow \gamma X$ occurs only 
through higher-order loops, resulting in a branching fraction that 
is only $\sim{10^{-4}}-{10^{-1}}$~\cite{Bergstrom:1997fh,Matsumoto:2005ui,Ferrer:2006hy,Gustafsson:2007pc,Profumo:2008yg}. 
If a DM particle annihilates to $\gamma X$ the photons are 
monochromatic in the rest frame with rest-frame energy

\begin{equation}\label{eq:LineEn}
E_{\gamma} = \mchi \left( 1-\frac{m^2_{X}}{4m^2_{\chi}} \right). 
\end{equation}

This would result in a line-like feature in the \gray spectrum.  
Possible evidence of such a line-like feature at 130\GeV strongly correlated with the Galactic center  region was reported~\cite{Bringmann:2012vr,Weniger:2012tx,Tempel:2012ey,Su:2012ft},
and also seen in nearby galaxy clusters~\cite{2013EPJC...73.2578H}, 
and unassociated LAT sources~\cite{Su:2012zg,Hektor:2012jc}. 
The feature was not seen in the vicinity of nearby dwarf 
galaxies~\cite{GeringerSameth:2012sr}.  However such a signal 
is expected to be much fainter than in the Galactic center.  Potential 
instrumental effects and a similar feature detected in the bright 
\gray emission from cosmic-ray interactions in Earth's upper 
atmosphere (the Limb) have also been
discussed~\cite{Whiteson:2012hr,2013EPJC...73.2578H,Finkbeiner:2012ez}.  
A systematic investigation of the spatial morphology of the 130\GeV 
feature and other line-like features in the Galactic plane is
presented in~\cite{2013PDU.....2...90B}.  

\subsubsection{Spectral Lines: Current Status}
\label{sec:dm_lines_status}

Since these reports, the LAT Collaboration has searched for 
spectral lines using both the Pass~7REP~\cite{2013arXiv1304.5456B} and 
Pass~8~\cite{2013arXiv1303.3514A} data sets and found that the original putative signal
has faded to less than one standard deviation significance ($ < 1\sigma$) once the 
trials factors associated with scanning in DM particles mass from 
1\GeV to 500\GeV and for several different \newText{ROIs}
associated with different DM radial profiles, which are shown in 
Fig.~\ref{fig:line_ROIs}.

\begin{figure}[!htbp]
  \begin{center}
    \includegraphics[width=0.80\columnwidth]{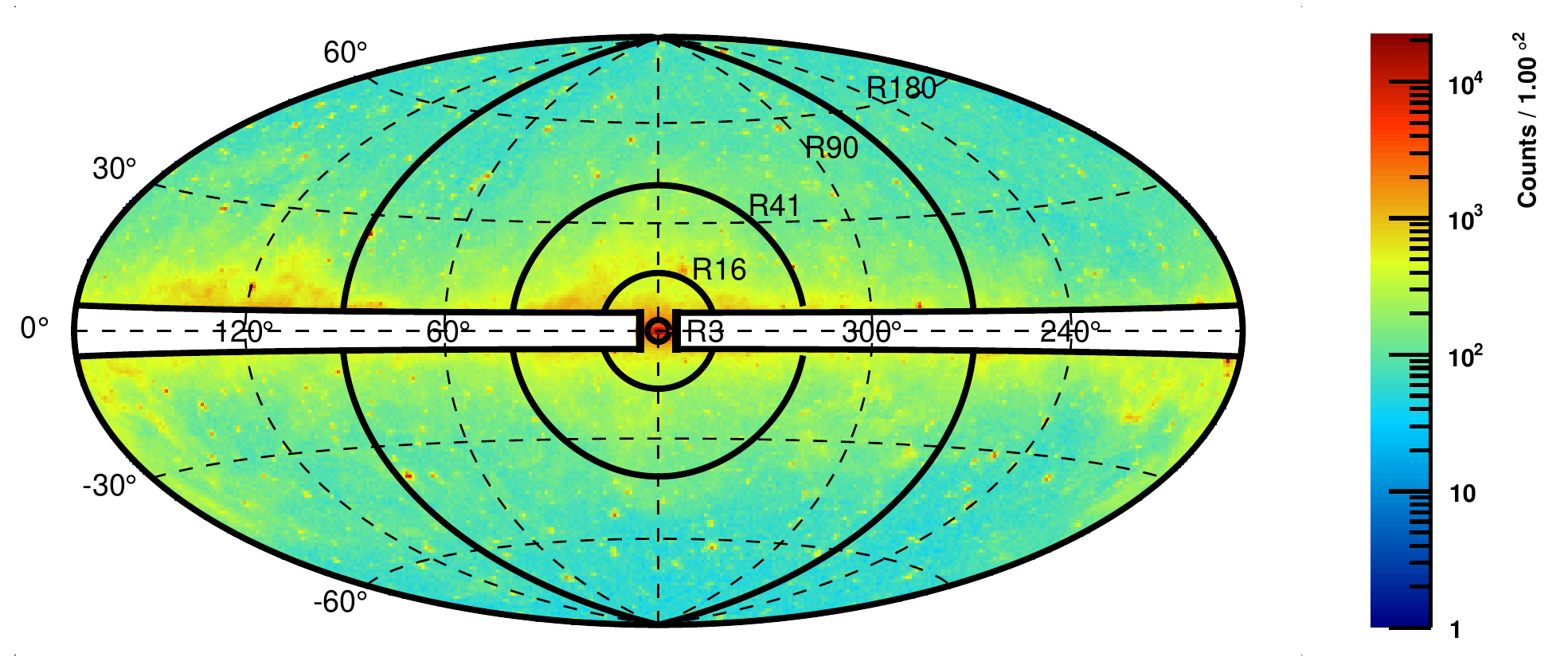}
  \end{center}
\caption{Counts map for six years of \evtclass{P8R2\_CLEAN} data
  binned in $1\degree\times1\degree$ spatial bins plotted in
  Galactic coordinates in Hammer-Aitoff
  projection.  The energy range is 1 to 750\GeV.  Also shown are the outlines of
  the ROIs used in Ref.~\cite{2015PhRvD..91l2002A}.   The Galactic plane 
  region with longitude greater than $6\degree$ from the Galactic center and
  latitude smaller than $5\degree$ is removed from all signal ROIs.
  The R3, R16, R41, and R90 ROIs are optimized for the gNFW ($\gamma=1.3$), NFW, Einasto, and Isothermal 
  profiles, respectively.   \newText{Refs.~\cite{Bringmann:2012vr,Weniger:2012tx} 
    used somewhat different ROIs that were created by selecting a 
    set of pixels with the largest expected signal-to-noise given the 
    observed data and a similar set of DM density profiles; however
    the difference in ROIs does not significantly alter the results.}
This figure is from Ref.~\cite{2015PhRvD..91l2002A} (reproduced by permission of the APS and the authors).
\label{fig:line_ROIs}}
\end{figure}

It is important to note that almost all of the fitting for spectral lines
has been performed in the energy domain only.   That is to say, 
the fitting used the \gray spectrum averaged over each of the ROIs,
rather than simultaneously fitting spatial and spectral templates.
This method greatly reduced the systematic uncertainties associated
with modeling the astrophysical backgrounds, as we need only consider
the potential biases in modeling the background with relatively simple
spectral form rather than all of the potential biases from modeling
the complicated spatial distribution of the Galactic diffuse emission.  
As shown in Fig.~\ref{fig:fractional_signal_lines}, by scanning
control regions including the Galactic plane away from the Galactic
center, the Vela pulsar and the Earth's limb, we have found
that that the systematic uncertainties are at the level of 
$f_{\rm syst} \sim 0.015$ 

\begin{figure}[!htbp]
  \begin{center}
    \includegraphics[width=0.49\columnwidth]{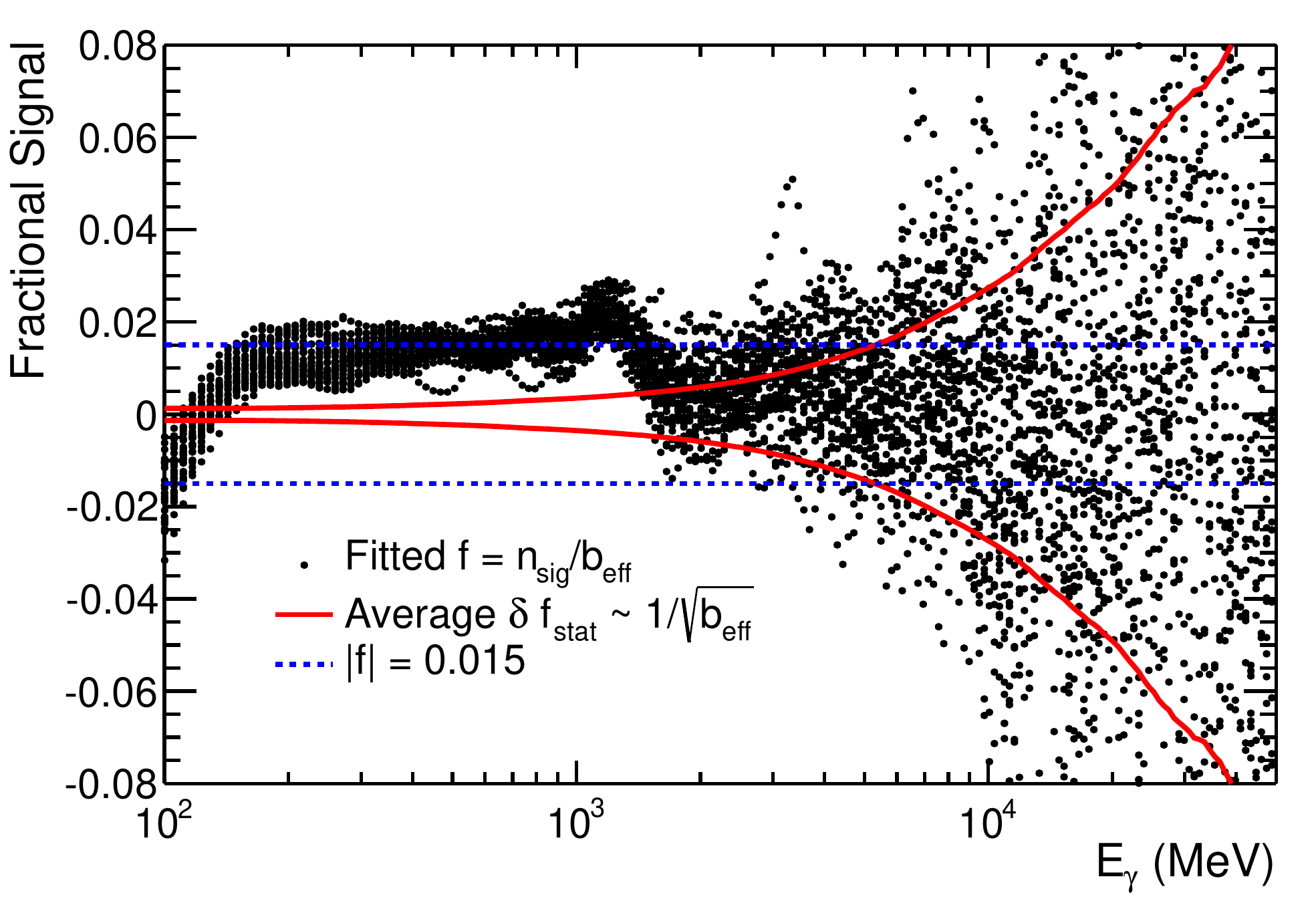}
    \includegraphics[width=0.49\columnwidth]{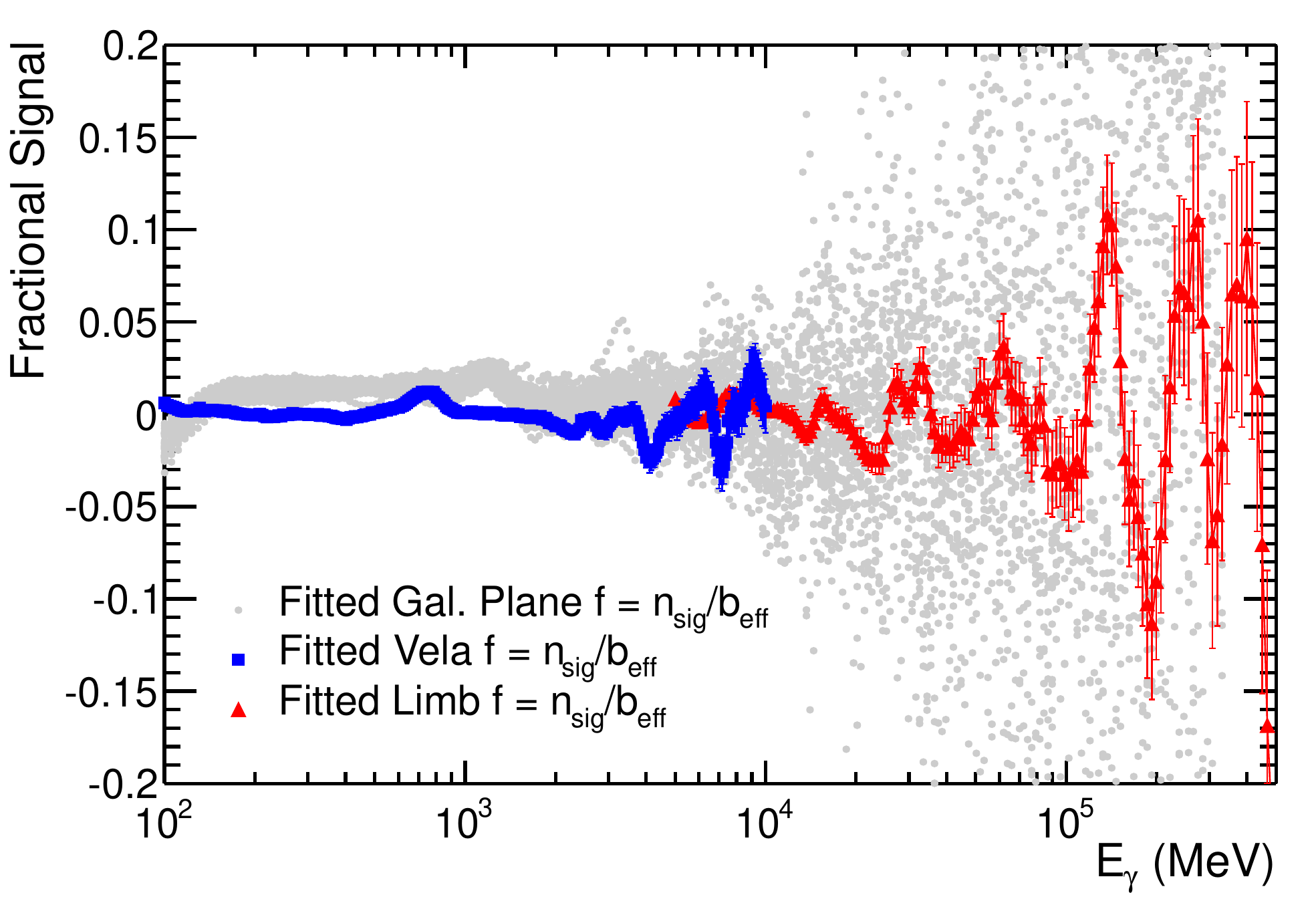}
  \end{center}
\caption{ Fractional signals (\fsig) in a scan of the Galactic plane.  
  (Left) dots show observed \fsig in 31 $10\degree\times10\degree$ boxes along the Galactic
  plane. The solid red line is the average of the statistical
  uncertainties of the individual boxes. The blue dashed line
  is the value we chose to characterize \fsyst  modeling biases.  (Right)
  The same, but with values of \fsig from the Vela pulsar and Earth Limb
  control samples added and the Galactic plane points shown in gray.
  These figures are from Ref.~\cite{2015PhRvD..91l2002A} (reproduced by permission of the APS
  and the authors); see that reference for additional details.
\label{fig:fractional_signal_lines}}
\end{figure}

\subsubsection{Spectral Lines: Sensitivity Projections}
\label{sec:dm_lines_projections}

To project the effect of additional data on searches for spectral lines, we use
largely the same methodology as for the DM searches targeting the
Galactic center and halo (\S\ref{sec:dm_milky_way}).    The main
difference is that instead of calculating the effective background by
summing over pixels and energy bins, here we only sum over the energy
bins.   We model the background with a simple spectral form (typically 
the product of a power law and the exposure as a function of energy),
model the signal with the LAT's energy dispersion function (which has
a resolution of $\Delta E/E < 0.10$ from 1\GeV to 500\GeV), use 
narrow energy bins (typically $0.01$~dex, or $\Delta E/E = 0.023$),
and assign $\fsyst = 0.015$.  

\begin{figure}[!htbp]
  \begin{center}
    \includegraphics[width=0.50\columnwidth]{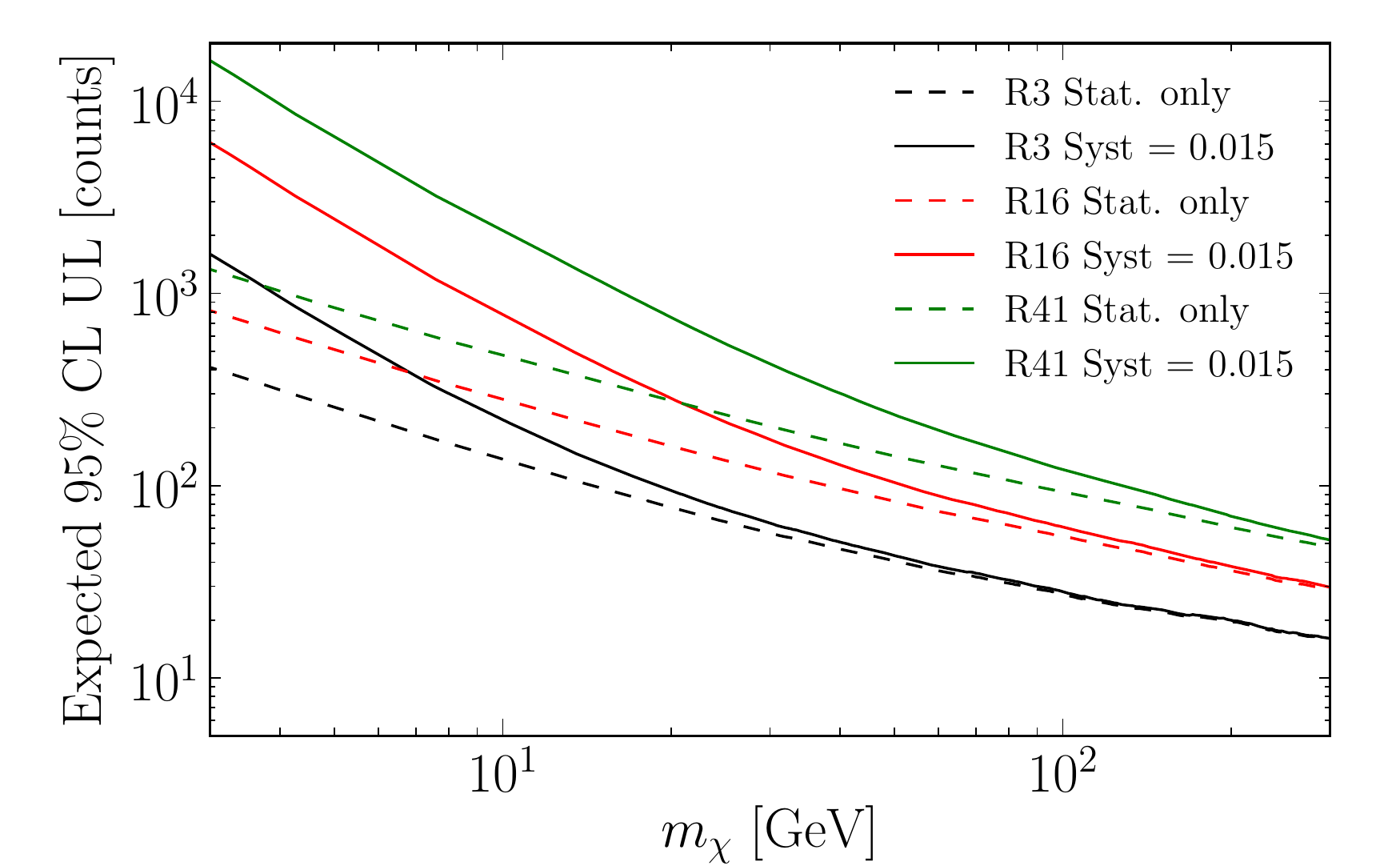}
  \end{center}
\caption{ Projected upper limits in signal counts for 15~years of LAT
  data for three different ROIs.  Dashed lines show projected limits from statistical uncertainties
  only, while the solid lines show projected limits including both statistical
  and systematic uncertainties, with the level of systematic uncertainty 
  assigned at $\fsyst = 1.5\%$ of the effective background.  See text for details.
\label{fig:lines_limit_projections}}
\end{figure}

From the projections shown in Fig.~\ref{fig:lines_limit_projections}
it is clear that searches for spectral lines are systematics limited up
to \roughly100\GeV (\roughly10\GeV) for the R41 (R3)
region, and background limited up to hundreds of \GeV (i.e., the
expected upper limit is $> 10\photons$).

\subsection{Axions and Axion-Like Particles}
\label{sec:dm_axion_targets}
Axions and ALPs may be detected through either their decay to photons 
or through spectral signatures in high-energy sources induced by photon-ALP oscillations. 

Heavy axions decaying into \grs would be very similar to 
decaying DM and could potentially be constrained with observations of
the IGRB (see \S\ref{sec:dm_cosmo}).\footnote{However, these
  heavy axions cannot constitute the DM since their lifetime is too short. 
  ALPs in this mass range, on the other hand, could constitute all of the 
  DM, if additional degrees of freedom above the electro-weak scale at
  high energies are assumed~\cite{2012JCAP...02..032C}.
  Otherwise, the ALPs would exceed the critical density of the Universe.}
Decaying light axions and ALPs could be detected at \gray energies 
if they are produced relativistically.   This is the case in the
production, e.g., through nucleon-nucleon Bremsstrahlung in the
interiors of neutron stars~\cite{PhysRevD.93.045019}.

A second approach is to search for signatures of photon-ALP oscillation
in the spectra of \gray sources.  For \grs propagating
through a magnetic field $B_{\mu\mathrm{G}} = B / \mu\mathrm{G}$
and a plasma with plasma frequency $\omega_\mathrm{neV} = \omega /
\mathrm{neV}$, the oscillation probability between photons and ALPs 
is maximal and independent of energy between $E_\mathrm{crit} \lesssim E \lesssim E_\mathrm{max}$, 
where~\cite[e.g.][]{2007PhRvL..99w1102H,bassan2010}
\begin{eqnarray}
E_\mathrm{crit} &=& 2.5\,\mathrm{GeV}\,\frac{|m_\mathrm{neV}^2 - \omega^2_\mathrm{neV}|}{g_{11} B_{\mu\mathrm{G}}},
\label{eq:Ecritaxions} \\
E_\mathrm{max} &=& 2.12\times10^6\,\mathrm{GeV}\,g_{11}B_{\mu\mathrm{G}}^{-1}, 
\label{eq:Emaxaxions}
\end{eqnarray} 
for ALP masses $m_\mathrm{neV} = m_a / \mathrm{neV}$ and photon-ALP 
coupling $g_{11} = g_{a\gamma} / 10^{-11}\,\mathrm{GeV}^{-1}$.~\footnote{In 
  these equations, we have neglected a possible photon dispersion
  term~\cite{dobrynina2015}, since it will not be relevant for the energies and magnetic fields 
  considered in the following.}
The magnetic field strengths in galaxies and clusters of galaxies are
found to be of the order of the $mG$ and $\mu\mathrm{G}$, respectively~\cite[e.g.][]{govoni2004}.
For the intergalactic medium, the field strength is constrained to be
 $B\lesssim 10^{-9}\,\mathrm{G}$ \cite[e.g.][]{widrow2002}.
Thus, from these numbers and
Eqs.~\ref{eq:Ecritaxions}--\ref{eq:Emaxaxions}, it is evident
that for \grs propagating in the magnetic fields of galaxies, galaxy
clusters and the intergalactic medium, only an oscillation into light 
ALPs is possible (i.e., $m_a \lesssim \mu\mathrm{eV}$) whereas the 
conversion probability to axions is vanishingly small.

The photon-ALP oscillations could manifest themselves in two ways 
in the spectra of \gray sources:
\begin{enumerate}
\item{Above $E_\mathrm{crit}$, up to 50\,\% of the photons of  an
    initially un-polarized pure photon beam 
    can convert into ALPs, leading to a decrease of the source photon flux. 
    Depending on the (often turbulent) ambient magnetic field, this
    decrease is accompanied by chaotic oscillations (irregularities)
    in the energy spectrum~\cite{ostman2005,wouters2012,Wouters:2013hua}. These
    spectral irregularities are not time dependent; thus, for each
    target, a particular, stable-in-time spectral pattern is expected.} 
\item{ALPs could enhance the \gray fluxes from cosmological 
    objects that are otherwise attenuated in interaction with
    photons of the extragalactic background light (EBL)~\cite[e.g.][]{meyer2013,Simet:2007sa,2007PhRvD..76l1301D,sanchezconde2009,2011JCAP...11..020D}. 
    The EBL spans from UV to far-infrared and is
    composed of the integrated starlight emitted over the entire 
    history of the Universe and starlight absorbed and re-emitted 
    by dust in galaxies (e.g.,~\cite{hauser2001}).  A fraction of the
    photons converting to ALPs during propagation could lead to a 
    secondary component in the spectrum at high energies given that 
    the ALPs can reconvert into \grs before reaching Earth.} 
\end{enumerate}
Photon-ALP oscillations could also be used to search for a \gray 
signal from core-collapse supernovae \cite{1996PhLB..383..439B,1996PhRvL..77.2372G,payez2015}.  
Axions and ALPs would be produced in such explosions in a burst
lasting several seconds, similar to the production of neutrinos. 
ALPs produced in such a way could subsequently convert into 
\grs in the magnetic field of the Milky Way.   On the other hand, 
\grs produced in the dense core of the supernovae would not escape to
the surface, and the \gray emission is expected to be delayed with
respect to the neutrino emission.  Thus, the observation of a pulse 
of \grs coincident with the expected neutrino signal from a
Galactic supernova would be evidence for the existence of ALPs.

\subsubsection{Axions and ALPs:  Current Status}
\label{sec:axions_status}

\begin{figure}[!htbp]
  \begin{center}
    \includegraphics[width = 1\linewidth]{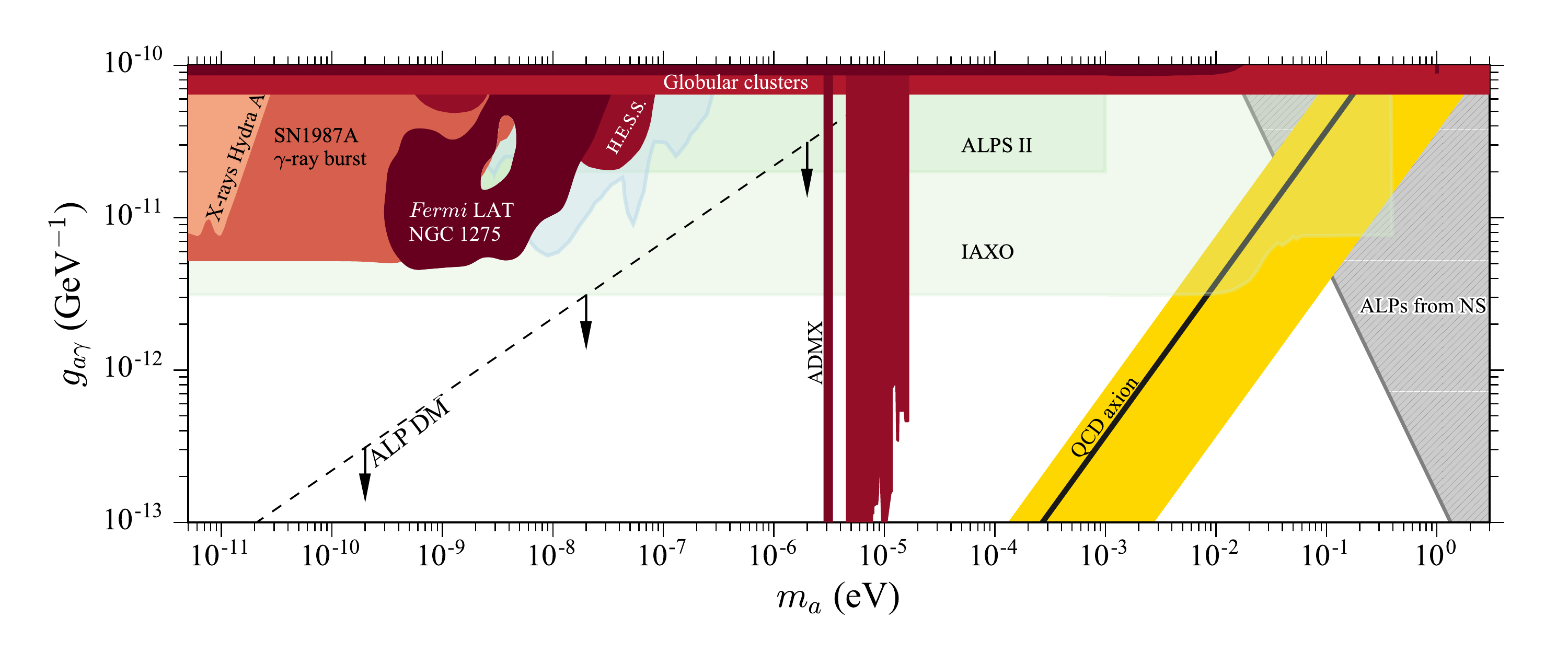}
  \end{center}
\caption{\label{fig:alps}Current status of the limits imposed on the
  ALP parameter space by different experiments and targets. 
  Limits derived with LAT observations are shown as \newText{dark red}
  (NGC\,1275 in the Perseus galaxy cluster~\cite{ajello:2016zue}) and gray hatched 
  (neutron stars~\cite{PhysRevD.93.045019}) regions. Limits from other
  experiments are shown in red.  The parameter space where ALPs
  could explain a low \gray opacity is shown in light blue.
  The parameter space where ALPs could explain \newText{hints for} a low \gray
  opacity are shown in \newText{blue}.  \newText{Sensitivity estimates for
  future laboratory experiments are shown in green.}
  The QCD axion line is shown in yellow. ALP parameters below 
  the dashed line could account for all the DM. See
  Refs.~\cite{2013arXiv1311.0029E,meyer2014cta} and references therein.}
\end{figure}

The current limits on the ALP parameters are summarized in
Fig.~\ref{fig:alps}  (red shaded regions). The search for \grs from
axion or ALP decays produced in neutron star interiors lead 
to the limits labeled ``ALPs from NS'' (gray hatched region, assuming 
the model-dependent factor $\mathcal{N} = 1$)~\cite{PhysRevD.93.045019}.
The analysis used five years of Pass~7REP LAT
data and combined the results of four nearby neutron stars which are not 
observed at \gray energies.   The limits constrain the QCD axion 
mass (black line and yellow band in Fig.~\ref{fig:alps}) to be below $7.9\times10^{-2}\eV$.
It should be noted that these limits not only depend on the axion 
and ALP coupling to photons but also on the axion and ALP production 
through nucleon-nucleon Bremsstrahlung.

Limits depending only on the coupling between ALPs and photons were 
derived by searching for spectral irregularities around $E_\mathrm{crit}$ 
in the \gray spectrum of the radio galaxy NGC\,1275~\cite{ajello:2016zue}.
In total, six years of Pass~8 data were analyzed. NGC\,1275 is the most 
viable source for such searches as it is (a) a bright \gray
emitter leading to high signal-to-noise ratio and a well determined
spectrum and (b) it is located at the center of the cool-core Perseus
galaxy cluster.  Such galaxy clusters have strong magnetic fields of 
the order of tens of $\mu$G  at their centers~\cite[e.g.][]{govoni2004}.   
The limits are the strongest reported so far in the mass range 
$0.5\,\mathrm{neV} \lesssim m_a \lesssim 20\,\mathrm{neV}$ and surpass 
the sensitivity of the planned laboratory experiment ALPS~II~\cite{alpsII}
in that mass range. Together with the bounds from the non-observation 
of a Gamma-Ray Burst (GRB) from the SN\,1987A~\cite{payez2015} and the non-observations 
of irregularities with the H.E.S.S. Cerenkov telescopes in the
spectrum of the blazar PKS\,2155$-$304~\cite{hess2013:alps} (see also~\cite{2015ApJ...813L..34D}), 
ALPs are now almost ruled out as a viable solution for the (now questioned) reported
evidence of a reduced opacity of the Universe to \grs~\cite{meyer2013} 
(blue shaded region in Fig.~\ref{fig:alps}).   However, the limits do
not yet constrain models where ALPs could account for all the DM, 
indicated by the dashed line labeled $\theta_1\mathcal{N} = 1$ in Fig.~\ref{fig:alps}.

\subsubsection{Axions and ALPs:  Sensitivity Projections}
\label{sec:axions_projections}

The flux limits derived from the observation of neutron stars
in~\cite{PhysRevD.93.045019} depend on the precision of the measurement 
of the astrophysical backgrounds and are therefore expected 
to improve with $\sqrt{t}$, similarly to
the dSphs for lower WIMP masses.    

On the whole, the LAT sensitivity to ALPs from the measurement of
distortions of the spectrum of NGC\,1275 appears to be systematics
limited.   The major, dominant uncertainties in the limits derived from the spectrum of NGC\,1275 in~\cite{ajello:2016zue}
stem from the unknown magnetic field structure in the Perseus galaxy
cluster over large spatial scales. 
Future radio observations with the Square Kilometer Array will provide a wealth 
of rotation measures~\cite{gaensler2004}. 
These will improve understanding of the strength and morphology 
of the cluster magnetic field~\cite{bonafede2015}, and thus lower the uncertainties.

On the other hand, the bounds on low-mass ALPs ($m_a \lesssim
1\,\mathrm{neV}$) could be significantly improved if a Galactic
supernova occurred during the operational lifetime of the LAT: its
sensitivity is superior to that of the Gamma-Ray Spectrometer of
the {\it Solar Maximum Mission} that was used to derive the current bounds 
from the non-detection of \grs from  SN\,1987A (light-red 
shaded region labeled ``SN \gray burst'' in
Fig.~\ref{fig:alps}). 

Future LAT observations of distant AGNs could also provide an
additional step forward to testing ALPs as a possible solution to the
suggested reduced opacity of the Universe to \grs
\cite{2011PhRvD..84j5030D,2011JCAP...11..020D,meyer2013,2014JETPL.100..355R}. 
In particular,
a careful statistical study of those \grs suffering from the strongest
EBL attenuation should shed light on the role of ALPs as possible
``boosters'' of the cosmological \gray flux received at Earth.   The
LAT is the ideal instrument not only because of its point-source
sensitivity, which enables observation of high-redshift AGNs up 
to very high energies, but very especially thanks to its continuous 
scanning of the entire \gray sky. Indeed, any violent and transient 
cosmological phenomena (such as \gray flares from blazars or GRBs)
that could be potentially used for these opacity studies will very
likely be observed by the LAT (consequently enhancing the chances to 
detect some of the few photons expected at the highest optical
depths).  Note that this is less likely for Cerenkov telescopes 
given their much narrower fields of view and reduced duty cycles.

% LocalWords:  DM NFW kpc baryonic galacto extrema P8R2 dSphs Multi spectrally
% LocalWords:  Eq dSph dispersions WIMPs discoverable SDSS DES LSST ROI CDM ULs
% LocalWords:  subhalos Zwicky unassociated 7REP ROIs Aitoff axion 3FGL gNFW un
% LocalWords:  redshift Anisotropy Axions Axion datasets Unid annLOW ALLGeV Ang
% LocalWords:  Extragal parametrization systematics mis triaxiality P7REP LMC
% LocalWords:  hydrodynamical observationally kinematically luminosities SMC R3
% LocalWords:  PanSTARRS spectroscopically dex blazars isotropically GeV 3FGL
% LocalWords:  indices Lactea subhalo 2FGL multi 1FGL intra ICM HIFLUCGS Fornax
% LocalWords:  AGN unmodeled HIFLUGCS rescaling 7REP Einasto IGRB EGB PBHs 2FGL
% LocalWords:  extragalactic astro anisotropies anisotropy APS LSS lensing MAGN
% LocalWords:  multipoles redshifts R16 R41 ALPs axions electro crit
% LocalWords:  relativistically neV Eqs vanishingly EBL NGC QCD GRB 1987A 1FGL
% LocalWords:  blazar PKS AGNs GRBs  3FGL AAS 2FGL 1FGL 7REP R90 Isothermal

% LocalWords:  1987A

\section{Discussion}\label{sec:discussion}

Here we review the projected limits for the different search targets
(\S\ref{sec:disc_summary_limits}), and consider the results
in the context of other indirect-detection DM searches (\S\ref{sec:disc_other_efforts})
as well as direct-detection and collider-production searches
(\S\ref{sec:disc_indirect}).    This discussion will be focused on
WIMP searches, as the results for Axions and ALPs are summarized
in Fig.~\ref{fig:alps} and \S\ref{sec:dm_axion_targets}.

\subsection{Summary of Projected Limits}
\label{sec:disc_summary_limits}

With the categories of limiting factors presented in 
\S\ref{sec:dm_limiting_factors} in mind, we can broadly categorize the 
search targets as follows: 

\begin{enumerate}
\item{Searches targeting objects with substantial astrophysical backgrounds,
  such as the Galactic center, large Galaxy clusters, and the LMC are
  systematics limited up to high energies (e.g., $\mchi \lesssim 10\TeV$ 
  for the Galactic center, $\mchi \lesssim 1\TeV$ for the LMC).}
\item{Searches for spectral lines from the Galactic halo benefit from
  additional discriminating power in the spectral domain, but are still
  systematics limited at low energies (e.g., $\mchi \lesssim 100\GeV$ for
  the R16 ROI optimized for the Einasto profile).}
\item{Searches targeting relatively small objects at high Galactic latitudes,
  in particular the dSphs, are background limited at lower energies, and signal
  limited at higher energies.  For the combined analysis of 
  dSphs the cross-over point comes at $\mchi \sim 500\GeV$ $(100\GeV)$ for the
  \bb (\tautau) channel.}
\item{Although the precision of the measurement of the IGRB is systematics
  limited, statistical analyses of the IGRB, such as fluctuation analysis or
  cross-correlation analysis, are, for the most part still signal limited in
  much of the LAT energy range.  This is because above a few GeV the combined
  brightness of the isotropic and Galactic diffuse backgrounds is low enough that on average
  less than a single background \gr is expected within the LAT
  angular resolution.}
\item{Some searches are limited by our understanding of a source population. This
  could be a population of background sources, such as the various AGN classes
  in analyses of the EGB.  Or it could be the signal population, such as in
  the search for the dark satellites among the unassociated cataloged sources.
  Although the source sensitivity generally scales as $\sqrt{t}$, because of
  the typically steeply falling distribution of fluxes,
  $dN/dS \propto S^{3/2}$, the sensitivity of these searches scales as
  $\sim t^{3/4}$.}
\end{enumerate}

In short, the most notable benefits from additional data taking
come from the improved sensitivity to signals from dSphs for higher 
DM masses $(\mchi \gtrsim 100\GeV)$, from the improved sensitivity 
of statistical methods of disentangling contributions to the IGRB by
virtue of the continued improvement in the point-source sensitivity,
and from increased sensitivity to a population of dark satellites
whose number rises quickly at decreasing flux thresholds.

Furthermore, searches targeting the dSphs have \newText{smaller} modeling
uncertainties and \newText{more} robust determinations of the astrophysical 
$J$~factors than searches targeting other objects.  Additionally these searches 
have among the best sensitivities across much of the LAT energy band.    Overall these
considerations suggest that, looking forward, searches targeting
the dSphs will continue to have the best sensitivity.  
This can clearly be seen in Fig.~\ref{fig:limit_sample_proj}.   

\begin{figure}[!htbp]
  \begin{center}
    \includegraphics[width=0.49\columnwidth]{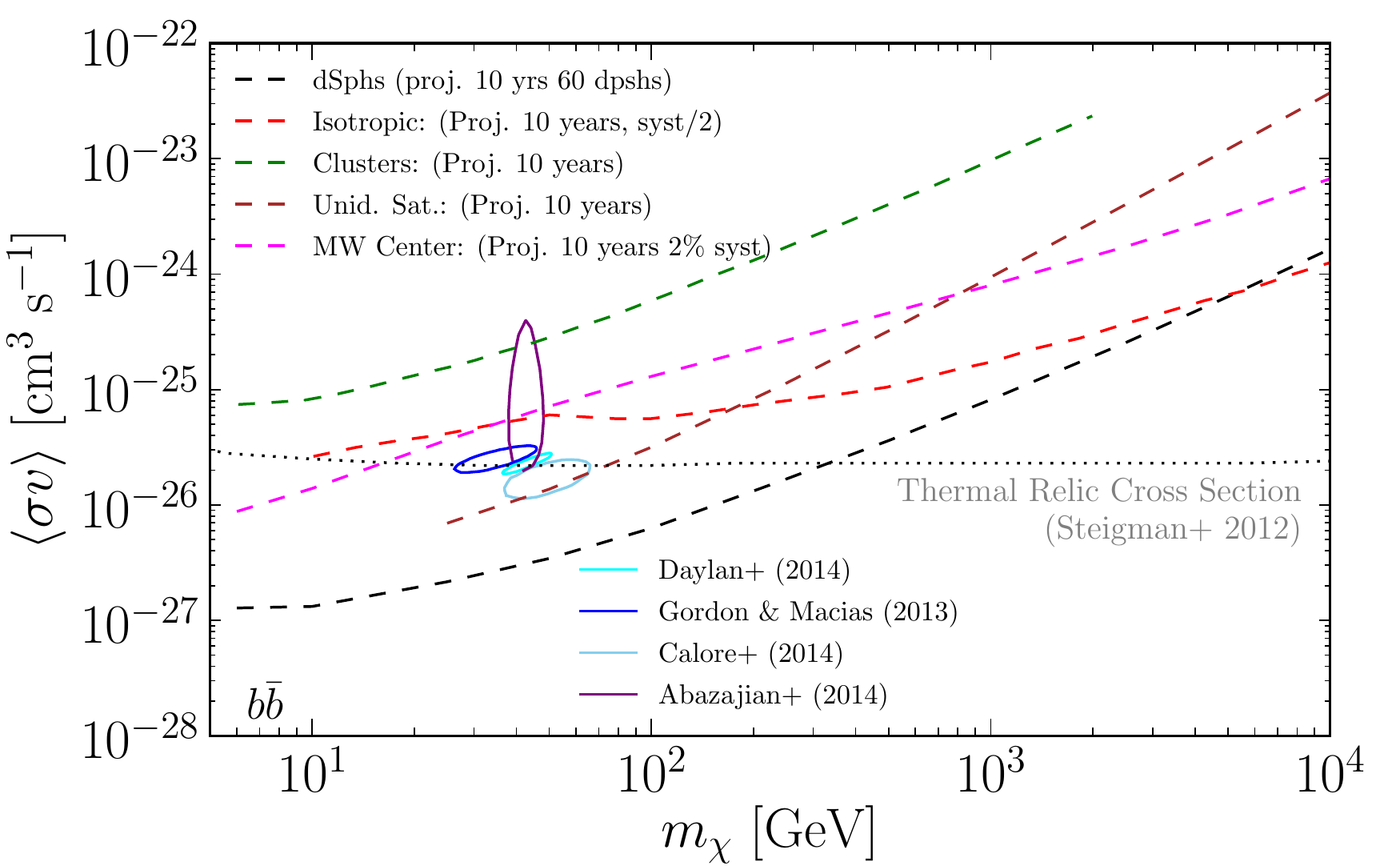}
    \includegraphics[width=0.49\columnwidth]{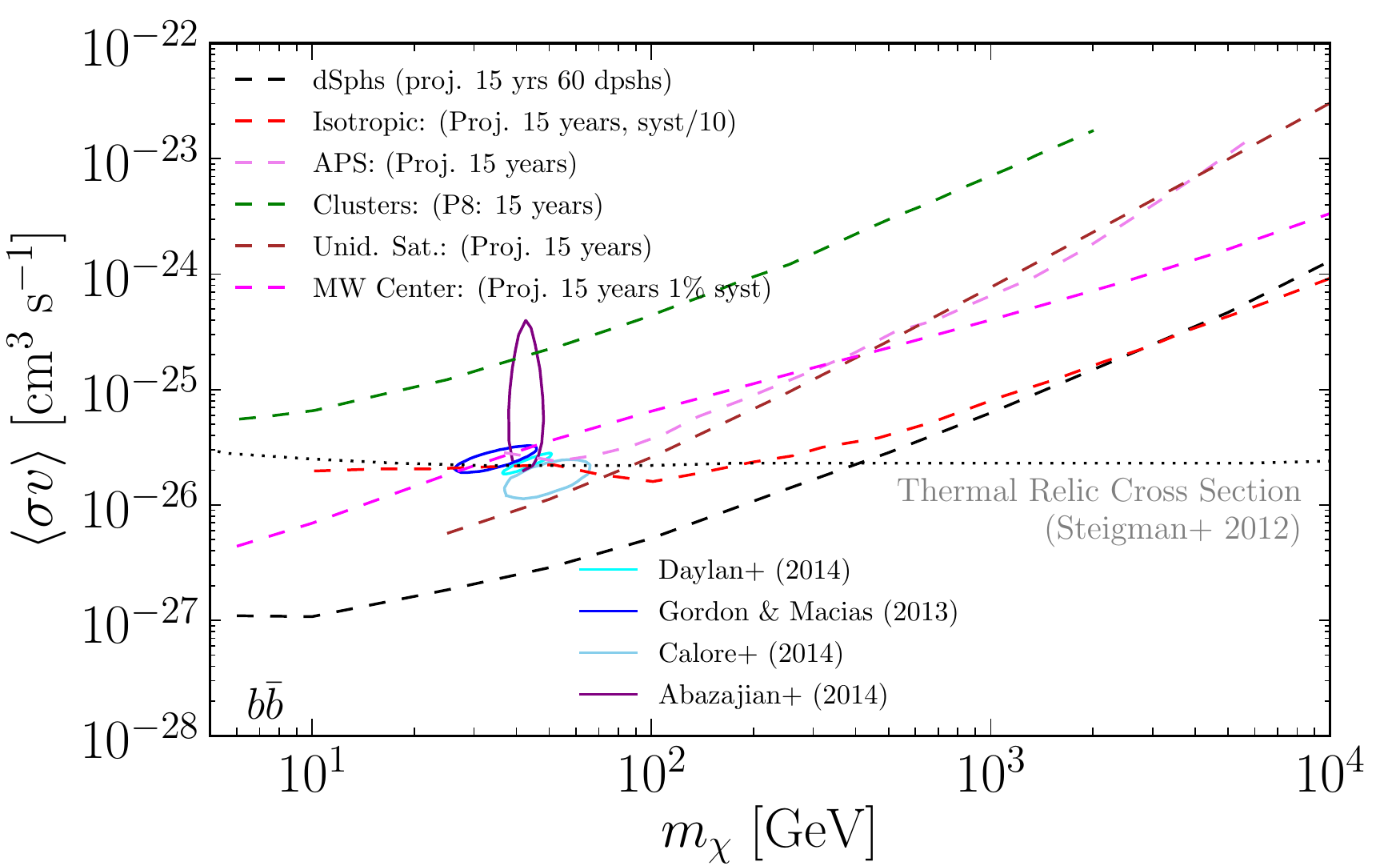}
  \end{center}
\caption{Comparison of projected LAT limits for 10~years (left) and 15~years (right) 
  of data for the search methods described in \S\ref{sec:dm_targets}.
  Favored contours for several Galactic center analyses are also 
  included for comparison.\label{fig:limit_sample_proj}}
\end{figure}

\subsection{Relation to Other Indirect-Detection Efforts }
\label{sec:disc_other_efforts}

Several other kinds of experiments can undertake indirect-detection searches for DM:
(a) instruments
that are sensitive to GeV-to-TeV cosmic rays, (b) imaging atmospheric 
Cerenkov telescopes (IACTs), (c) water Cerenkov telescopes, (d) cosmic-ray 
air shower arrays, (e) neutrino telescopes, and (f) instruments
sensitive to fluctuations in the cosmic microwave background (CMB).
Of these, cosmic-ray detectors,  IACTs, and the {\it Planck} measurements of the CMB are all
capable of probing WIMP annihilation at or near the thermal relic
cross section in some mass ranges.   In this section we compare the sensitivity of 
DM searches using LAT data with results from these 
other instruments.    

In Fig.~\ref{fig:gce_iact_limits} we show the most sensitive
limits projected for searches using LAT data and compare them 
with other indirect-detection limits derived from IACTs and {\it Planck}.
The sensitivity of future IACTs, and in particular the Cerenkov
Telescope Array (CTA), are somewhat uncertain, owing to
uncertainties of the instrument performance and the effects of
astrophysical backgrounds.   We show projections that include the
effects of conservatively estimated systematic
uncertainties~\cite{2015JCAP...03..055S}, as well as projections made with
smaller systematic uncertainty and updated instrument performance
estimates~\cite{2015arXiv150904123L}.

\begin{figure}[!htbp]
  \begin{center}
    \includegraphics[width=0.49\columnwidth]{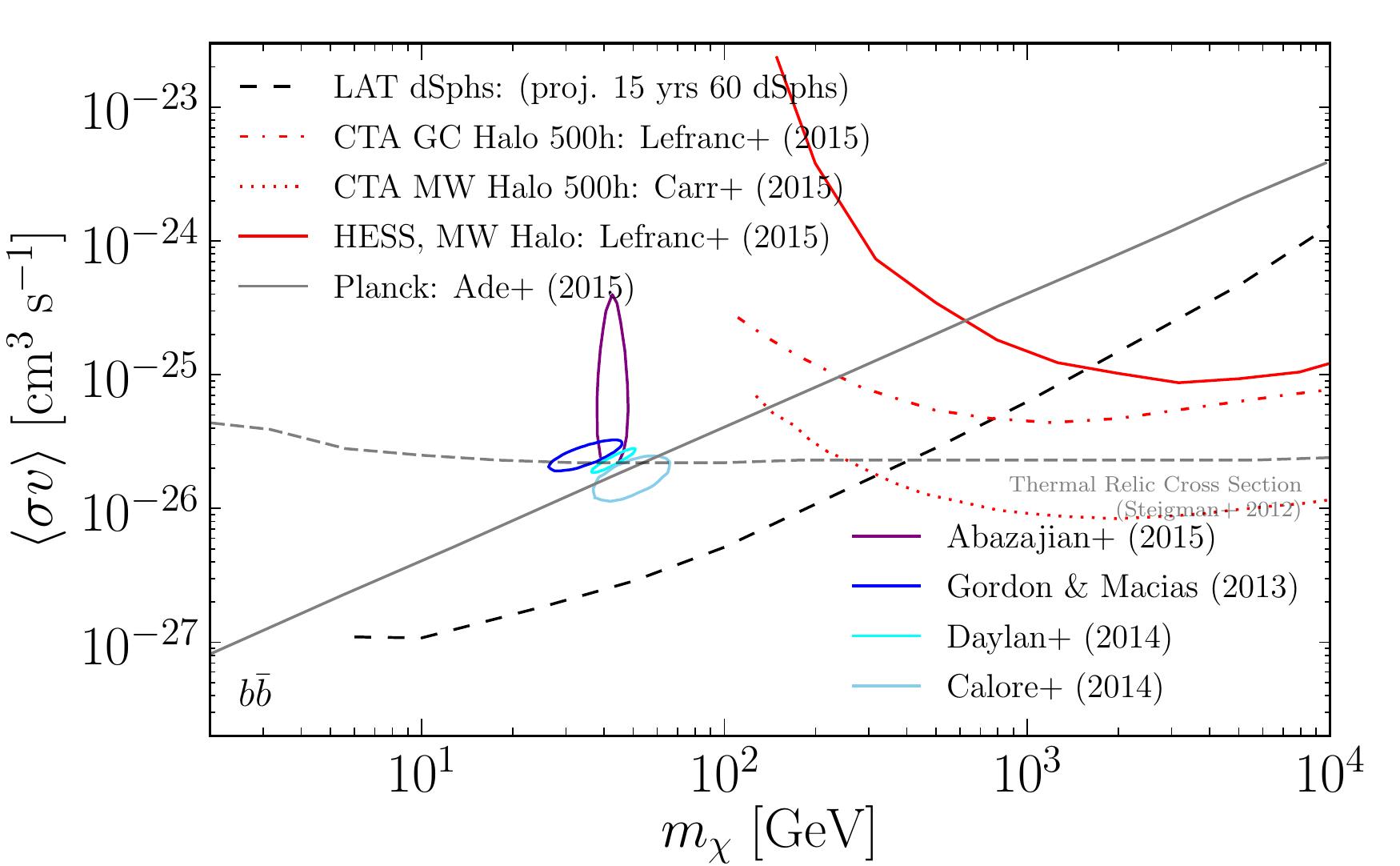}
    \includegraphics[width=0.49\columnwidth]{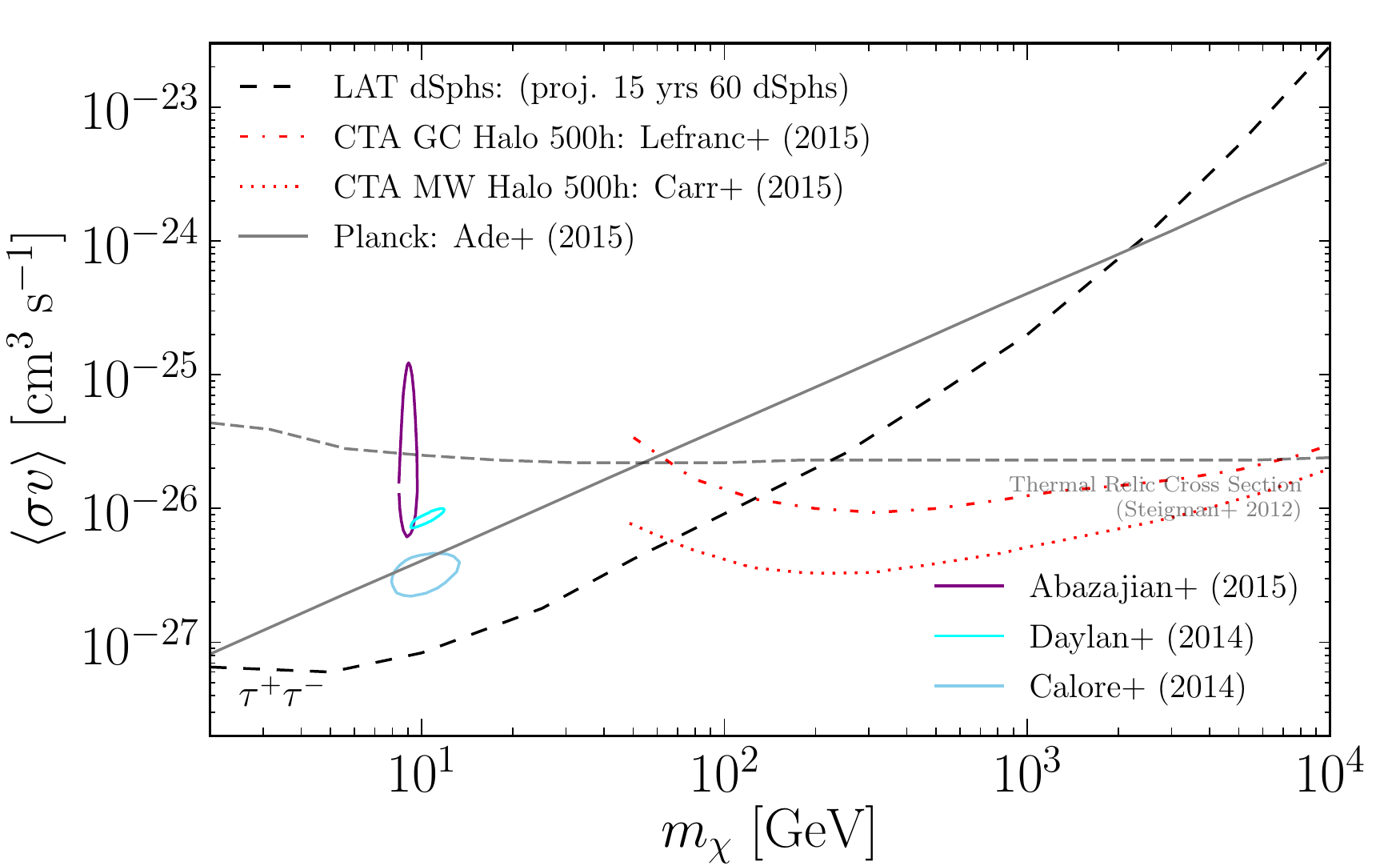}
  \end{center}
\caption{Comparison of projected dSph stacking limits with current and
  future IACT limits from CTA for the \bb (left) and \tautau (right) channels.  The dashed black curve
  shows the expected limit from the analysis of the artificially expanded target described in
  \S\ref{sec:dm_satellites_projections} for the 15-year data set.  IACT limits are in red and
  taken from~\cite{2015arXiv150904123L,2015JCAP...03..055S}.   The limits derived from the
  {\it Planck} data \cite{2015arXiv150201589P} are in gray.  Finally, favored contours 
  for several Galactic-center analyses are included for comparison.\label{fig:gce_iact_limits}}
\end{figure}

In principle, fluxes of cosmic-ray anti-particles (in particular
positrons, anti-protons, and also possibly anti-deuterons or
anti-helium) could be a signature
of DM interactions.   Although anti-particles can be produced in 
astrophysical accelerators, the fluxes and spectra expected from such 
accelerators are quite different from the expectations for DM interactions.
However, comparing results from cosmic-ray measurements from instruments such as PAMELA and AMS-02
with results from \gray data is complicated as the constraints on
the DM annihilation are dominated by systematic modeling
uncertainties.  As an example, the measurement of the
ratio of anti-protons to protons, $\Phi(\bar{p})/\Phi(p)$, could in
principle be used to probe cross sections below the thermal relic
level.  In practice, however, the constraints based on cosmic-ray
data have large modeling uncertainties and 
are quite model dependent (see Figs.~\ref{fig:cosmic_ray_errors}
and~\ref{fig:cosmic_ray_limits}).

\begin{figure}[!htbp]
  \begin{center}
    \includegraphics[width=0.50\columnwidth]{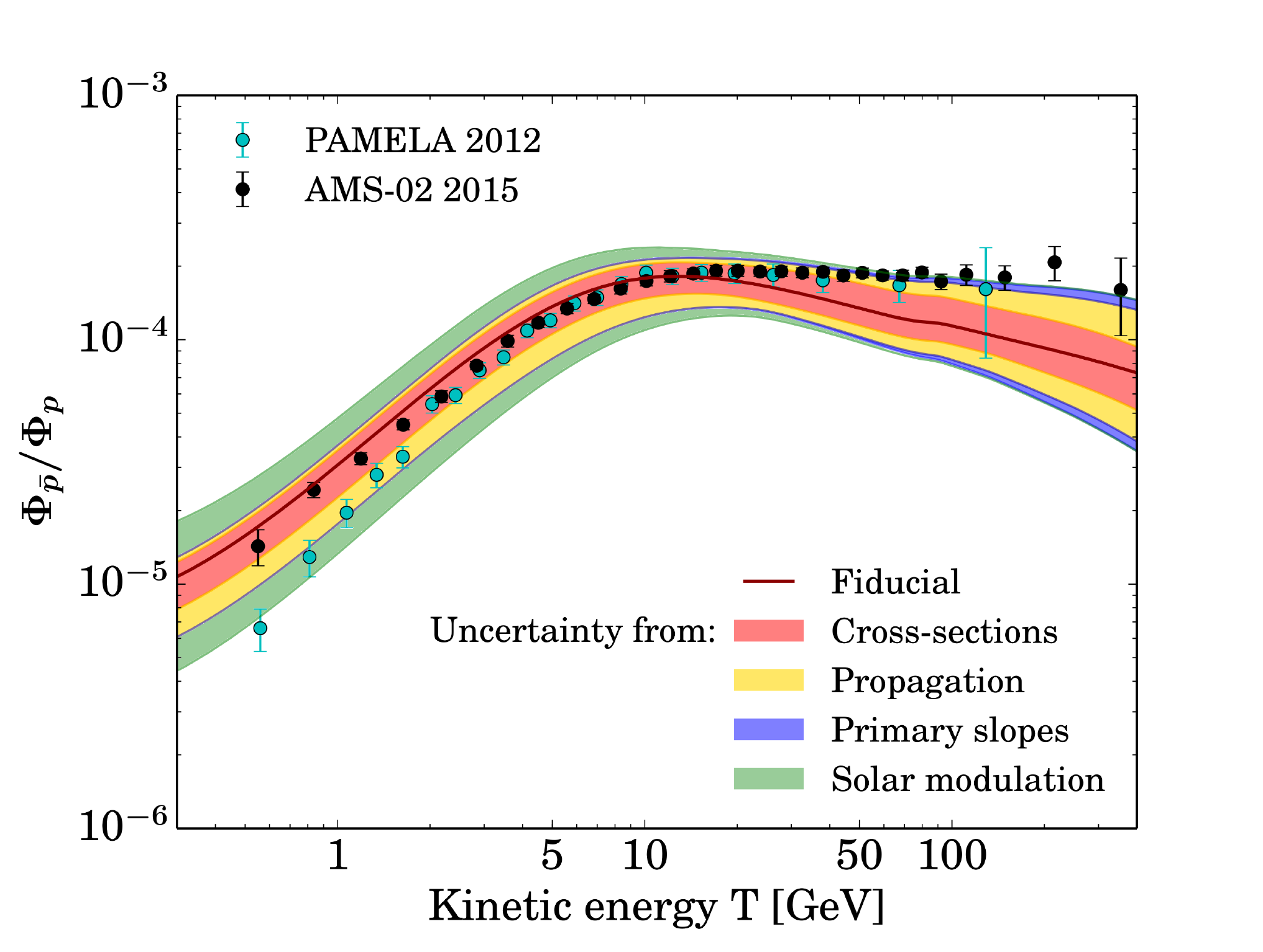}
  \end{center}
\caption{Combined total uncertainty on the predicted
secondary $\bar{p}/p$ ratio, superimposed on the PAMELA~\cite{2010PhRvL.105l1101A}
and AMS-02~\cite{AMS_Talks,2015PhRvL.114q1103A} data.   This figure
appeared as Fig.~2 of Ref.~\cite{2015JCAP...09..023G}; additional details about the
uncertainty bands may be found in that work;  reproduced under the
Creative Commons attribution license.
\label{fig:cosmic_ray_errors}}
\end{figure}

\begin{figure}[!htbp]
  \begin{center}
    \includegraphics[width=0.50\columnwidth]{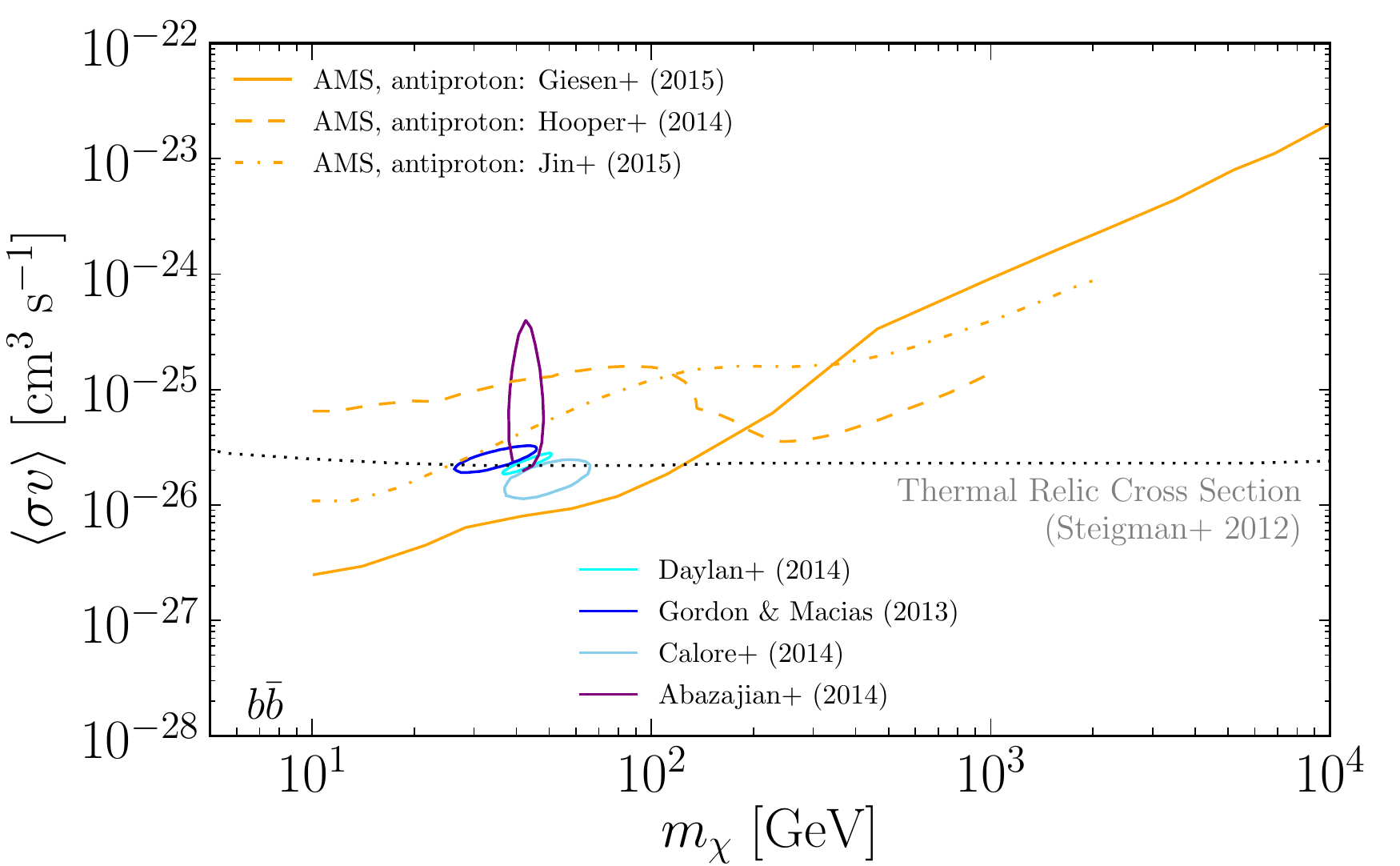}
  \end{center}
\caption{Comparison of three sets of WIMP annihilation cross-section limits derived from the
AMS-02 data for $\Phi(\bar{p})/\Phi(p)$, but with different assumptions regarding the comic-ray 
behavior.\cite{2015JCAP...09..023G,2015JCAP...03..021H,2015PhRvD..92e5027J} 
Favored contours for several Galactic center analyses are included for comparison.
\label{fig:cosmic_ray_limits}}
\end{figure}

Similarly, the ratio of positron to electron fluxes has been
measured by the LAT~\cite{2012PhRvL.108a1103A},
AMS-02~\cite{Aguilar:2013qda,Accardo:2014lma} 
and PAMELA~\cite{2013PhRvL.111h1102A} and
is potentially sensitive to DM interactions.   The observed positron
to electron flux ratio rises steadily from \roughly 5\% at 1\GeV to
\roughly 15\% above 100\GeV, suggesting the injection of high-energy positrons into the
interstellar medium.   Similarly to the situation with anti-protons, 
the interpretation of the rising positron fraction and implied constraints on
DM annihilation are dominated by systematic modeling
uncertainties, see, e.g., Refs~\cite{2009PhRvD..79b1302S,2013ApJ...772...18L,2013PhRvD..88b3013C,{2014PhRvD..89f3539I}} for
discussion of the interpretation of the positron excess.

In summary, the LAT data, and in particular the analysis of the dSphs
provide the best current constraints on the indirect detection of WIMP 
annihilation over a wide range of masses and channels, \roughly 15\GeV to 1.6\TeV
(\roughly 10 to 400 \GeV) for the \bb (\tautau) channel.
Furthermore, the constraints will continue to improve with additional
data taking.  

No instruments currently under development will improve markedly on the 
LAT sensitivity to \grs in the 1\GeV to 50\GeV energy band that
effectively sets the sensitivity to WIMP annihilation in the \roughly
10\GeV to 500\GeV mass range.   For 15~years of LAT data taking
and estimates of the CTA sensitivity that include systematic uncertainties,  
we project that the LAT would set the deepest constraints on \sigmav 
up between 300 and 700\GeV (between 80 and 120\GeV) in the \bb (\tautau) channel 
for the foreseeable future.

\subsection{Role of Indirect-Detection Searches for Dark Matter }
\label{sec:disc_indirect}

Indirect-detection, direct-detection, and production (i.e., collider) searches for DM have different advantages.
The complementarity between the methods
has been reviewed in a number of recent works~\cite[e.g.,][]{2013arXiv1310.8621A,2013arXiv1311.0299G,2015PDU.....7...16B}.
As discussed in \S\ref{sec:dm_theory}, indirect-detection searches  
measure the rate of DM annihilation (or decay) into Standard Model particles.   
When combined with estimates of the DM density, these observations can be used
to estimate or constrain the DM annihilation cross-section averaged over the velocity distribution, 
\sigmav, (or decay lifetime,
$\tau_{\chi}$).

Direct-detection searches measure the rate of scattering of DM particles with 
Standard Model target particles, typically by measuring the recoil of nuclei 
(although some experiments are also sensitive to electron
recoil, e.g., ~\cite{2015arXiv150707747T,2015arXiv150707748T}).
With knowledge of the scattering target, and estimates of the DM density 
and velocity distribution, these measurements can be used to estimate or 
constrain the scattering cross-section of DM particles with Standard Model 
particles, $\sigma_{{\rm SM},\chi}$.   Direct-detection searches are
typically $10^3$ to $10^4$ times more sensitive to spin-independent
couplings (where the DM particle coherently interacts with the
entire nucleus) than to spin-dependent couplings 
(where the DM particle effectively scatters off a single unpaired nucleon).

Collider searches look for the missing energy and momentum 
carried away by stable DM particles that were produced in the collisions, 
but escape the detector volume without interacting.
Because stable DM particles are not seen in the detectors, most searches at colliders 
do not directly measure the cross-section to produce the DM particles by
colliding Standard Model particles, $\sigma_{\rm SM SM \to \chi\chi}$, but rather the cross-section 
to produce heavier exotic particles that then cascade down to stable DM final state 
particles before reaching the detector, $\sigma_{\rm SM \to exotic}$. 
  
Additionally, the flux of ultra high-energy neutrinos from the 
Sun is sensitive to the DM-proton scattering cross-section (also $\sigma_{{\rm SM},\chi}$, which determines 
the rate at which DM is captured by the Sun).  For values of 
DM annihilation cross-sections near $\sigmav = 3 \times 10^{-26}
\cm^3\second^{-1}$ the DM capture rate would be in equilibrium 
with the DM annihilation rate.   The density of DM in the sun
would be dependent on \sigmav, but the observable neutrino flux
would not.  The limits on the spin-dependent scattering cross-sections
inferred from limits on ultra high-energy neutrino fluxes from the sun
are competitive with the best direct-detection measurements~\cite{2015PhRvL.114n1301C,2015APh....62...12A,2016JCAP...04..022A}.

There is broad consensus that the three approaches probe
different and complementary regions of the DM parameter 
space~\cite{2013arXiv1310.8621A,2013arXiv1311.0299G,2015PDU.....7...16B}.
Because they measure very different 
observables, direct comparisons between them depend on 
a number of assumptions about the nature of DM particles
and their interactions.  These kinds of comparisons are most
often presented in the $m_\chi$, $\sigma_{{\rm SM},\chi}$ space, which is most
appropriate for discussing direct-detection results.  
In Fig.~\ref{fig:dm_complemetarity} we present a comparison
in $m_\chi$, \sigmav space using a representative set of assumptions 
to convert the direct-detection and collider observables into 
constraints on \sigmav.

\begin{figure}[!htbp]
  \begin{center}
    \includegraphics[width=0.49\columnwidth]{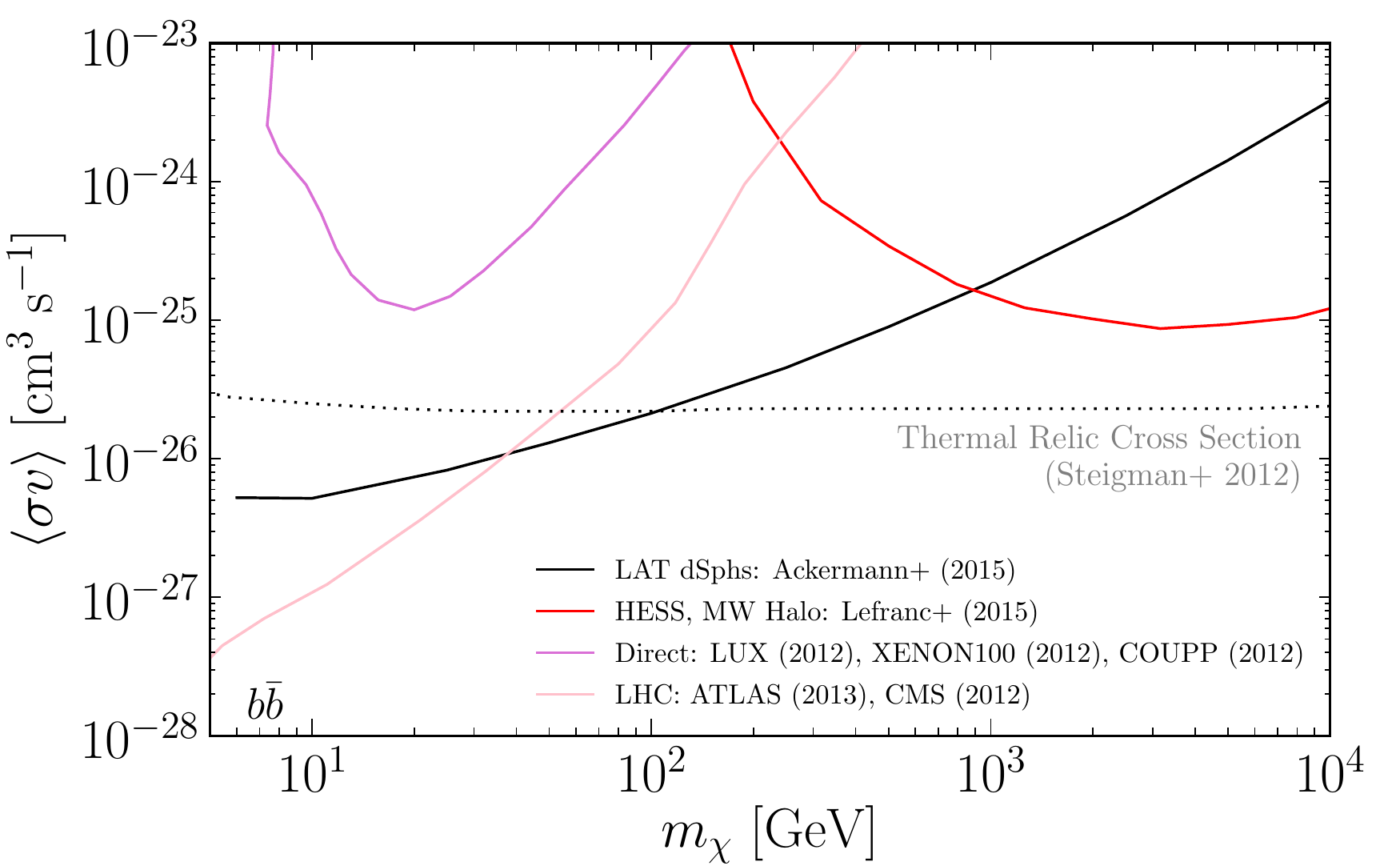}
    \includegraphics[width=0.49\columnwidth]{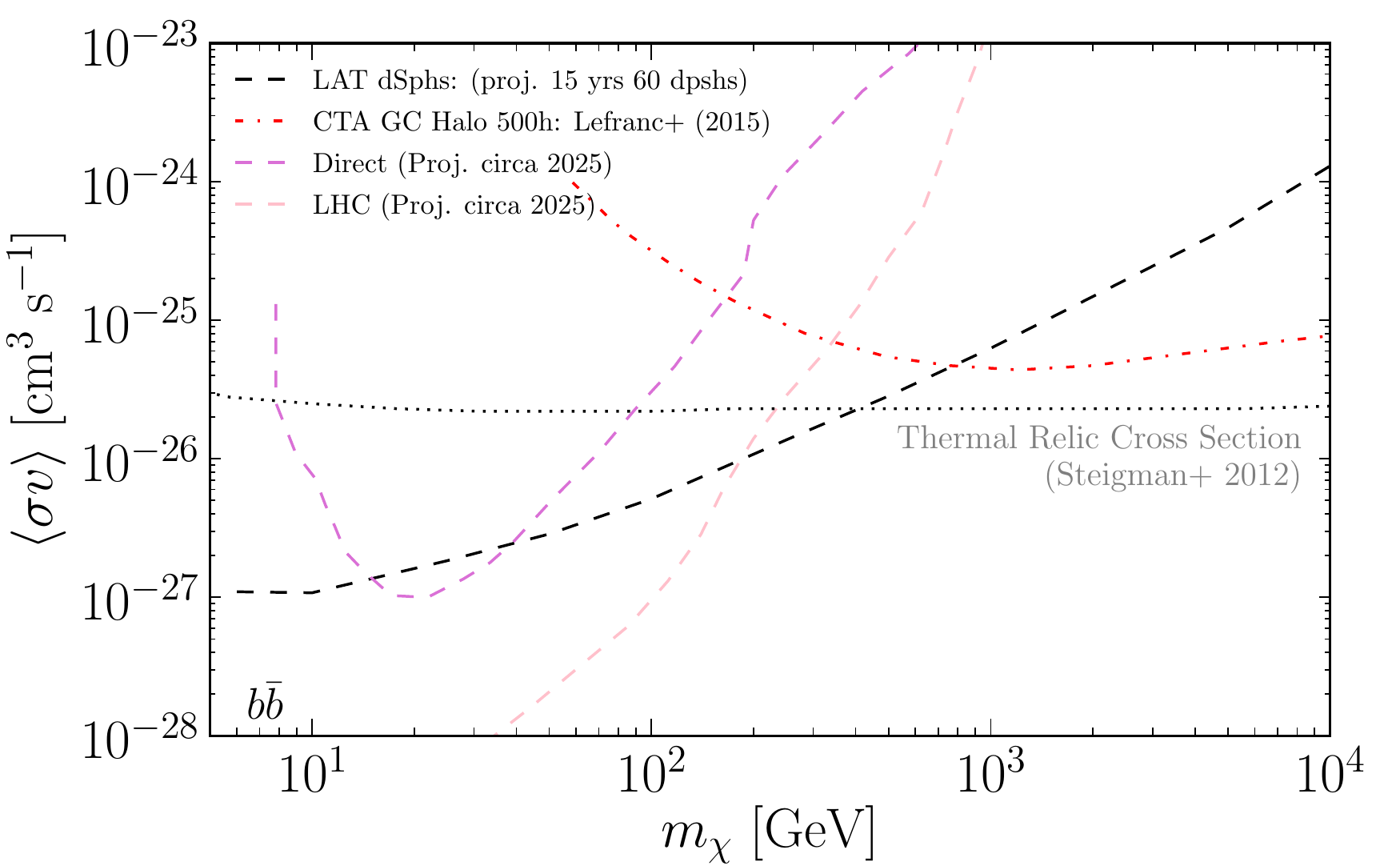}
  \end{center}
\caption{Comparison of best current (left) and projected (right)
  indirect-detection, direct-detection (spin-dependent) and collider-production limits
  on \sigmav in the \bb channel.  Conversion of direct-detection and collider limits to the \sigmav 
  scale is based on the assumption of four particle contact interactions for the production/annihilation of DM.  As noted in the text, this
  assumption is {\em quite uncertain}  (potentially by orders of
  magnitude) and the comparisons shown here should be considered
  schematic.   The current IACT limits are taken from Ref.~\cite{2015arXiv150904123L}.
  Following Ref.~\cite{2013arXiv1310.8621A}, the
  direct-detection limits are taken from Ref.~\cite{COUPP}, and the projection was
  made using the expected LZ sensitivity.  The collider-production limits are
  taken from Refs.~\cite{2013JHEP...04..075A,2012JHEP...09..094C}, 
  and the projection was made for $300\fb^{-1}$ of data at 13\TeV \cite{2010PhRvD..82k6010G}.
  \label{fig:dm_complemetarity}}
\end{figure}

Conversions from $\sigma_{{\rm SM},\chi}$ and $\sigma_{\rm SM \to exotic}$ to \sigmav (i.e., direct-detection and collider 
results to indirect detection results respectively) require assumptions 
about the particle interactions.  In Fig.~\ref{fig:dm_complemetarity} we show results 
derived using an Effective Field Theory (EFT) framework developed for the 
Community Summer Study (Snowmass) 2013 by the Cosmic Frontier Working Groups~\cite{2015PDU.....7...16B}.
The EFT framework is a simple and fairly model independent assumption; however, it
implies that the dark matter interacts with the Standard Model via a four-particle contact interaction. 
The interaction particle mass must be much larger than the transfer of momentum of the physical process of interest. 
Additional details on the representative operators and the model parameter space can also be found 
in~\cite{2010PhRvD..82k6010G, 2011PhLB..695..185G}.

In the context of searches for particle DM, we summarize key points regarding indirect detection.

\begin{itemize}
\item{Indirect-detection searches look for DM where it has already been determined to exist, i.e., 
	in large-scale astrophysical objects. A discovery of a new particle at the LHC would require follow-on 
	studies of astrophysical data to confirm that it does in fact account 
	for the observed properties of DM. Furthermore, if the mass and cross-section of the DM 
	particle were to be measured at the LHC, indirect searches 
	could be used to directly measure 
	the distribution of the DM particle in astrophysical objects. The results 
	from indirect-detection experiments, given a potential candidate from direct-detection 
	experiments, would measure 
	the distribution in the cosmos.} 
\item{The canonical thermal relic cross-section of $\sigmav \sim
    	3\times10^{-26}$ cm$^3$ s$^{-1}$ applies to an entire class of DM particle
    	candidates.  
	Testing for signals at that cross section tests the entire class of
        models that predict such candidates.}
\item{Indirect-detection searches
        generally have better sensitivity for high DM particle masses,
        (see, e.g., Fig.~\ref{fig:dm_complemetarity}).  
        Compared to collider searches, which must use the beam energy to produce 
        exotic particles that then decay to DM particles, indirect-detection searches 
        target final state particles with masses $\sim 0.1 m_{\chi}$ from the interactions of 
        already existing DM particles. Similarly, direct-detection searches lose sensitivity
        when the mass of the DM particle is larger than the mass of the target nuclei.}
\item{Indirect-detection searches have the additional benefit that, in the
        event of a detection, they measure the mass of the DM
        candidate.  If the DM mass is large, an indirect-detection
	measurement could determine the energy reach needed for the next collider, or
        influence the choice of the best nuclear target for direct detection.}
\item{LHC limits on new physics do not translate directly to limits on \sigmav
        (Fig.~\ref{fig:dm_complemetarity}).  Inferred cross-section limits
    	vary by orders of magnitude depending on assumptions about
    	the coupling and the mediator of the DM annihilations.
    	For example, unless a specific model is imposed (e.g., Supersymmetry), the
    	collider limits are based on an effective field theory. This requires the total 
    	energy of the event to be much lighter than the mediator for the field theory 
	to be valid ~\cite{2010JHEP...09..037B}.}
\item{Relative to direct-detection and collider searches, indirect-detection searches
    	generally have better sensitivity for DM particles coupling to leptons. 
    	Direct-detection searches are typically more sensitive to nuclear
        recoil than electron recoil (though see, e.g.,
        \cite{2015arXiv150707747T,2015arXiv150707748T}),
        and the highest-energy lepton collider (LEP2) had a maximum center of mass energy of $\sim$200 GeV. }
\item{Direct and indirect searches probe complementary DM interactions
    with the Standard Model.  In some models, the scattering cross
    section is velocity suppressed while the annihilation cross
    section is not, making indirect searches more 
    sensitive for those models~\cite{2011PhLB..695..185G}.}
\end{itemize}

% LocalWords:  dSph IACT CTA dSphs DM collider LMC systematics R16 ROI Einasto
% LocalWords:  IGRB GeV AGN EGB unassociated TeV IACTs CMB AMS complementarity
% LocalWords:  LHC Supersymmetry LEP2 LZ colliders observables deuterons Axions
% LocalWords:  Snowmass ALPs

\section{Summary}\label{sec:summary}

In this paper we have examined current strategies for 
searching for DM signals in the LAT data, presenting
existing search results and projecting how their sensitivities
will improve with additional LAT data.   

Although many LAT search targets are becoming systematics- or background-limited with the several years 
of LAT data already obtained, the most promising targets are still
signal limited in much of the LAT energy range and 
for these cases the LAT sensitivity scales as better than $\sqrt{t}$.
  
\begin{enumerate}
  \item{The most promising target, the dSphs
      (\S\ref{sec:dm_satellites_status}), are signal-limited 
      above \roughly 500\GeV (\roughly 100\GeV) in the \bb
      (\tautau)  channel.  Furthermore, the searches targeting
      the dSphs have already set the most constraining limits on DM
      signals.   Thus we are in the fortunate situation that the best
      target will also provide the most improvement as additional data
      are analyzed.}
    \begin{enumerate}
    \item{Ongoing and planned deep, optical surveys are expected to discover numerous 
        additional dSphs in the coming years, possibly by the hundreds once LSST is operational.}
    \item{With 15 years of data the LAT sensitivity for the dSph search would allow
        excluding the thermal relic cross section for masses up to
        $> 400\GeV$ ($> 200\GeV$) in the \bb (\tautau) channels.}
    \item{Projections also indicate that with 15 years of data the sensitivity
        for the dSphs would reach below $2\times 10^{-27} \cm^3\second^{-1}$
        for masses around 50\GeV in the  \bb channel, 
        allowing the DM interpretation of the
        Galactic center excess to be confirmed or refuted.}    
    \item{With 15 years of data the volume over which
        the LAT would be able to detect 100\GeV DM annihilation to \bb
        from a typical dwarf galaxy at $5\sigma$ significance would approximately double
        from $\sim 2100\kpc^3$ presently to $> 4200\kpc^3$.}
    \end{enumerate}
  \item{The second approach for which the LAT sensitivity will continue to improve faster
      than $\sqrt{t}$ is in constraining the population of the highest
      $J$ factor Galactic DM subhalos (\S\ref{sec:dm_unid}).   Although
      the source detection sensitivity will improve
      only as $\sqrt{t}$, combined with an estimated $J$~factor distribution
      that scales as $dN/dJ \propto J^{-3/2}$, the sensitivity to
      Galactic DM subhalos scales as $t^{3/4}$.}     
  \item{The third approach for which the LAT is, in effect, signal limited is in
      characterizing the populations of sources that are 
      currently below the detection threshold. By reducing the uncertainties on estimates of the
      contributions of unresolved sources to the EGB, and thus 
      reducing the uncertainty on what fraction of the EGB can
      be accounted for from the known \gray emitting source populations, 
      limits on cosmological DM annihilation will tighten accordingly.
      We estimate that both the fraction of the EGB not attributable to
      detected sources and the uncertainties of the total contribution
      of known source classes will decrease as $t^{3/4}$ or better.}
  \item{A fourth search strategy for which the LAT results are largely signal limited is the measurement of the
      attenuation of emission from high-energy AGN.   Observation of
      even a few \grs from these sources with energies above the optical depth limit for $\gamma\gamma$ attenuation on the EBL could be 
      evidence of photon-axion oscillations.}
  \item{A fifth approach for which the LAT is  purely signal limited is in the observation
      of spectacular transient phenomena (such as Galactic supernovae, or bright, ultra-distant GRBs)
      that can also be used to search for signals from photon-axion oscillations.} 
  \end{enumerate}    

Indirect detection is the only approach that searches for DM 
in the astrophysical targets where it is known to exist.
The LAT is uniquely exploring an important range of cross-section-WIMP 
mass space, with a cross section sensitivity that is commensurate with 
constraining theories, or indirectly detecting dark matter at theoretically 
motivated cross sections.  The advances expected with continued LAT
data taking will be significant and will not be superseded 
by another experiment in the foreseeable future.   Furthermore, the variety of search 
strategies possible with LAT data allows consistency cross checks, which are important, e.g., for
interpreting the GeV excess near the Galactic center.

% LocalWords:  DM systematics dSphs LSST dSph subhalos EGB AGN EBL axion GRBs
% LocalWords:  GeV

\section*{Acknowledgments}

The \textit{Fermi} LAT Collaboration acknowledges generous ongoing support
from a number of agencies and institutes that have supported both the
development and the operation of the LAT as well as scientific data analysis.
These include the National Aeronautics and Space Administration and the
Department of Energy in the United States, the Commissariat \`a l'Energie Atomique
and the Centre National de la Recherche Scientifique / Institut National de Physique
Nucl\'eaire et de Physique des Particules in France, the Agenzia Spaziale Italiana
and the Istituto Nazionale di Fisica Nucleare in Italy, the Ministry of Education,
Culture, Sports, Science and Technology (MEXT), High Energy Accelerator Research
Organization (KEK) and Japan Aerospace Exploration Agency (JAXA) in Japan, and
the K.~A.~Wallenberg Foundation, the Swedish Research Council and the
Swedish National Space Board in Sweden.

Additional support for science analysis during the operations phase is
gratefully acknowledged from the Istituto Nazionale di Astrofisica in
Italy and the Centre National d'\'Etudes Spatiales in France.

\newText{Brandon Anderson and Manuel Meyer have been supported by  
a grant of the Knut and Alice Wallenberg foundation, PI : Jan Conrad.
Miguel S\'anchez-Cond is a Wenner-Gren Fellow and acknowledges the 
support of the Wenner-Gren Foundations to develop his research.}

We thank Daniel Hooper and Bridget Bertoni for useful
discussions about projecting the sensitivity for searches targeting
potential dark satellites among the LAT unassociated sources.  

We thank Leslie Rosenberg and Gray Rybka for providing the 
contours from the ADMX axion search used in preparing Fig.~\ref{fig:alps}.

We thank Tim Tait and William Shepherd for their discussions on 
dark matter complementarity with direct-detection experiments and collider searches. 

We thank David Williams and Emmanuel Moulin for discussions on the
CTA sensitivity and for providing data used in Fig.~\ref{fig:gce_iact_limits}.

Finally, we thank Francesca Calore and Matthieu Schaller for preparing
Fig.~\ref{fig:lat_gc_radial} based on Fig.~4 in~{\cite{2016MNRAS.455.4442S}.

% Appendices
\appendix

\section{Dark Matter Density Profiles}
\label{app:dark_matter_profiles}

In the studies described in this paper we used several models of the spatial 
distribution of the DM.   A generalized Navarro-Frenk-White (gNFW) profile~\cite{1998ApJ...502...48K},
given by 
\begin{equation}
\label{eq:gNFW}
  \rho(r) = \frac{\rho_0}{(r/r_s)^{\gamma}(1+r/r_s)^{3-\gamma}}
\end{equation}
models adiabatic contraction in the core of the
distribution, i.e., within the scale radius $r_s$.  
Typical values of $\gamma$ for the gNFW profile are in the range 1.0 to 1.4. 
The $\gamma = 1$ case gives the canonical NFW
profile~\cite{1996ApJ...462..563N}.   
The Einasto profile~\cite{2010MNRAS.402...21N} is defined as:
\begin{equation}
  \rho(r) = \rho_0 \exp\{ -(2/\alpha)[(r/r_s)^\alpha - 1]\},
\end{equation}
(we typically set $\alpha = 0.17$), and is similar to gNFW profiles.  
The Burkert profile~\cite{1995ApJ...447L..25B} is defined as
\begin{equation}
  \rho(r) = \frac{\rho_0}{[r/r_s + 1][(r/r_s)^2 + 1]},
\end{equation}
and is much flatter inside the scale radius, and is similar
to a cored, isothermal profile given by
\begin{equation}
  \rho(r) = \frac{\rho_0}{1+(r/r_s)^2}.
\end{equation}
For practicality, we typically terminate the profile at some distance $r_{\rm max}$ that
is several times larger than $r_s$.   The values of $(r_s,r_{\rm max},\rho_0)$ depend
on the target in question.   For the Milky Way halo we set $\rho_0$
such that the local DM density (at $8.5\kpc$ from the Galactic Center)
is $0.4\GeV\cm^{-3}$.

\section{Calculating Gamma-ray Spectra from WIMP Annihilation}
\label{app:dark_matter_spectra}

Owing to the cascade of particles produced as unstable DM annihilation
products decay or hadronize, the \grs from DM annihilation in
most channels do not feature a sharp line at $E_\gamma = \mchi$, but rather a continuous 
spectrum with characteristic energies significantly lower than the DM
mass.  Indeed, most indirect searches for DM do not involve searches for \gray spectral
lines from 
DM annihilating directly into \grs, but rather consider the DM annihilation channels:
\begin{equation}
\chi\chi \to s\bar{s},~\bb,~t\bar{t},~gg,~W^-W^+,~\ee,~\mumu,~\mbox{and}~\tautau. \label{eq:channels}
\end{equation}
Annihilation into pairs of $u$ or $d$ quarks produces a similar spectrum 
as annihilation into gluon pairs, $c\bar{c}$ is similar to $s\bar{s}$,  
and $ZZ$ to $W^-W^+$, so bounds on such channels can be extrapolated approximately 
from the subset of channels we analyze in detail.   Channels of DM annihilating 
to massive particles are open only above the mass threshold, i.e., when the DM particle
mass is equal to that of the heavy Standard Model particle in the final state.
In our searches we typically scan over DM masses between 5\GeV and 10\TeV, though the range
depends somewhat on the search target (e.g., for spectral lines we
have scanned from 100\MeV to 500\GeV).   

For each final state, we calculate the spectrum of \grs
per annihilation as a function of DM mass (i.e., $dN_\gamma/dE_\gamma$)
using the {\tt DMFIT} software package~\cite{Jeltema:2008hf}.
We note that this formulation does not yet include electroweak 
corrections~\cite{2007PhRvD..76f3516K,2002PhRvL..89q1802B,2009PhRvD..80l3533K,2010PhRvD..82d3512C,Ciafaloni:2010ti}.  
The electroweak corrections are expected to be important (assuming s-wave annihilation) when the DM mass is heavier than 1\TeV, 
and for this mass range would alter the spectra substantially for the $W^+W^-$, \ee, \mumu and \tautau 
channels, increasing the number of expected \grs per DM annihilation 
below \roughly 10\GeV~\cite{Ciafaloni:2010ti,Cirelli:2010xx}.   However, the bounds on DM annihilation fluxes in 
the high-mass regime come primarily from the highest-energy \grs, which are statistics-limited.  Even for 10\TeV dark 
matter masses in the most-affected channels, including the electroweak corrections 
improves the limits on \sigmav by $\lesssim 20$\%
for typical analyses \newText(see e.g., \cite{Cirelli:2010xx,Buch:2015iya,2015PhRvD..91j2001B})}.

\section{The \FermiLAT and LAT Data}
\label{sec:lat}

The LAT is a pair-conversion telescope that converts \grs
to $e^{+}e^{-}$ pairs that are tracked in the instrument.   The data 
analysis is event-based; individual pair conversion events are reconstructed and 
their energies and directions are estimated from the reconstructed data.  
Rates of charged-particle backgrounds exceed the \gray rates by
factors of up to $10^4$, requiring powerful event selection criteria 
to obtain relatively pure \gray samples.

The LAT consists of three detector subsystems:  a
tracker/converter to promote pair conversion and measure the
directions of the resulting particles, a calorimeter composed of 8.6
radiation lengths of CsI(Tl) scintillation crystals that provides an
energy resolution of $\Delta E / E \sim 10\%$ at 100\GeV, and an
anti-coincidence detector of plastic scintillator tiles that
surrounds the tracker and is key in charged-particle background rejection.   
The tracker comprises 18 $x$-$y$ layers of silicon-strip detectors; 
the front 12 layers are interleaved with thin (3\% radiation
length) tungsten converter foils and the next four layers are interleaved with
thick (18\% radiation length) foils, and the final two layers have no converter foils.  
Detailed descriptions of the LAT and of its performance can
be found elsewhere~\cite{Atwood:2009ez,2012ApJS..203....4A}.

Iterations of the LAT event reconstruction and classification algorithms 
have been grouped into so called ``Passes''.  A number of iterations occurred 
before launch and were informed by simulations and beam test data of prototypes.  
For the first three years of the \Fermi mission, data were processed with the last pre-launch versions
(``Pass 6'').
Since then, the data have been reprocessed three times.  
The first was with ``Pass~7''~\cite{2012ApJS..203....4A}, which
consists of the same event reconstruction algorithms, 
but updated event classification criteria to account for knowledge 
gained about the LAT performance and charged-particle backgrounds 
since launch.  The second reprocessing, designated ``Pass~7
Reprocessed'' or ``Pass~7Rep'', used almost exactly the same 
algorithms as Pass~7 but incorporated updated calibration
constants.   The most recent reprocessing has been with 
``Pass~8'', which included entirely new reconstruction algorithms that were 
designed to account for effects observed in on-orbit data that had not
been anticipated in simulations and affected the performance of the
LAT, in particular the pile-up of out-of-time signals into event readouts~\cite{2013arXiv1303.3514A}.
Pass~8 also included new reconstruction algorithms that had not been
fully implemented prior to launch, such as improved pattern
recognition in all three sub-systems~\cite{2013arXiv1303.3514A}, and 
improvements to the energy fitting algorithm
that push the upper edge of the LAT energy range up to 3\TeV~\cite{2012JPhCS.404a2033B}.

Each reprocessing has included all of the data from the entire mission
to date.  The LAT Collaboration released the 
entire \gray data sets, and corresponding analysis files (see below), for each,
to present a single, coherent, data set. 

Each Pass of the algorithms implements several different event selection criteria that 
are optimized for different types of analyses.  For Pass~8, the LAT Collaboration 
redeveloped the nested event selections that provide varying levels of cosmic-ray 
background rejection~\cite{2013arXiv1303.3514A}.  The nomenclature convention
for the various event selections is to provide the Pass version and the name of
the event selection criteria (e.g., \evtclass{P8R2\_SOURCE}).   
Associated with each event selection are instrument response functions (IRFs)
that parametrize the LAT performance.   As our understanding of the
instrument improves, from time to time the LAT Collaboration updates
the IRFs for the various event selections.   The IRF names indicate
the analysis Pass they are associated with, as well as a version number
(e.g., \irf{P8R2\_SOURCE\_V6}).  More details about the event reconstruction, event 
selection criteria, and IRFs can be found in Refs.~\cite{2012ApJS..203....4A} and~\cite{2013arXiv1303.3514A}.
\footnote{Performance details for each of the iterations of the event reconstruction and classification
  algorithms used since launch are available at 
  http://www.slac.stanford.edu/exp/glast/groups/canda/lat\_Performance.htm}

At the time of writing, the most recent published versions of many LAT analyses were
performed with approximately six years of data, running from 2008 August to 2014 August.  
Throughout this paper we will specify both the Pass version and data interval used for 
the LAT results we discuss.   Most of the current results presented in this paper
were obtained with either the Pass~7 Rep or Pass~8 data sets.  Unless
stated otherwise, all of projections that we present were made with
simulations of the Pass~8 data using the corresponding IRFs.

% LocalWords:  calorimeter CsI scintillator P8R2 IRFs parametrize IRF V6 htm
% LocalWords:  7Rep

\section{Monte Carlo Simulations}
\label{app:method_monte_carlo}

We performed three types of Monte Carlo simulations to make the data
realizations we used to project the expected sensitivity of the DM
searches described in this paper.  In practice all three types of simulations 
give results that agree to better than a few percent and to well within the analysis
uncertainties.  The choice of simulation method is primarily dictated by 
convenience and technical details such as available computer processing time 
and data storage.  

\begin{enumerate}
\item{{\em All-sky photon simulations}: we used the {\tt
      gtobssim}\footnote{See, e.g., http://fermi.gsfc.nasa.gov/ssc/data/analysis/scitools/obssim\_tutorial.html
      for more details about the {\tt gtobssim} tool.} tool to
    simulate a total of 15 years of LAT data taking over
    the entire sky.  The {\tt gtobssim} tool generates individual \grs from each source
    in the model, applies accept-reject sampling to simulate the instrumental acceptance, 
    and finally convolves each \gr with parametrized representations of the 
    LAT spatial and spectral resolution.  We simulated data for most of the 
    DM targets described in \S\ref{sec:dm_targets}, as well as for models of all of the
    astrophysical backgrounds described in \S\ref{sec:backgrounds}.}

\item{{\em ROI-specific photon simulations}: we also used the {\tt
      gtobssim} tool to simulate a total of 15 years of \FermiLAT data for much 
    smaller regions covering specific targets, such as the individual
    dwarf spheroidal galaxies.  These simulations included the
    ``baseline'' astrophysical background model described in
    \S\ref{sec:backgrounds}, and in some cases also included the 
    specific model of the unresolved source populations.}

\item{{\em Binned model map simulations}: we used the {\tt
      gtmodel} tool to make predicted counts maps (i.e., maps of 
    the expected number of photons in each pixel) of small 
    regions of the sky.  We then simulated Poisson fluctuations with respect 
    to the predicted counts maps.  These simulations generally included only 
    the ``baseline'' astrophysical background model described in \S\ref{sec:backgrounds}.}
\end{enumerate}

\section{Modeling the Galactic Diffuse Emission}
\label{app:galactic_diffuse_model}

The construction of the Galactic diffuse emission model used for
these projections is described in considerably more detail in 
Ref.~\cite{2016ApJS..223...26A}.   Here we will briefly summarize some  
key points, and discuss the uncertainties that most directly affect
DM searches.

Determination of the distribution of interstellar gas relies mostly on the 21-cm line
of atomic hydrogen, which traces neutral atomic hydrogen, and on the 2.6-mm
line of carbon monoxide, which is a surrogate tracer of neutral
molecular hydrogen. The Doppler shifts of the lines due to the motion of gas
around the center of the Galaxy enables building quasi three-dimensional
models of the gas distribution~\cite[e.g.,][App.~B]{2012ApJ...750....3A}.  This method suffers from limitations
due to pile up along the line of sight of many structures, internal velocity dispersions of gas clouds,
systematic deviations from circular motion about the Galactic center, and from
kinematic ambiguity (for each velocity there are two possible
positions) for the region of the Galaxy within the orbit of the solar system.
For directions within about $10\degree$ longitude of the Galactic center or anti-center, 
the Doppler shifts due to Galactic rotation are smaller than the internal velocity 
dispersions and the distribution of gas cannot be determined via this method.

Additional interstellar gas components are the dark neutral medium,
neutral gas that is not traced properly by the combination of the
21-cm and CO lines, and ionized gas.  The dark neutral medium can be
traced by infrared emission from interstellar dust grains, since they
are sufficiently well mixed with neutral
gas~\cite{2015A&A...582A..31A,2005Sci...307.1292G}. 
Thermal emission from dust, however, does not provide information on the location along the
line of sight of the emitting gas. The distribution of ionized gas can
be evaluated approximately using tracers like dispersion measures of pulsar
emission, radio recombination lines (observations are available only
in limited regions of the sky), and free-free emission in the
microwave band (which does not carry distance information).

Target radiation fields span wavelengths from radio to
optical/UV. Aside from the CMB, the radiation
field is often estimated using numerical codes to calculate the
radiation transfer of stellar radiation~\cite[e.g.,][App.~C]{2012ApJ...750....3A}. The distribution of
optical/UV light is taken from simplified yet realistic models of the
stellar populations in the Milky Way informed by observations. From these stellar 
radiation fields, combined with models of the dust distribution and properties in the Milky Way (built
similarly to gas models described above), the numerical codes are used to derive the spectrum and
spatial distribution of the
reprocessed light in the IR domain.

The distribution of cosmic rays for the purpose of modeling
interstellar \gray emission has been determined mainly with two 
complementary strategies. On one hand, it can be calculated using
numerical codes that solve the cosmic-ray transport equations given a 
distribution of sources, injection spectra and transport parameters
that are informed by direct cosmic-ray measurements and observations 
of their putative sources at various wavelengths~\cite{2012ApJ...750....3A}.  On the other hand,
it is reasonable to assume that variations in the cosmic-ray nuclei 
spectra are mild on the characteristic scale of interstellar gas complexes 
due to the moderate energy losses in propagation, so that the \gray emission 
from their mutual interactions can be assumed to be proportional to 
gas column densities derived as discussed above, and appropriately 
split into regions along the line of sight using the Doppler shift information~\cite{2010ApJ...710..133A,2016ApJS..223...26A}.

An unexpected facet of Galactic interstellar emission that was
highlighted by the LAT data is the presence of localized enhancements
of cosmic-ray densities; i.e., of diffuse \gray emission on top 
of the large-scale, relatively smooth distributions of cosmic rays in the Milky Way that are evaluated by
the propagation codes or by the fit of gas templates to LAT data.
The most important example is the {\it Fermi} bubbles, two lobe-like structures 
seen on opposite sides of the Galactic center and extending up to
Galactic latitudes $\sim 50^{\circ}$~\cite{2010ApJ...724.1044S,2013A&A...554A.139P,2008ApJ...680.1222D}. 
Other excesses were seen along the Galactic
plane, for example one that was characterized as a cocoon of freshly-accelerated 
cosmic rays in the cavities of the interstellar medium in the
massive-star forming region of Cygnus X~\cite{2011Sci...334.1103A}. Such excesses are not
completely understood yet; thus their modeling is only purely
phenomenological based on the LAT data themselves.

\section{Likelihood Formalism and Effective Background}
\label{app:method_effective_background}

In LAT analysis we typically optimize our models to maximize the 
binned Poisson likelihood, defined as:
\begin{equation}\label{eq:binned_likelihood}
  \mathcal{L}({\boldsymbol \mu},{\boldsymbol \theta} | \mathcal{D}) = \prod_{k}
  \frac{\lambda_{k}^{n_{k}}e^{-\lambda_{k}}}{n_{k}!},
\end{equation}
which depends on the \gray data $\mathcal{D}$, signal
parameters ${\boldsymbol \mu}$, and nuisance ({\it i.e.}, background) 
parameters ${\boldsymbol \theta}$. The number of observed counts in
each energy and spatial bin, indexed by $k$, depends on the data
$n_{k}(\mathcal{D})$, while the model-predicted counts depend on 
the input parameters $\lambda_{k}({\boldsymbol \mu},{\boldsymbol
  \theta})$.  The likelihood function includes information about the 
observed counts, instrument response, exposure and model components. 
The nuisance parameters are the scaling coefficients and spectral 
parameters of the astrophysical backgrounds discussed in \S\ref{sec:backgrounds}.  

When projecting the performance of future searches it is important to
consider both statistical and systematic uncertainties.  It is also
important to distinguish between a search parameter space that is
background limited (i.e., where the sensitivity is dominated by
either statistical or systematic uncertainties of the astrophysical
backgrounds) from parameter space that is signal limited (i.e.,
where the sensitivity is dominated by the small expected signals).

A useful concept that addresses both these considerations is the 
``effective background'', i.e., the background that is correlated with the signal,
or is, so to say, ``under'' the signal.    

For a binned analysis, given normalized models of the signal and
background components: $P_{{\rm sig,}i}({\bf \mu})$ 
and $P_{{\rm bkg,}i}({\bf \theta})$, 
we can estimate the effective background by calculating the likelihood 
fit covariance matrix element for the signal size ({\it e.g.},
starting from Eq.~28 in Ref.~\cite{2011EPJC...71.1554C}) in the approximation
that the background is much larger than the signal, giving:
\begin{equation}
\beff = \frac{N}{ \left( \sum_{k}\frac{P_{{\rm sig,}k}^{2}({\bf \mu})}{P_{{\rm bkg,}k}({\bf \theta})} \right) - 1},\label{eq:beff}
\end{equation}
where the summation runs over all pixels in the ROI and all the
energy bins and $N$ is the total number of events in the ROI and energy range. 
This definition has a few useful properties.  First, if the only free
parameters in the fit are the overall normalizations of the signal and 
background components, then in the limit that the signal is much 
smaller than the background the statistical uncertainty on the number 
of signal counts will be $\delta \nsig \simeq \sqrt{\beff}$. 
Second, if the signal and background models are totally degenerate 
($P_{{\rm sig,}k}({\bf \mu}) = P_{{\rm bkg,}k}({\bf \theta})$ for all
$k$), then the term in the summation will be equal to 1 and \beff 
will diverge, indicating that we have little power to
distinguish signal from background. If this were the case, the
statistical errors for the likelihood fit would be extremely large, 
corresponding to an upper limit on the cross section sufficient 
to generate all of the observed events through DM
annihilation.  Finally, if the signal and background models differ 
significantly the term in the summation will be much greater than 
1 and \beff will be proportionally less than $N$. That is, the 
statistical uncertainty on the signal will correspond to an effective 
background that is much less than the total number of background 
events in the ROI.   The LAT Collaboration
has found empirically that by quantifying the systematic uncertainties 
of the astrophysical background modeling as a percentage 
of \beff, we are often able to account for those uncertainties in
the likelihood fitting procedure and include them in DM
constraints.

In the best-case scenario, where the background is well modeled, 
the expected uncertainty of the number is signal events is simply $\delta \nsig = \beff^{1/2}$.
We can then define the ratio of the signal to the
effective background as \fsig, and the estimate of the
statistical uncertainty $\delta \fstat$ in terms of the effective background:

\begin{align}
\fsig = & \frac{\nsig}{\beff}, \label{eq:fsig} \\
\delta \fstat = & \frac{\delta \nsig}{\beff} \simeq  \beff^{-1/2} \label{eq:fstat}.
\end{align}

\noindent Using the exposure, $J$ factor, and the per-annihilation \gray spectra
we can convert from $\delta \nsig$ to the uncertainty on the
thermally averaged annihilation cross section $\delta \sigmav$.
Furthermore, when the number of background events is large and 
well into the Gaussian regime, we can estimate the median 
upper limits ($1.645 \delta \sigmav$), the 68\% containment band 
($0.925 \delta \sigmav$ to $2.645 \delta \sigmav$) and the 95\% 
containment band ($0.590 \delta \sigmav$ to $3.645 \delta \sigmav$)
in terms of $\delta \sigmav$.   

In practical terms, we note that in many cases systematic biases 
could either induce a fake signal or mask a true signal much 
larger than the expected statistical uncertainties.
From the width of the distribution of \fsig for the trials in
control regions, where we do not expect to detect any signal, 
we can estimate the total (statistical + systematic) uncertainty.  
When $\fstat < \fsig$, the total error is dominated by
systematic uncertainties.  When our fitting procedure allows for
negative signals (such as in the searches for spectral lines) we can 
take a simple measure of the width such as the root-mean-square of the 
distribution.   However, in many cases our fitting procedure only allows for
positive signals and approximately half of the trials have $\fsig =
0$.   In those cases we assign as our estimate of the systematic 
error the difference (taken in quadrature) of the $1\sigma$ 
(84\% CL) enclosure of the total error estimate and the statistical 
error estimate from the control region:
\begin{equation}
\delta \fsyst^2 = \fsig^2(84\%) - \delta \fstat^2(84\%). 
\label{eq:fsyst}
\end{equation}

\section{Projecting the Uncertainty of Contributions to the
  Extragalactic Gamma-Ray Background}
\label{app:method_egb_sensitivity}

We have estimated the point source sensitivity map for 15~years of
observations by following the procedure described in the 2FGL
catalog~\cite{2012ApJS..199...31N}.  
A four year sensitivity map is shown in \S\ref{fig:detection_map}.

For 15~years of data, the average high latitude ($|b|>20\degree$) integrated energy flux 
detection threshold for a source with a photon index of $\Gamma = 2.2$ 
is $\sim 1.1\times10^{-9}\photon\cm^{-2}\second^{-1}$ ($>$100 MeV).

We have modeled the intrinsic source count distribution of blazars (i.e., 
distribution of the true fluxes), $dN/dS$, where
$S$ is the photon flux for $E>100$ MeV, of the 1FGL 
catalog~\cite{2010ApJ...720..435A}  as a broken power law with a flux 
break at $7\times10^{-8}\photon\cm^{-2}\second^{-1}$ and a slope above
(below) the break equal to 2.63 (1.64).  We modeled the fluxes 
of the individual blazars with single power laws with index $\Gamma$.
Given this flux distribution the predicted number of detected blazars
 with 15~years of data is about 3300.

In Fig.~\ref{fig:fluxindexdistr} we show the distributions of $S^2 dN/dS$ 
and photon index, $dN/d\Gamma$, for this sample of simulated blazars.

\begin{figure}[!htbp]
  \begin{center}
    \includegraphics[width=0.49\columnwidth]{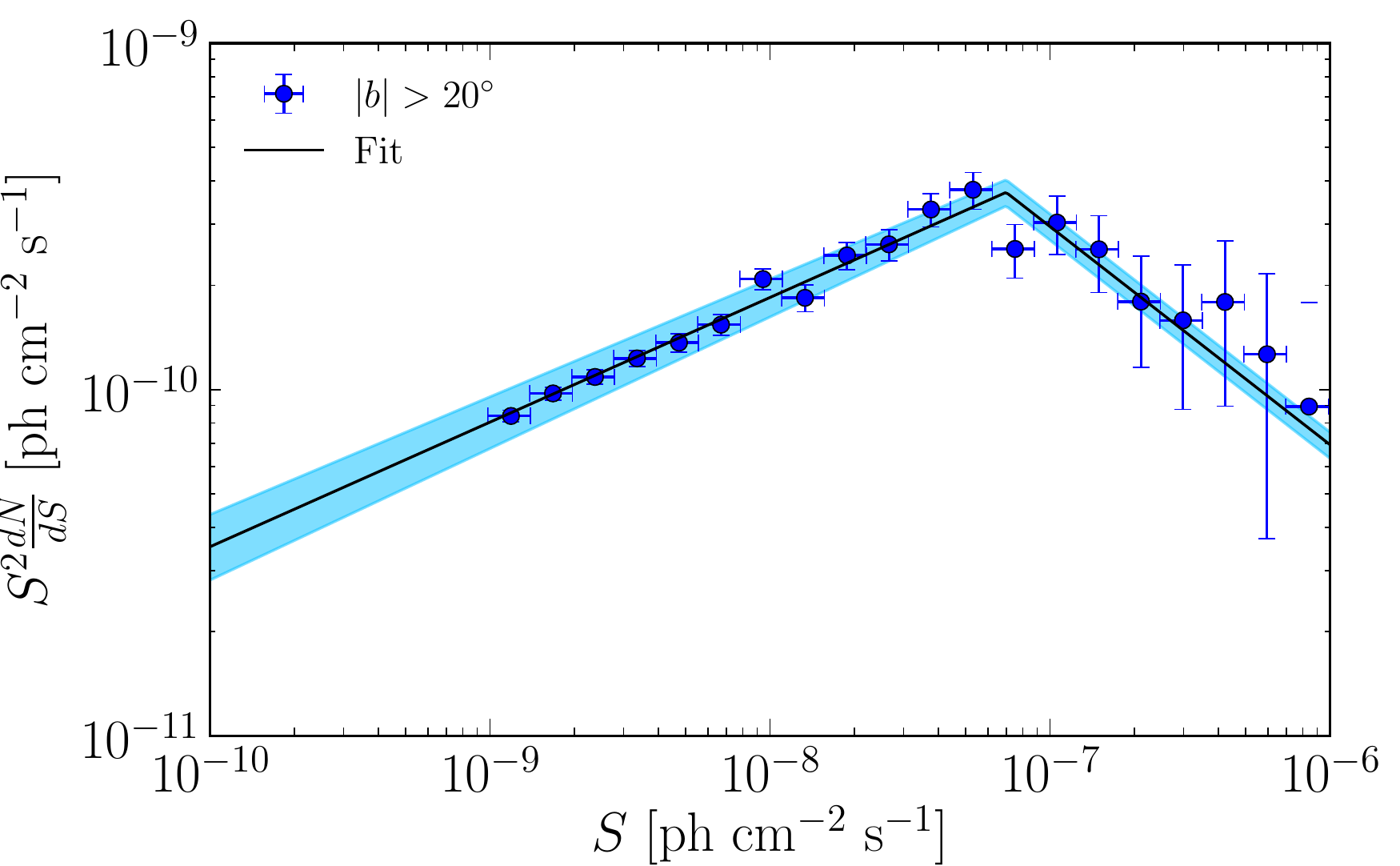}
    \includegraphics[width=0.49\columnwidth]{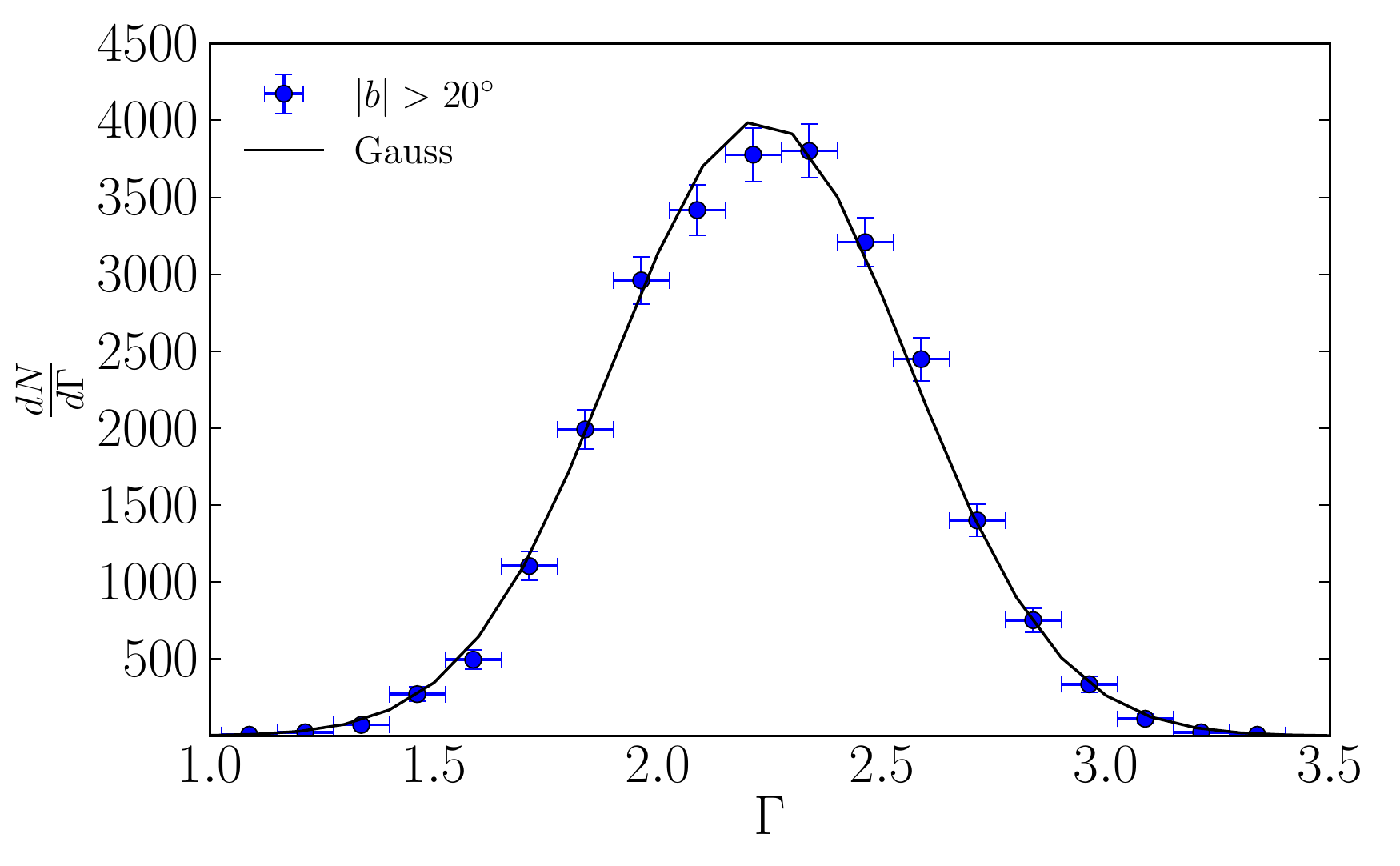}
  \end{center}
\caption{Left: Flux distribution of simulated blazars using the
  intrinsic $dN/dS$ of the 1FGL catalog~\cite{2010ApJ...720..435A}
  (blue points) together with a fit (black solid line) and the
  $1\sigma$ uncertainty band (cyan).  Right: photon index
  distribution of simulated blazars from the observed distribution of 
  sources in the 3FGL catalog~\cite{2015ApJ...810...14A}.  The
  black curve is the best-fit Gaussian distribution, with an average of 
  2.23 and a standard deviation of 0.33.}
\label{fig:fluxindexdistr} 
\end{figure}

We can then estimate the contribution of blazars, which 
are the most numerous source population in \gray catalogs 
(see, e.g., \cite{2015ApJ...810...14A}), to the EGB. 
Specifically, for each source (indexed by
$i$), the flux $S_i$ and index $\Gamma_i$ are sampled from the
distributions shown in Fig.~\ref{fig:fluxindexdistr} and the 
energy spectrum $dN/dE_i$ is derived from the definition 
of the of photon flux ($S=\int dN/dE dE$). Finally,
we sum of all of the simulated spectra and average over the solid
angle, $|b|>20^{\circ}$, to obtain the contribution of blazars to the
EGB.  The result for blazars is displayed in the left-hand panel
of Fig.~\ref{fig:astrocontr} where their average contribution to the
EGB is 65\% with an uncertainty of about 10\%.
\begin{figure}[!htbp]
  \begin{center}
    \includegraphics[width=0.49\columnwidth]{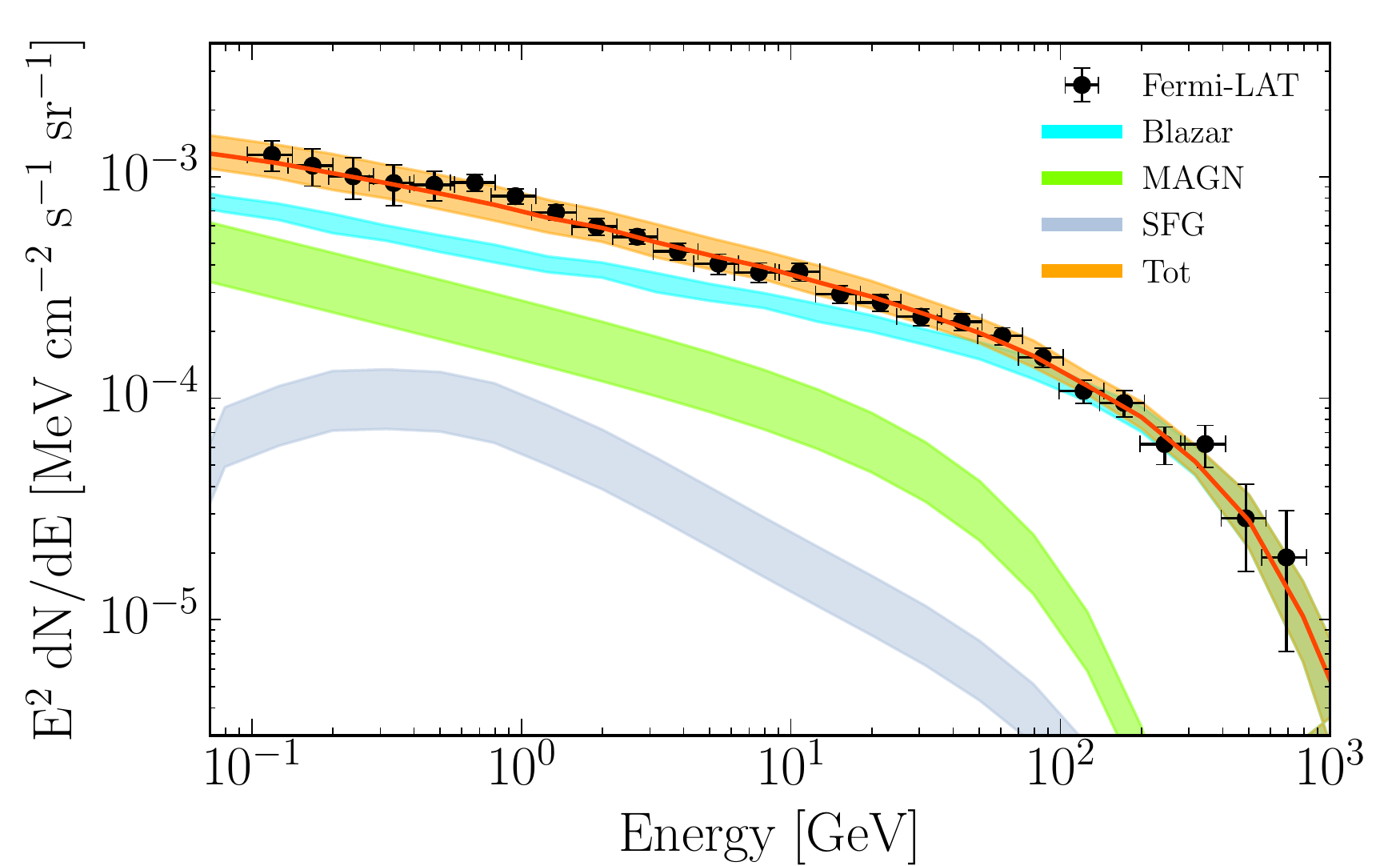}
    \includegraphics[width=0.49\columnwidth]{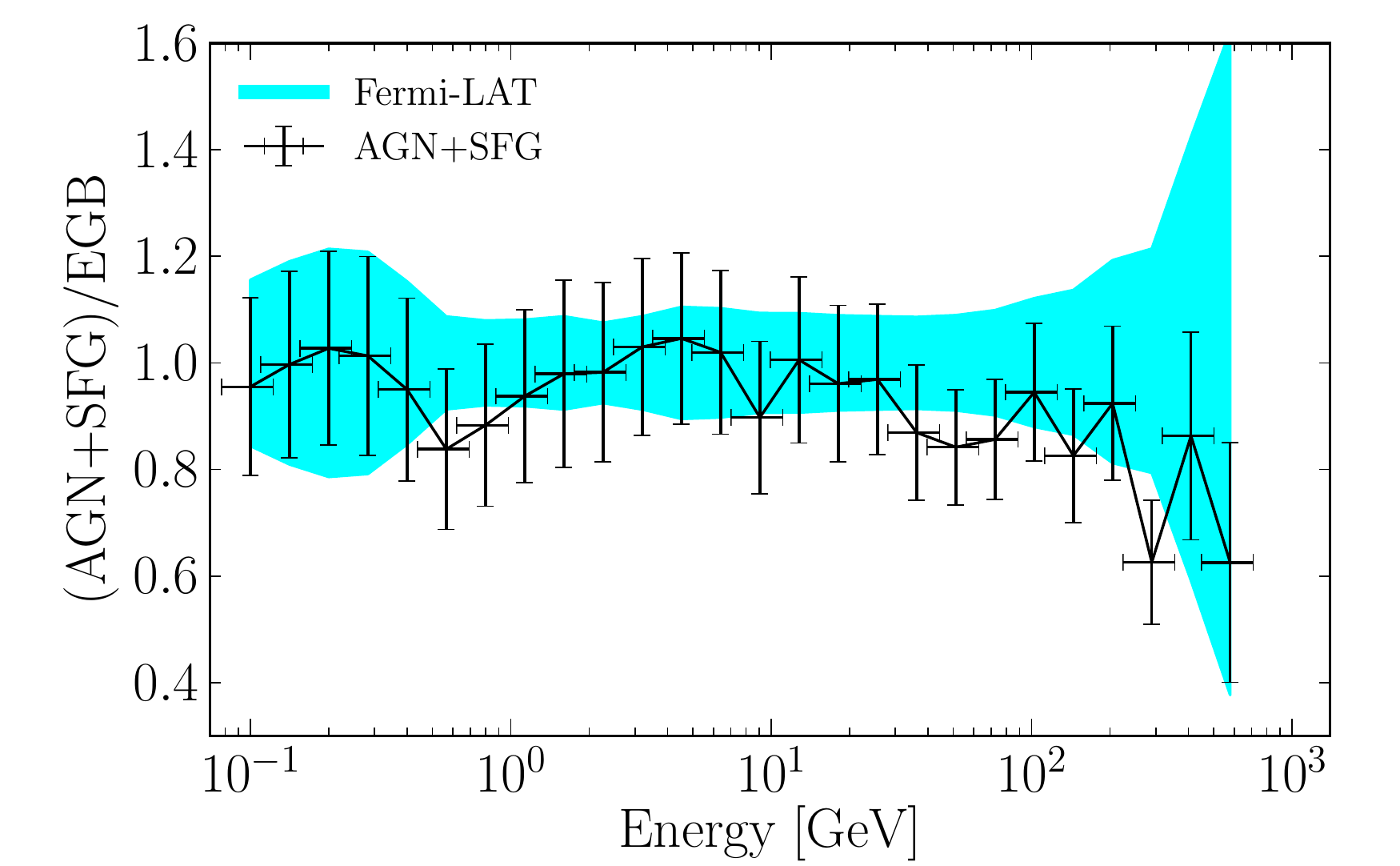}
  \end{center}
\caption{Left: contribution of blazars (cyan band), MAGN (green
  band) and SFGs (gray band) to the {\it Fermi}-LAT EGB data (black
  points).  The total best fit of the summed contributions from 
  AGN (blazars and MAGN) and SFG is depicted with
  a solid red line, with the orange uncertainties given by the orange band.
  Right:  fraction of the AGN plus SFG contribution 
  relative to the EGB (black points) together with the
  statistical EGB errors (cyan band).}
\label{fig:astrocontr} 
\end{figure}

We have followed the same procedure to estimate the uncertainty
for the MAGN contribution to the EGB.  We used the same intrinsic source
count distribution, but rescaling the number of sources 
with a factor of $1/50$, to match the ratio of MAGN to blazars
observed in the 3FGL catalog.  We predict a total of
$\sim$70 MAGN detectable in 15 years and we predict an uncertainty 
of the order of 30\% for the contribution of this source population to the EGB.
We consider the same level of uncertainty to apply to the contribution of 
Star-Forming Galaxies (SFG) as well.

The contribution of blazars, MAGN, and SFG populations to the EGB are
shown in Fig.~\ref{fig:astrocontr} (left panel).  The total
\gray emission from these source populations is consistent with 
the EGB measurement derived with 50 months of
data~\cite{2015ApJ...799...86A}  (see right panel of Fig.~\ref{fig:astrocontr}).
The uncertainties of the contributions of MAGN and SFG are a factor of
$\sim$4 smaller than the uncertainties of their contributions from the
analyses of the current LAT
catalog~\cite{Ajello:2015mfa,2015PhRvD..91l3001D} 
due to the increased number of detected sources with 15 years of data.

\section{Projecting the Sensitivity of the Angular Power Spectrum Analysis}
\label{app:method_aps_sensitivity}

In order predict to the sensitivity of the LAT to DM
signals in the APS we have used the same formalism as
in Ref.~\cite{DiMauro:2014wha}.  That work considered the APS
produced by the unresolved flux of populations of unclustered point 
sources, which is given by:
\begin{equation}
     \label{Cpdef}
        C_{\rm P}(E_0 \leq E \leq E_1) = \int_{\Gamma_{\rm min}}^{\Gamma_{\rm max}} d\Gamma \int^{S_{\rm t}(\Gamma)}_0 S^2 \frac{d^2N}{dS d\Gamma} dS,
    \end{equation} 
where $S$ is the photon flux of the source integrated in the range $E_0 \leq E \leq E_1$,
$S_{\rm t}(\Gamma)$ denotes the flux detection threshold as a function
of $\Gamma$ and $d^2N/(dSd\Gamma)$ is
the differential number of sources per unit flux $S$, unit photon index $\Gamma$ and unit solid angle.
We take $S_{\rm t}(\Gamma)$ to be constant in $\Gamma$
because we restrict the analysis to energies $ > 1\GeV$ 
where the bias for the source detection with respect to the photon
index is small.   We evaluate $S_{\rm t}(\Gamma)$ as the average value of the
point source sensitivity for 15 years of exposure at
$|b|>30^{\circ}$ and $E>1\GeV$:  $S_{\rm t} \approx 1\times10^{-10}\photon\cm^{-2}\second^{-1}$.
Using this model we have derived the $C_P(E)$ in the same energy 
bins as in Ref.~\cite{2012PhRvD..85h3007A}.  
The results are displayed in Fig.~\ref{fig:Cp15years}.
\begin{figure}[!htbp]
  \begin{center}
    \includegraphics[width=0.50\columnwidth]{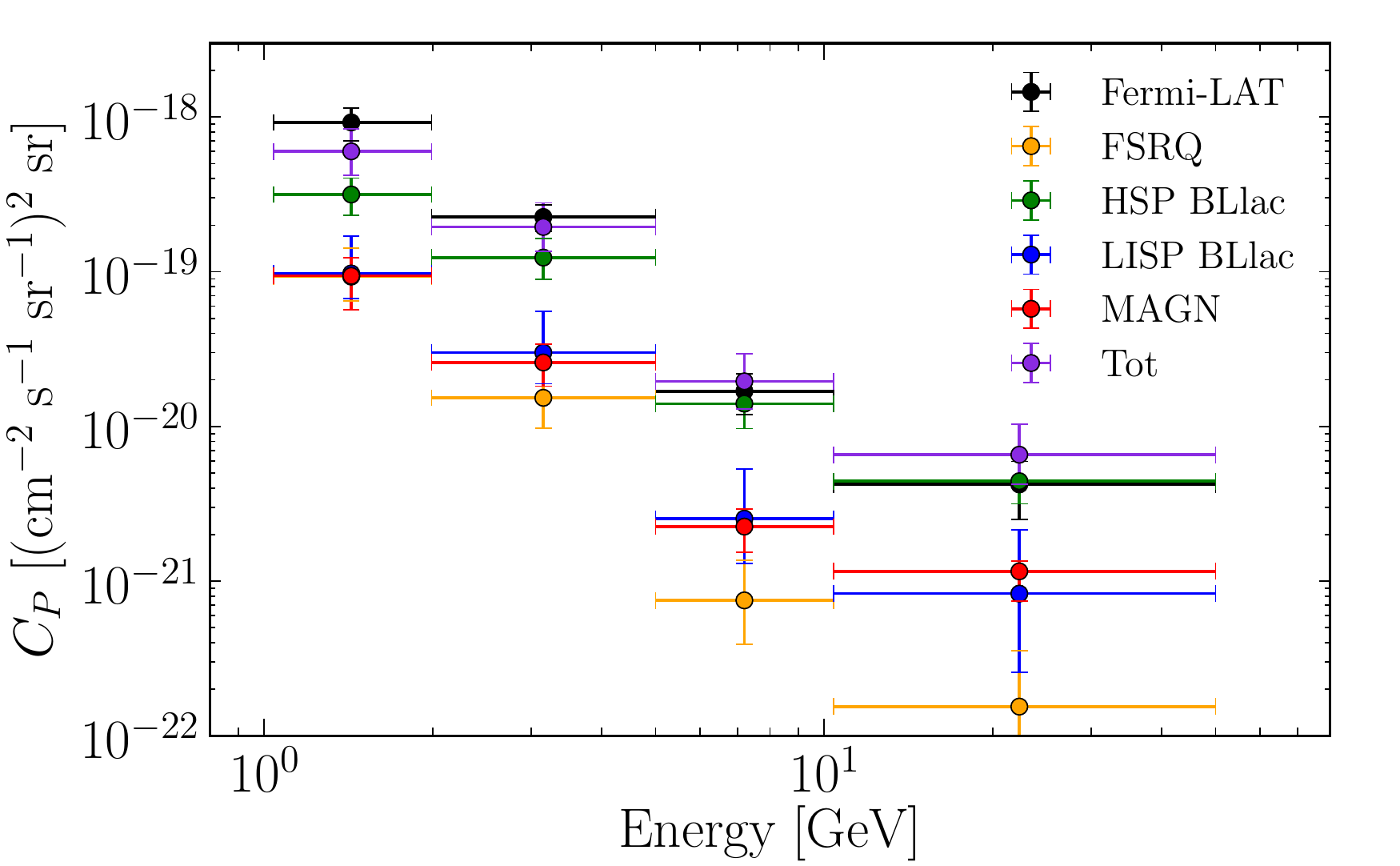}
  \end{center}
\caption{Angular power $C_P(E)$ for MAGN (red long-dashed points),
  LISP (blue short-dashed), HSP (green dotted), FSRQ (yellow dot-dashed), 
  and the total anisotropy (violet solid) from
  all the radio-loud AGNs. The data are a rescaled version, with a
  factor 1/5, of the measured APS in Ref.~\cite{2012PhRvD..85h3007A} (black solid points).}
\label{fig:Cp15years} 
\end{figure}
With 15 years of data the threshold flux is lower by about a
factor of 5 with respect to the threshold from two years of data, 
therefore the APS from unresolved AGN is also lower. 
However, different source populations have different intrinsic flux distribution, 
resulting in different scaling behaviors for $C_P(E)$.
The $C_P(E)$ for MAGN, for example, decreases by about a factor of
3, while for Flat Spectrum Radio Quasars (FSRQ) it decreases by a factor of 7. 
MAGN and low-intermediate synchrotron peak BL Lac objects
(LISP) have steeper source count distributions, below the flux threshold, 
than High-Synchrotron Peak BL Lacertae objects (HSP) and FSRQ.
Therefore, with lower flux thresholds the angular power for unresolved
MAGN and LISP are reduced by a smaller factor
than for HSP and FSRQ sources.  Since the $C_P(E)$ 
of astrophysical sources are lower we have rescaled the LAT data
by a factor of 1/5 in Fig.~\ref{fig:Cp15years} so as to 
match the theoretical predictions from AGN.  

We used the results shown in Fig.~\ref{fig:Cp15years} 
to project the LAT sensitivity to a DM contribution of
extragalactic subhalos in the case of a \bb annihilation channel.
We used the APS data in four energy bins~\cite{2012PhRvD..85h3007A} 
independently, and derived $95\%$ CL limits on \sigmav by requiring that the 
sum of the DM-induced intensity APS averaged in the multipole range $155 < l < 204$ 
and our prediction for the Radio Loud AGN contribution to $C_P$ not exceed the 
measured $C_P$ in the $155 < l < 504$ multipole range plus 1.64 times the uncertainty.  
We show the projected limits for each energy bin as well as the envelope produced by 
the limits for all four energy bins in Fig.~\ref{fig:Cp15years_limits} in S\ref{sec:dm_cosmo_aps_projections}.

% LocalWords:  DM gtobssim ROI gtmodel Eq sig Frenk gNFW NFW Einasto Burkert
% LocalWords:  isothermal hadronize gluon ScienceTools DMFitFuction html 2FGL
% LocalWords:  electroweak convolves dispersions normalizations Extragalactic
% LocalWords:  MeV blazars 1FGL 3FGL EGB blazar MAGN SFGs AGN SFG rescaling APS
% LocalWords:  unclustered synchrotron Synchrotron Lacs HSP FSRQ anisotropy CMB
% LocalWords:  rescaled Lacertae extragalactic subhalos multipole DMFIT 2FGL

% LocalWords:  2FGL 1FGL 3FGL AGNs

\section{List of Acronyms and Abbreviations}\label{app:acronym}

\acronym{1FGL}{First \Fermi-LAT Source Catalog}
\acronym{2FGL}{Second \Fermi-LAT Source Catalog}
\acronym{3FGL}{Third \Fermi-LAT Source Catalog}
\acronym{AGN}{Active Galactic Nucleus}
\acronym{ALP}{Axion-Like Particle}
\acronym{APS}{Angular Power Spectrum}
\acronym{CMB}{Cosmic Microwave Background}
\acronym{DES}{Dark Energy Survey}
\acronym{DM}{Dark Matter}
\acronym{dSph}{Dwarf Spheroidal Galaxy}
\acronym{EBL}{Extragalactic Background Light}
\acronym{EGB}{Extragalactic Gamma-ray Background}
\acronym{EFT}{Effective Field Theory}
\acronym{FSRQ}{Flat-Spectrum Radio Quasar}
\acronym{GRB}{Gamma-Ray Burst}
\acronym{IACT}{Imaging Atmospheric Cerenkov Telescope}
\acronym{ICM}{Intra-Cluster Medium}
\acronym{IGRB}{Isotropic Gamma-Ray Background}
\acronym{IRF}{Instrument Response Function}
\acronym{LAT}{Large Area Telescope}
\acronym{LHC}{Large Hadron Collider}
\acronym{LMC}{Large Magellanic Cloud}
\acronym{l.o.s.}{line-of-sight}
\acronym{LSS}{Large-Scale Structure}
\acronym{MAGN}{Misaligned AGN}
\acronym{PBH}{Primordial Black Hole}
\acronym{pNGB}{pseudo-Nambu-Goldstone Boson}
\acronym{ROI}{Region of Interest}
\acronym{SDSS}{Sloan Digital Sky Survey}
\acronym{SFG}{Star-Forming Galaxy}
\acronym{SMC}{Small Magellanic Cloud}
\acronym{WIMP}{Weakly Interacting Massive Particle}

% Bibliography
\bibliography{DM_wp}

\end{document}